\documentclass[conference,compsoc]{IEEEtran}

\usepackage[title=normal, margins=normal, paragraphs=normal, tracking=normal]{savetrees}
\usepackage{cite}
\usepackage{amsmath,amssymb,amsfonts}
\usepackage{textcomp}
\usepackage{xcolor}
\usepackage{siunitx}
\usepackage{booktabs}
\usepackage{adjustbox}
\usepackage{subcaption}
\usepackage{graphicx} 
\usepackage{url}
\usepackage[hidelinks]{hyperref}
\usepackage{caption}
\usepackage{comment}
\usepackage{tcolorbox}
\usepackage{enumitem}
\usepackage{nicefrac}
\usepackage{titlesec}

\usepackage{anyfontsize} 
\newcommand{\mytiny}{\fontsize{6}{7}\selectfont}

\usepackage{algorithm} 
\usepackage{algpseudocode} 
\algrenewcommand\algorithmicrequire{\textbf{Input:}}
\algrenewcommand\algorithmicensure{\textbf{Output:}}

\usepackage{soul,xcolor}

\usepackage{xspace}
\newcommand{\ppddd}{\emph{Pixel Patrol 3D}\xspace}
\newcommand{\ppabv}{\emph{PP3D}\xspace}
\newcommand{\ppdis}{\emph{Pixel Patrol Discover}\xspace}
\newcommand{\ppdisabv}{\emph{$\text{PP}_{dis}$}\xspace}
\newcommand{\ppdet}{\emph{Pixel Patrol Detect}\xspace}
\newcommand{\ppdetabv}{\emph{$\text{PP}_{det}$}\xspace}
\newcommand{\ppdef}{\emph{Pixel Patrol Defend}\xspace}
\newcommand{\ppdefabv}{\emph{$\text{PP}_{def}$}\xspace}

\begin{document}

\title{PP3D: An In-Browser Vision-Based Defense\\Against Web Behavior Manipulation Attacks}


\author{
  \IEEEauthorblockN{Spencer King\IEEEauthorrefmark{1} (sdk81722@uga.edu), 
                    Irfan Ozen\IEEEauthorrefmark{1} (irfanozen747@gmail.com), \\
                    Karthika Subramani\IEEEauthorrefmark{2} (ksubramani@gatech.edu),
                    Saranyan Senthivel\IEEEauthorrefmark{3} (ssenthivel2@dxc.com), \\
                    Phani Vadrevu\IEEEauthorrefmark{4} (kvadrevu@lsu.edu), 
                    Roberto Perdisci\IEEEauthorrefmark{1}\IEEEauthorrefmark{2} (perdisci@uga.edu)}
  \\
  \IEEEauthorblockA{\IEEEauthorrefmark{1}University of Georgia, Athens, GA, USA \quad
  \IEEEauthorrefmark{2}Georgia Institute of Technology, Atlanta, GA, USA \quad
  \\
  \IEEEauthorrefmark{3}DXC Technology, New Orleans, LA, USA \quad
  \IEEEauthorrefmark{4}Louisiana State University, Baton Rouge, LA, USA}
}


\maketitle

    
    

\begin{abstract}
    Web-based behavior-manipulation attacks (BMAs)---such as scareware, fake software downloads, tech support scams, etc.---are a class of social engineering (SE) attacks that exploit human decision-making vulnerabilities. These attacks remain under-studied compared to other attacks such as information harvesting attacks (e.g., phishing) or malware infections. Prior technical work has primarily focused on measuring BMAs, offering little in the way of generic defenses.
    
    To address this gap, we introduce \ppddd (\ppabv), the first end-to-end browser framework for discovering, detecting, and defending against behavior-manipulating SE attacks in real time. \ppabv consists of a visual detection model implemented within a browser extension, which deploys the model client-side to protect users across desktop and mobile devices while preserving privacy.

    Our evaluation shows that \ppabv can achieve above 99\% detection rate at 1\% false positives, while maintaining good latency and overhead performance across devices. Even when faced with new BMA samples collected months after training the detection model, our defense system can still achieve above 97\% detection rate at 1\% false positives. These results demonstrate that our framework offers a practical, effective, and generalizable defense against a broad and evolving class of web behavior-manipulation attacks.
\end{abstract}

\section{Introduction}
\label{sec:intro}


Social engineering (SE) encompasses a broad spectrum of attacks aimed at deceiving users into performing actions or divulging sensitive information that may have important negative consequences for the security and privacy of the users themselves or their organizations~\cite{mann2008hacking}. These threats exploit weaknesses in human decision-making processes by using tactics such as pretext, baiting, impersonation, etc.~\cite{KROMBHOLZ2015}. 

Within the broad spectrum of SE attacks, we focus on attacks that include a significant web-based component. In this context, we can distinguish between two classes of attacks: \emph{information harvesting} and \emph{behavior manipulation} attacks. Information harvesting attacks aim to extract sensitive information from users, such as login credentials, financial data, or personal details. 
Web-based phishing websites that steal login credentials~\cite{zhang2021crawlphish,subramani2022phishinpatterns} and smishing messages that attempt to get sensitive data such as Social Security Numbers are some examples for this class of SE attacks.
On the other hand, behavior manipulation attacks seek to influence users into pursuing insecure actions, rather than giving up information, often leading to malware infections or immediate financial harm such as money transfer to criminals. Examples of behavior manipulation attacks include scareware, fake software downloads, tech support scams, fake sweepstakes, etc.~\cite{FakeAVEconomy,miramirkhani2017,KharrazRK18,liu2023understanding}.

Unfortunately, while information harvesting attacks (IHAs) have received significant attention from researchers~\cite{PhishingDetectionSurvey,AfrozG11,VisualPhishNet,lin2021phishpedia,liu2022inferring,liu2023knowledge,liu2024less,li2024knowphish}, to the best of our knowledge no comprehensive defense framework against web-based behavior-manipulation attacks (BMAs) has been proposed or evaluated. Yet, BMAs remain a formidable SE attack class in financial terms. For instance, US consumers reported losing more than a billions dollars to tech support scams alone in 2024~\cite{fbi_int_crime_ann_report}. Prior works have shown that a large number of these scams originate from fake tech support websites which display glaring warning messages to coax victims into making telephone calls to scam call centers~\cite{miramirkhani2017,liu2023understanding}. 

Previous studies on web-based BMAs are either limited to discovering and measuring attack campaigns~\cite{Vadrevu_IMC19,WebPushAds} or to studying only specific subclasses of attacks~\cite{miramirkhani2017,FakeAVEconomy,KharrazRK18,liu2023understanding,ms_edge_scareware_blocker}, with limited or no concrete defense. Furthermore, existing practical defenses against malicious web pages mostly rely on reactive approaches that do not focus on the fundamental traits of BMAs and tend to lag behind new threats, thus leaving many users exposed to the latest attack iterations. For instance, URL blocklist services such as Google Safe Browsing (GSB)~\cite{GSB} focus on \emph{where} malicious content is hosted, rather than detecting and blocking the malicious content itself. Therefore, attackers are often able to evade blocking by simply relocating their attacks to a different hosting location, as reported by previous measurements~\cite{Vadrevu_IMC19}.

A recent work by Yang et al.~\cite{TRIDENT}, named TRIDENT, addresses the problem of detecting social engineering ads, which can in turn lead to behavior-manipulation attacks. However, TRIDENT narrowly focuses on detecting JavaScript code served by low-reputation ad networks that often use SE techniques, such as transparent overlays, to hijack users' clicks and further redirect them to a potentially malicious page. While TRIDENT is able to detect SE online ads, it can detect some BMA webpages only indirectly, because low-tier ad network code may or may not redirect to SE attacks. This can thus also generate false positives, namely legitimate (non-SE) web pages that may be blocked because a redirection originated from a low-tier online ad. Additionally, attacks that are not reached via malicious ads would be completely missed by TRIDENT, thus resulting in false negatives. Therefore, there is a need for a more generic defense that is able to directly detect web-based behavior-manipulation attack pages based on their content, rather than being limited to detecting malicious online ads that may in some cases redirect users to such attacks.



\vspace{3pt}
\noindent
\textbf{\em Problem Definition.}
\label{sec:problemdef}
Based on the taxonomy proposed by Zaoui \textit{et al.}~\cite{zaoui2024taxonomy}, we focus on {\em remote, psychologically triggered, web-based} social engineering attacks. Within this slice, we further distinguish between two classes of attacks based on their end-goal: 

\begin{description}[style=unboxed, leftmargin=0cm, labelindent=0cm]
  \item[\textbf{Information-Harvesting Attacks (IHAs):}] In IHAs, the adversary solicits secrets, such as credentials or sensitive personal identifiers. These attacks include traditional phishing of login credentials, banking accounts, tax-related information, etc., as well as modern crypto-currency phishing attacks.

  \item[\textbf{Behavior-Manipulation Attacks (BMAs):}] Unlike IHAs, in BMAs the adversary's primary goal is to induce an unsafe action, rather than directly harvesting information. For instance, downloading malware~\cite{FakeAVEconomy}, granting browser notification permissions to unwanted websites~\cite{WebPushAds}, calling a fraudulent tech support line~\cite{miramirkhani2017}, paying an advance fee for a bogus lottery or gift or for a fake service registration.
\end{description}

In this paper, our goal is to propose a {\em novel framework for in-browser defense against web-based behavior-manipulation attacks (BMAs) using visual webpage content analysis}. Specifically, we focus on detecting the following sub-categories of BMAs, which have been defined in past measurement studies~\cite{Vadrevu_IMC19}: \emph{fake software downloads, scareware, tech-support scams, notification stealing,} \emph{fake lotteries/sweepstakes}, and {\em service registration scams}. These BMAs have the following properties:


1)~\emph{Glaring visual deception.} BMA web pages rely on persuasive imagery to drive behavior. For instance, a notification-stealing site (Table~\ref{Tab:sepdif}) renders a fake video player behind a genuine permission dialog, claiming playback requires clicking \textit{Allow}, TSS pages embed counterfeit system windows to coerce phone calls~\cite{miramirkhani2017}, etc.  

2)~\emph{BMAs are orchestrated via attack campaigns.} BMA pages are often part of a larger campaign, where multiple attack pages are hosted under different domains. For example, a scareware campaign may include numerous scareware web pages that all use similar visual deception techniques (e.g., claiming that the machine is infected with multiple viruses) but are hosted on different websites to evade traditional URL-based defenses.

3)~\emph{Different from IHA.} Unlike information harvesting attacks, such as phishing pages, BMAs do not require mimicking a specific site/brand or harvesting credentials; they instead invent a believable scenario with the goal of manipulating users' actions, as shown in Table~\ref{Tab:sepdif}. Hence, phishing detectors, including recent systems that check for visual similarity to known benign pages~\cite{VisualPhishNet} or for input boxes~\cite{liu2022inferring,liu2023knowledge,liu2024less, li2024knowphish} are not suitable for detecting the behavior-manipulation attacks we focus on in this paper.

4)~\emph{Need for more effective BMA defenses.} Traditional defenses, such as URL block-lists~\cite{GSB}, are reactive and relocation-sensitive; measurements show many BMA pages remain active for days, before being blocked~\cite{Vadrevu_IMC19}. Also, previous initial work on BMA detection only attempt to defend against them indirectly, by blocking malicious online ads~\cite{TRIDENT}. Therefore, we need a content-based defense that can detect BMA pages directly, based on their attack content and regardless of their hosting location.

Motivated by these fundamental traits of BMA pages and the lack of existing solutions that address them comprehensively, in this paper we propose \ppddd: a novel framework for multimodal, in-browser detection of BMA pages. \ppddd consists of three components: a discovery module that identifies BMA campaigns, a multimodal detection model that analyzes the content of web pages to extract attack-related visual and text (via OCR) cues, and a browser extension that puts the detection model into practice, allowing it to defend web users from BMAs in real-time within the browser. Our experimental results show that \ppddd achieves a high detection rates even on previously unseen BMA campaigns (Table~\ref{Tab:table41}). Furthermore, our browser extension is able to detect and block BMA pages in real-time with limited overhead, providing users with a first practical content-based defense against these attacks.

\begin{table*}[ht]
  \captionof{table}{Visual comparison between Information-Harvesting Attacks (IHA) and Behavior-Manipulation Attacks (BMA). IHA pages mimic legitimate sites and solicit credentials, while BMA pages use visual deception to induce unsafe actions. We focus on developing a content-based defense against BMA pages, which are not addressed by existing defenses.}
  \label{Tab:sepdif}
  \centering
  \begin{adjustbox}{width=\textwidth}
  \small
  \begin{tabular}{@{}cccccc@{}}
  \toprule
  \rotatebox[origin=c]{90}{\textbf{IHA}} &
  \raisebox{-.5\height}{\frame{\includegraphics[width=.22\textwidth]{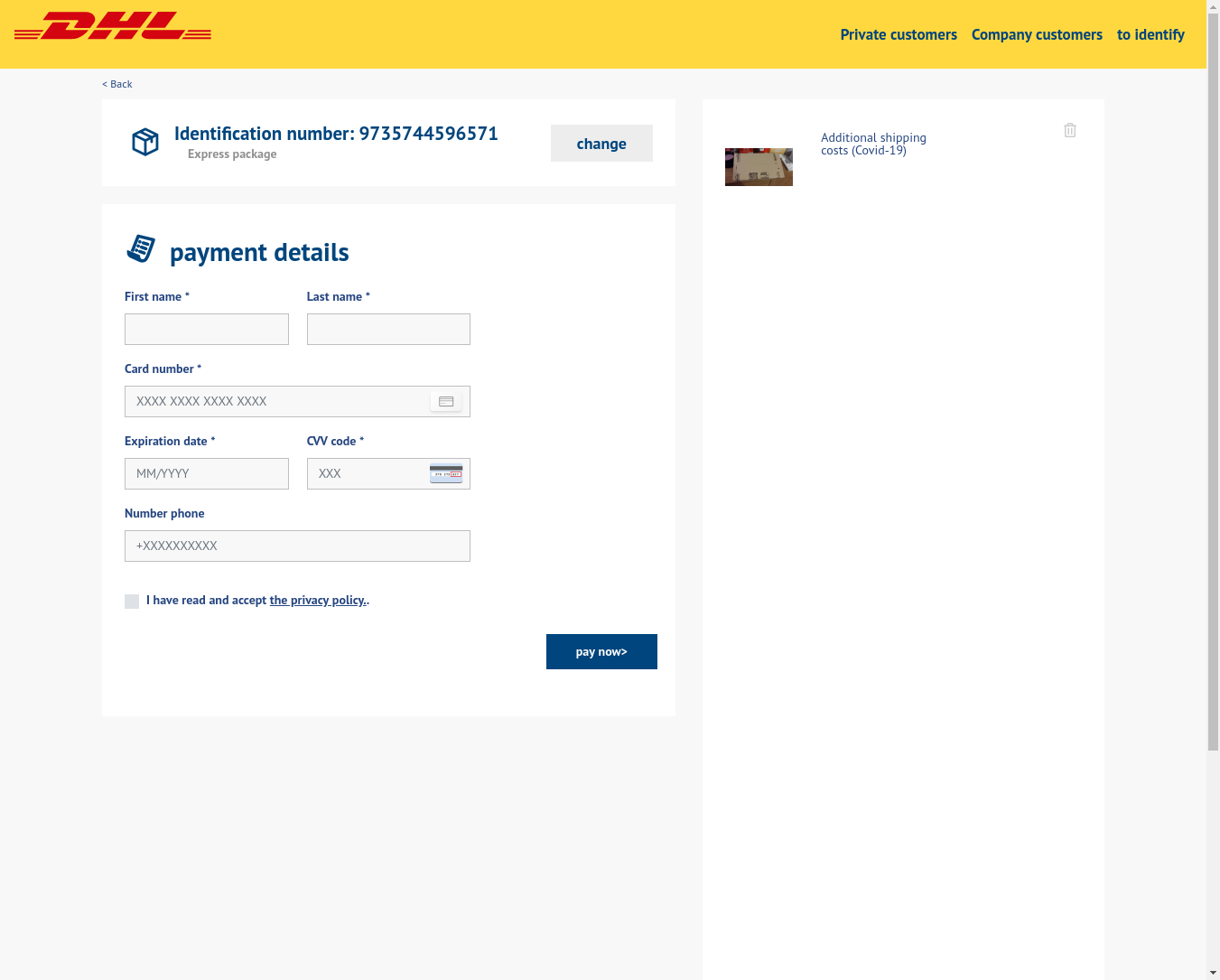}}} &
  \raisebox{-.5\height}{\frame{\includegraphics[width=.22\textwidth]{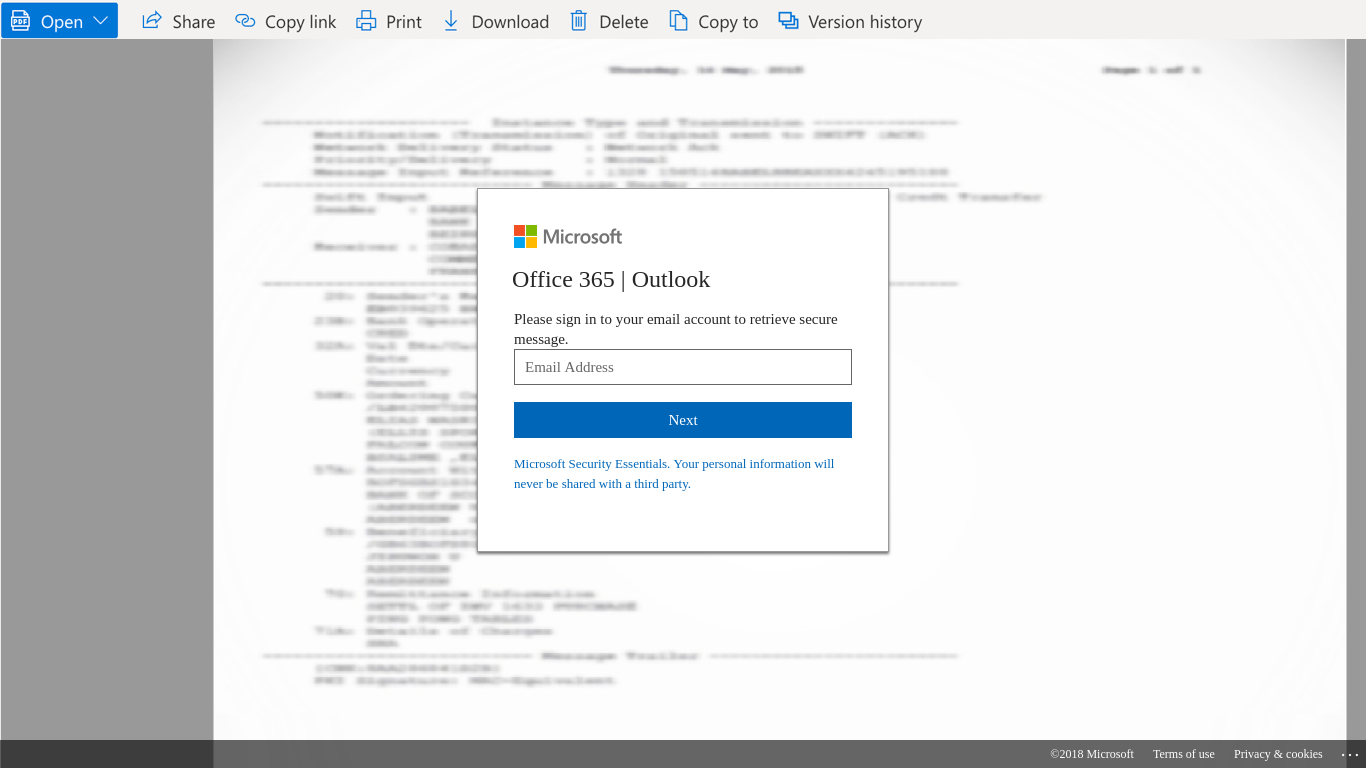}}} &
  \raisebox{-.5\height}{\frame{\includegraphics[width=.22\textwidth]{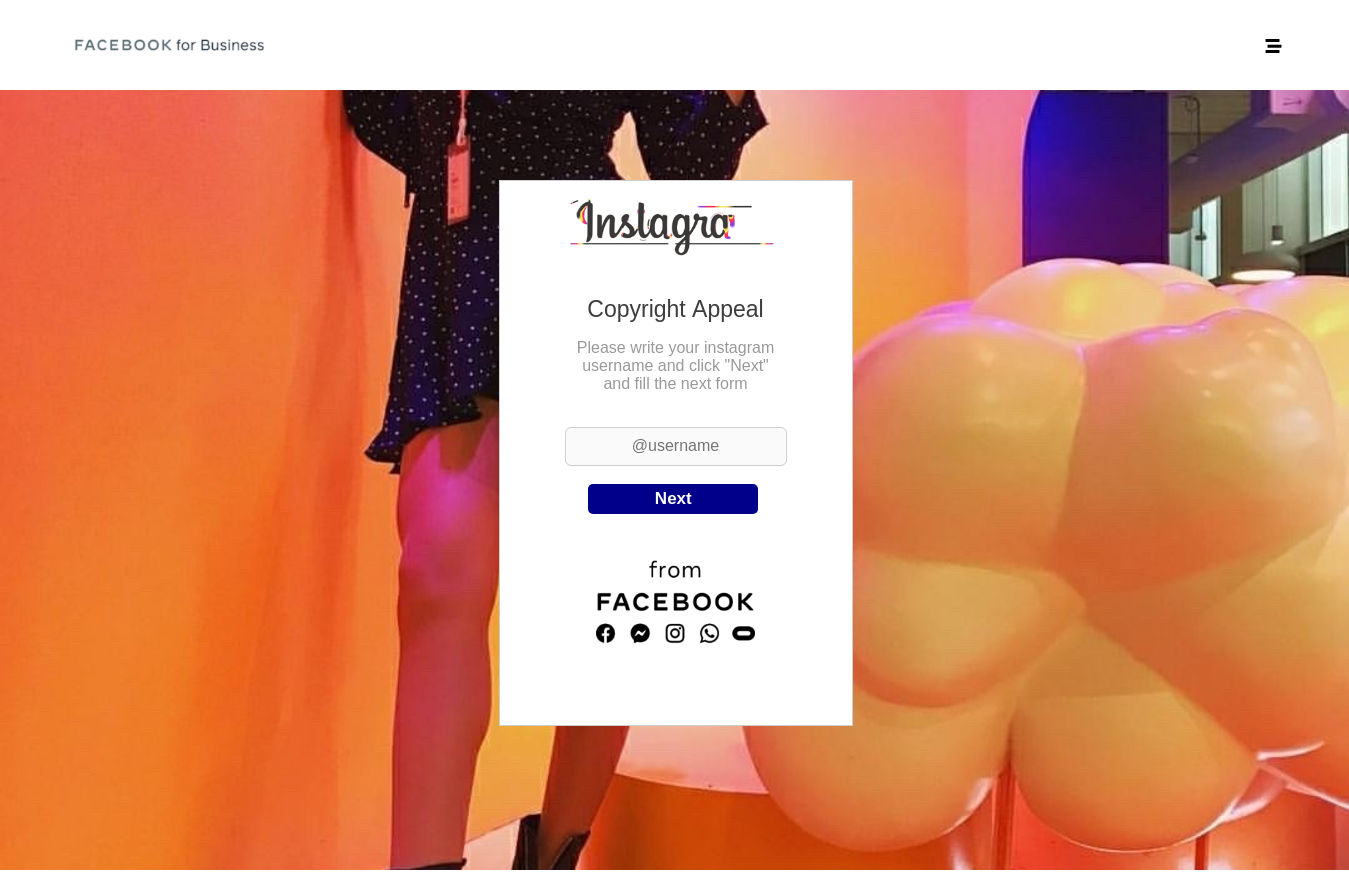}}} &
  \raisebox{-.5\height}{\frame{\includegraphics[width=.22\textwidth]{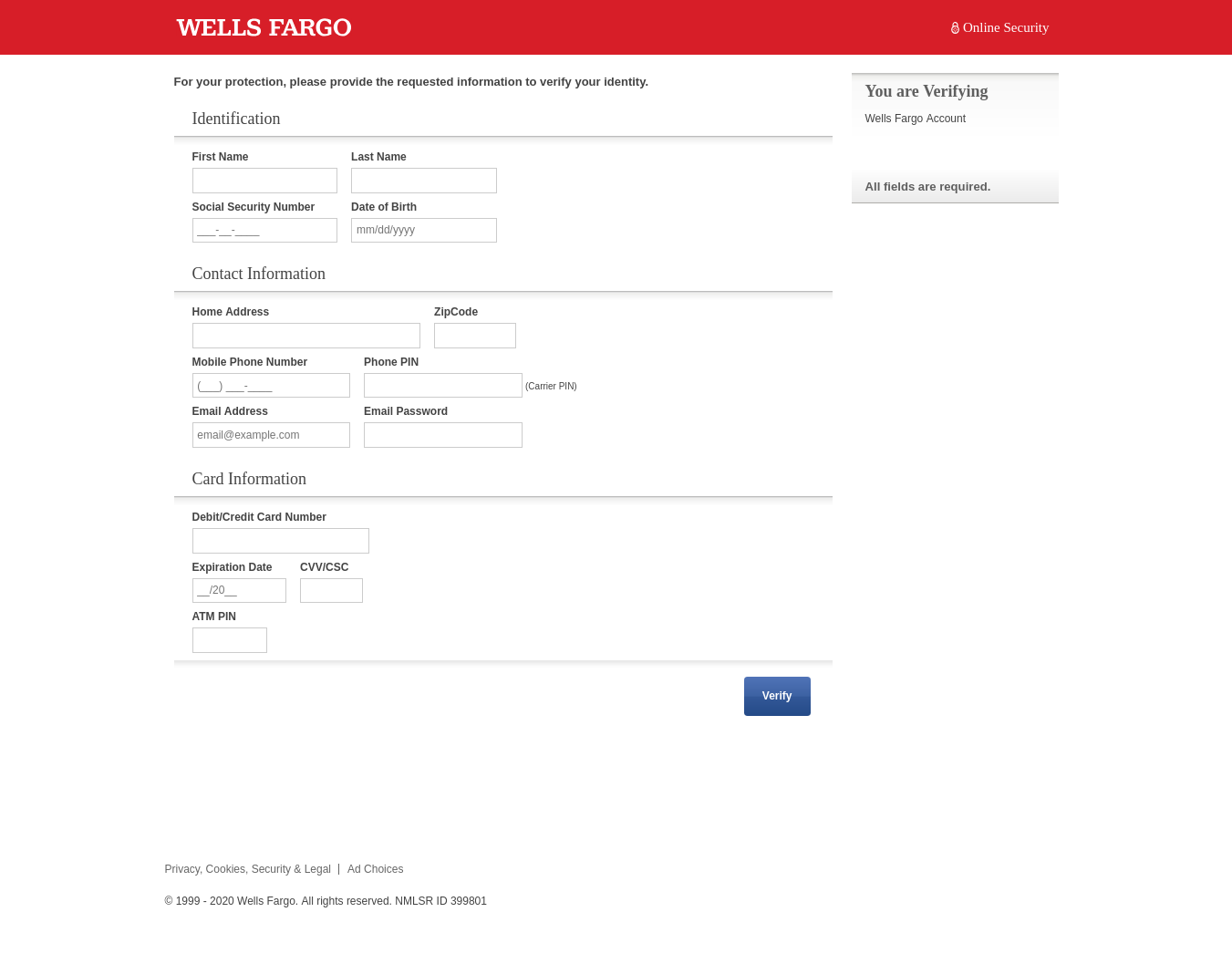}}} \\
  \addlinespace[0.5em]
  \hline
  \addlinespace[0.5em]
  \rotatebox[origin=c]{90}{\textbf{BMA}} &
  \raisebox{-.5\height}{\frame{\includegraphics[width=.24\textwidth]{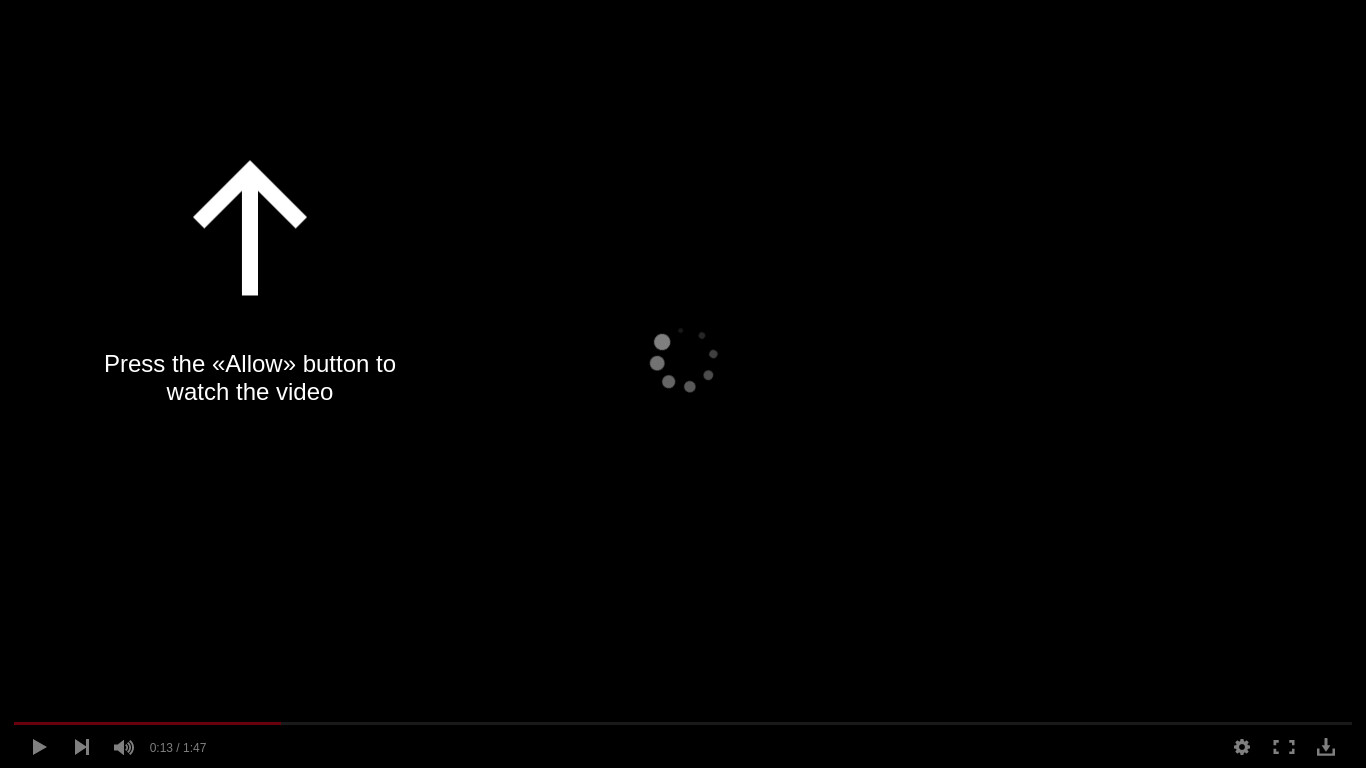}}} &
  \raisebox{-.5\height}{\frame{\includegraphics[width=.24\textwidth]{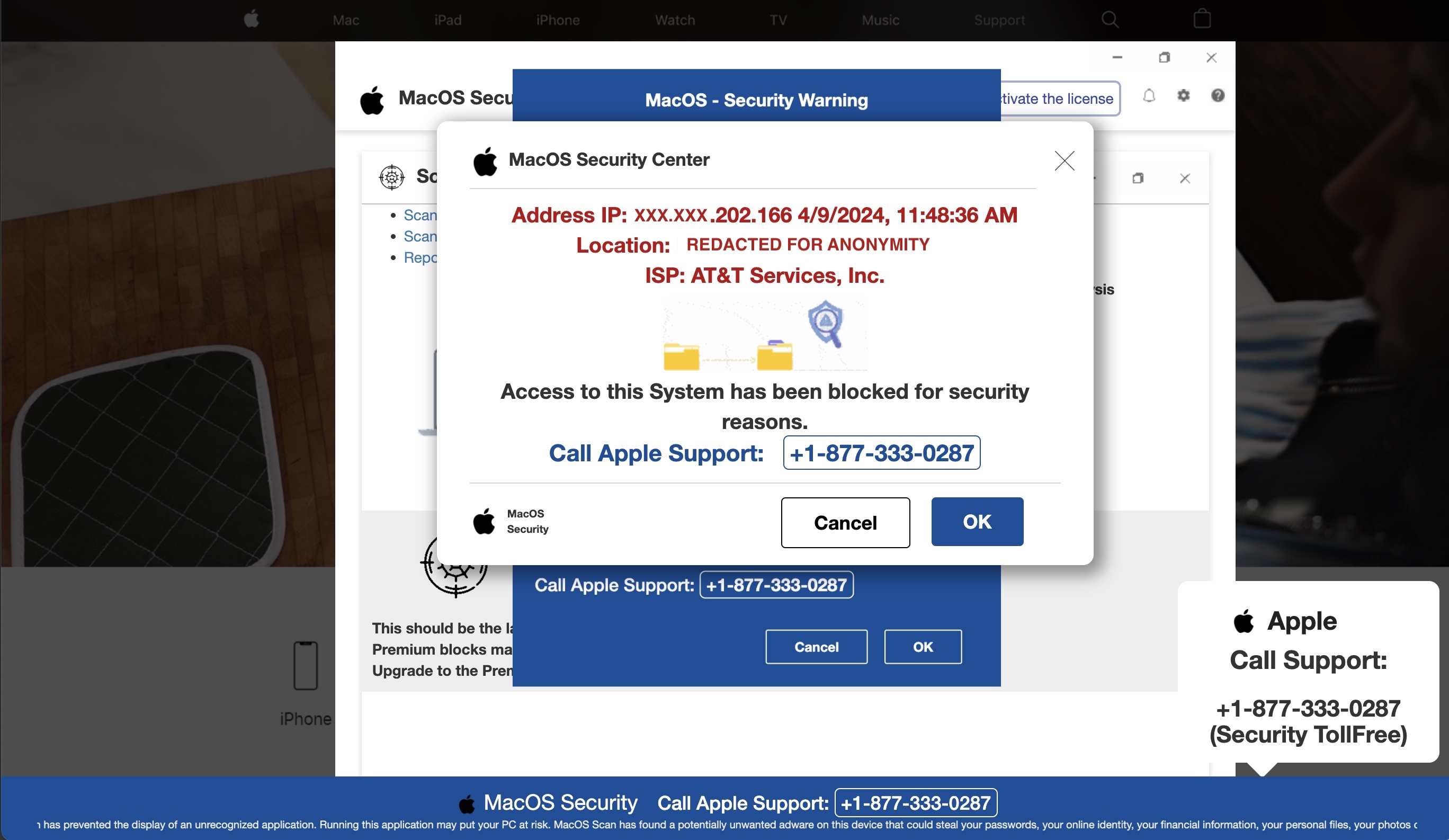}}} &
  \raisebox{-.5\height}{\frame{\includegraphics[width=.24\textwidth]{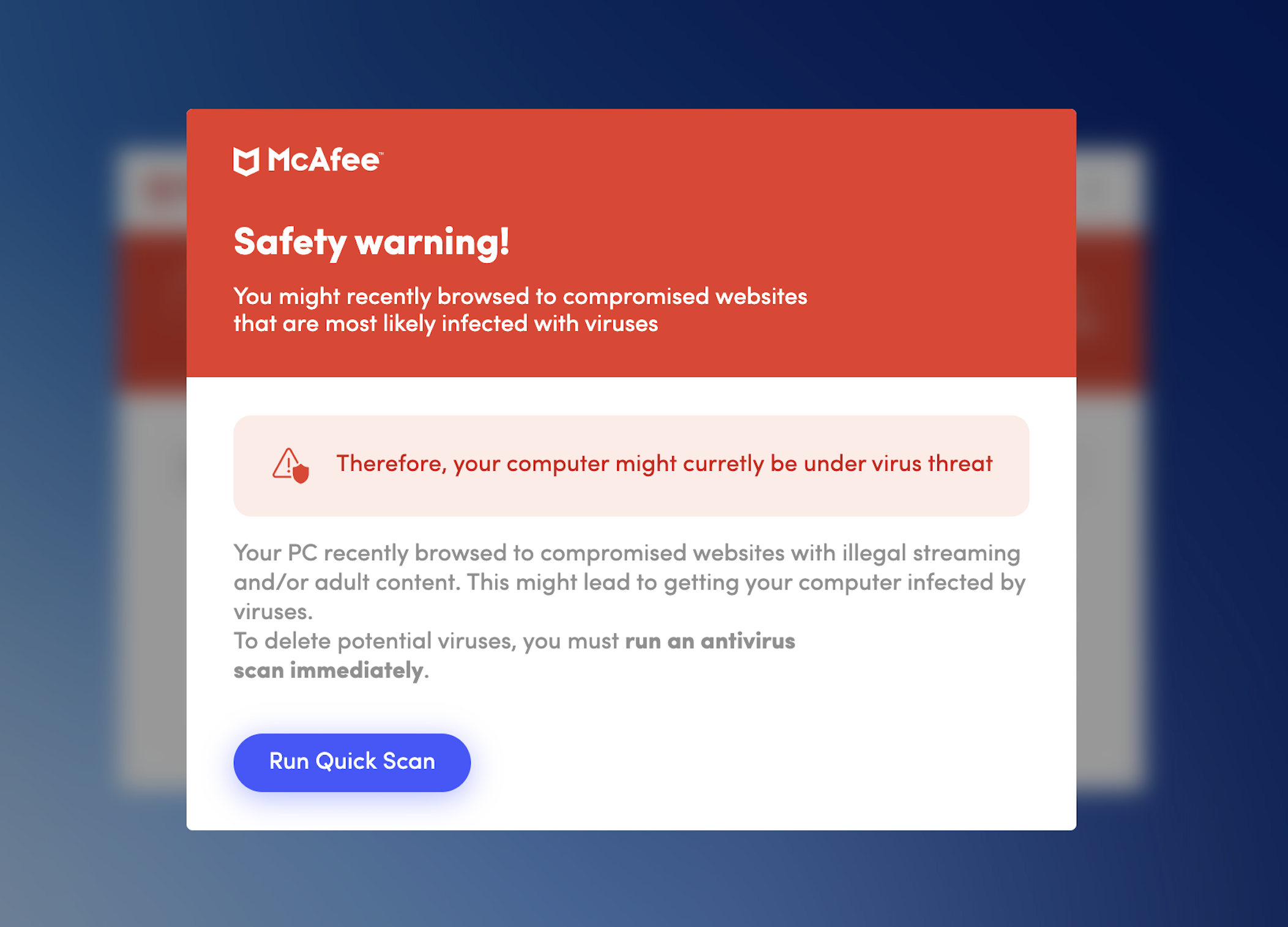}}} &
  \raisebox{-.5\height}{\frame{\includegraphics[width=.24\textwidth]{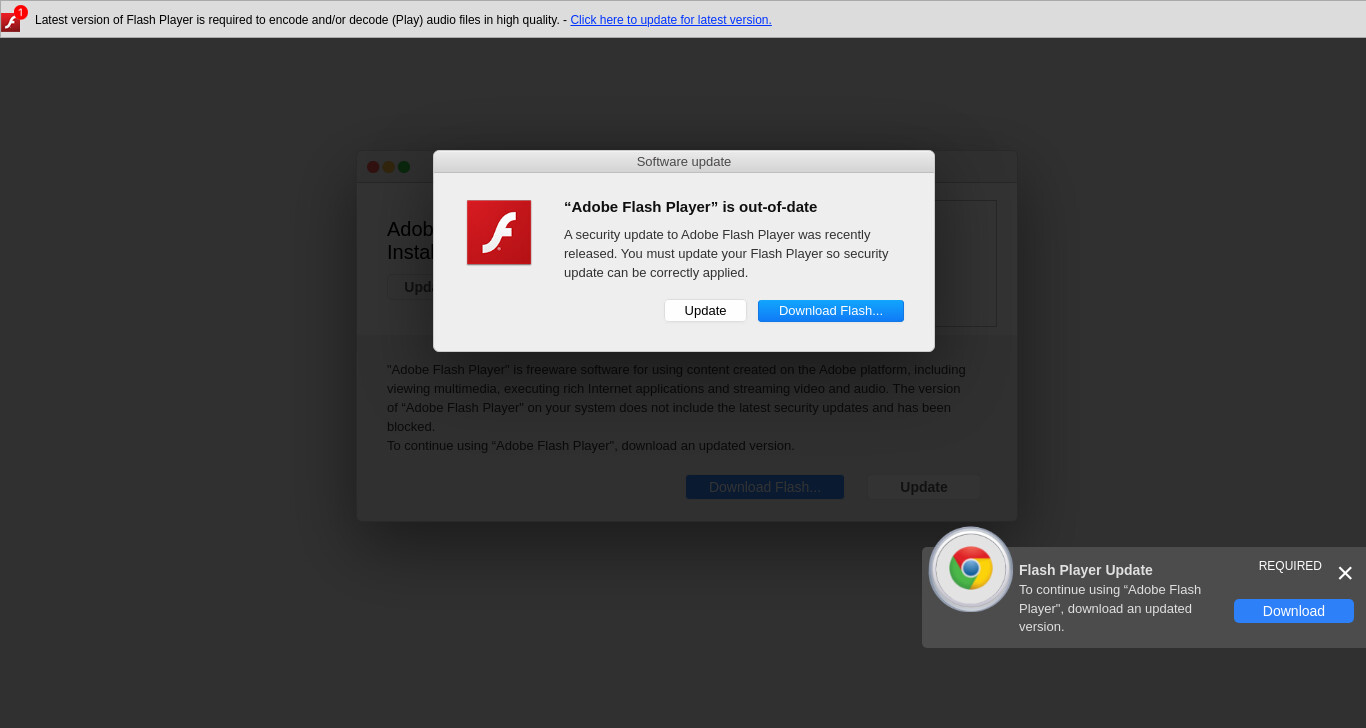}}} \\
  & Notification Permission Stealing & Technical Support Scam (TSS) & Scareware & Fake Software Download \\
  \bottomrule
  \end{tabular}
  \end{adjustbox}
\end{table*}



\vspace{3pt}
\noindent
\textbf{\em Contributions.}
\label{sec:contrib}
We make the following main contributions:


\begin{itemize}[leftmargin=*]
  \item \textbf{{\em Framework}}. We introduce \ppddd (\ppabv), the first end-to-end browser framework that discovers, detects, and defends against web-based \emph{behavior-manipulatating} social-engineering attacks. \ppabv can achieve a high detection rate of 99\% at a 1\% false positive rate, even on previously unseen BMA campaigns. When evaluated on new BMA samples collected months after our model was trained, the system still maintains a detection rate above 97\% at 1\% false positives. Furthermore, our browser extension is able to detect and block BMA pages with limited overhead by exploiting their visual cues, thus providing users with a first practical in-browser defense against these insidious attacks.

  \item \textbf{{\em Dataset}}. We use a specialized crawler to harvest a large corpus of 7,149 recent in-the-wild BMA pages belonging to 84 different attack campaigns. By combining our new dataset with historic samples of BMA campaigns collected by previous work, we build the largest and most comprehensive labeled dataset for evaluating BMA detection systems, which we will make available to the research community to foster future research in this area.

  \item \textbf{{\em Model}}. We design a novel multimodal detection model that fuses vision and text features extracted from web pages to accurately detect BMAs. Our model accepts screenshots of arbitrary size in input, to support deployment on a diverse set of devices, and can detect BMAs while generalizing across different screen resolutions and attack campaigns. We will open-source the full \ppabv code base\footnote{\url{https://github.com/NISLabUGA/PixelPatrol3D_Code}}, including our pretrained detection models and evaluation scripts.

  \item \textbf{{\em Deployment}}. To provide a practical detection solution, we deploy our detection model by implementing a lightweight browser extension for desktop, tablet, and mobile devices using the ONNX Web Runtime \cite{onnx} and Tesseract.js~\cite{tesseract}. The extension performs all inference locally to the user's browser, thus maintaining the privacy of users' browsing history. The complete extension codebase will also be open-sourced to facilitate future research.
\end{itemize}



\section{PP3D Framework}
\label{sec:pp3d_framework}

Our approach towards building a BMA detection system is informed by the following key observations:

{\em BMAs have a very significant visual component}: BMA pages exploit victims' decision-making processes via a number of deception and persuasion tactics. To attract users' attention and persuade them of the veracity of an invented scenario (e.g., the user's device is compromised, a software update is needed to watch a video, etc.), web-based BMA attacks make use of strong visual cues, including fake dialog boxes, flashy messages, abused brands, etc., as shown by the examples in Table~\ref{Tab:sepdif} and Figure~\ref{fig:MCLUSTERS}. Our classification approach aims to detect visual cues in such attacks.

{\em BMAs include text content that is often obfuscated}: In addition to visual cues, BMA pages also include attack messages that are often ``obfuscated'' via various techniques, such as text-to-image conversion, font remapping, or canvas rendering. These techniques aim to evade traditional DOM-based detectors that rely on the HTML structure of the page. At the same time, attack-related text in BMAs (e.g., ``Your machine is infected,'' ``Call Micosoft Tech Support,'' ``Click Allow to watch the video,'' and similar messages) is critical for attack success. To capture these cues, we leverage Optical Character Recognition (OCR) to extract the actual user-visible content from the rendered page. This allows us to recover attack-related text and detect BMA pages even when they employ HTML obfuscation.

{\em BMAs are typically orchestrated via campaigns}: To scale the attacks and facilitate their distribution to many potential victims, BMAs are orchestrated via campaigns. Specifically, a BMA campaign aims to ``advertise'' visually and semantically similar/same BMA attack content on multiple different websites, as noted in~\cite{Vadrevu_IMC19}. Thus, it may be sufficient to observe and learn from few examples of a BMA campaign to allow us to detect future instances of the same or similar BMA attacks on different websites.

\begin{figure*}[ht]
    \centering
    \begin{center}
        \includegraphics[width=18cm,keepaspectratio]{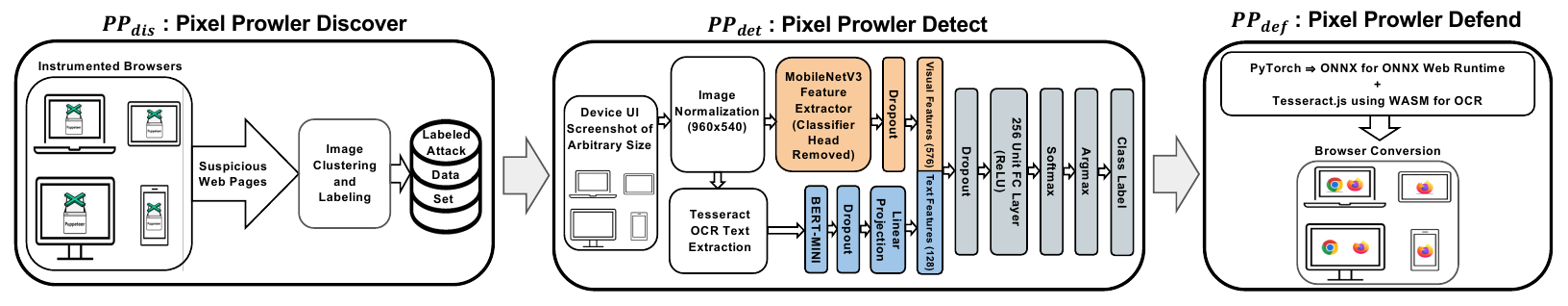}
        \caption{Overview of the \ppddd\ Framework. The \ppdetabv\ module includes both preprocessing steps (e.g., image normalization and OCR) and the core detection model. The colored blocks within \ppdetabv\ represent the core model components used for feature extraction and classification.}
        \label{fig:overview}
    \end{center}
\end{figure*}

Based on the above observations, our defense approach relies on (i) {\em continuously collecting examples of recent in-the-wild BMAs} from a variety of campaigns, (ii) {\em training a detection model} that can identify new BMA variants based on visual and textual cues, and (iii) {\em in-browser deployment of this detection model for real-time defense}. To this end, we design our \ppabv framework, shown in Figure~\ref{fig:overview}, with three main components: \ppdis (\ppdisabv), \ppdet (\ppdetabv), and \ppdef (\ppdefabv). In this paper, we focus most of our research efforts on \ppdetabv, and build an early prototype of \ppdisabv\ and \ppdefabv to enable the collection of a large BMA dataset and to demonstrate the viability of \ppdetabv and of the entire \ppabv defense framework.

\subsection{Pixel Patrol Discover}
\label{sec:ppdis_sum}
To automatically identify web-based BMAs, we first need to collect fresh examples of web pages associated with a variety of attack campaigns. To this end, we extend the crawler architecture proposed by Vadrevu et al.~\cite{Vadrevu_IMC19}, expanding the number of emulated devices and screen resolutions from 5 to 30, and logging the full interaction path taken to reach each potential BMA page. Additionally, we conducted a systematic labeling process using three expert human annotators to create a high-quality dataset of BMA examples. This dataset, which will be released to the research community, forms the foundation for training \ppdetabv. Details on \ppdisabv and the labeling process are provided in Appendix~\ref{sec:ppdis}.

\subsection{Pixel Patrol Detect}
\label{sec:ppdet}

\ppdetabv, the detection module within the \ppabv framework, constitutes the core technical contribution of this work. This section details its design, rationale, and implementation.

\subsubsection{Design Principles}

To ensure broad applicability and robustness, \ppdetabv is designed to meet several key requirements.

\emph{Resolution Agnosticism.} Webpages can be rendered on devices with vastly different screen sizes and aspect ratios, ranging from small mobile phones to ultra-wide desktop monitors. Traditional image classifiers typically assume fixed input dimensions, limiting their ability to generalize across diverse environments. In contrast, \ppdetabv must operate on screenshots of arbitrary resolution while maintaining high accuracy across a wide spectrum of BMA campaigns. This challenge is further compounded by our goal of deploying the model as an on-device browser extension, necessitating not only strong detection performance but also low-latency operation on resource-constrained mobile hardware.

\emph{Dual-Modality Reasoning.} Many BMA webpages share strong semantic similarities, even when their visual designs differ. To capture both superficial and latent patterns, \ppdetabv processes both visual and textual information. By combining visual features from webpage screenshots with OCR-extracted text, the model is able to recognize attacks based on both visual appearance and semantic similarity, thus improving generalization against diverse BMA variants.

\emph{Resilience to HTML Obfuscation.} Sophisticated attackers frequently obfuscate HTML content to bypass traditional DOM-based detection methods. A common tactic is embedding key phrases (e.g.,  ``Click here to claim your prize'') as images, rendering them invisible to systems that extract only text from the HTML. Beyond this, adversaries often use custom web fonts to remap visible characters to unrelated Unicode code points; although the user sees familiar phrases like ``Allow'' or ``Download,'' the underlying HTML contains meaningless sequences that evade keyword-based detectors. Another prevalent technique involves canvas-based rendering, where critical text is drawn dynamically using JavaScript and HTML5 \texttt{<canvas>} or SVG elements, completely bypassing the DOM. These and similar strategies highlight the inadequacy of DOM inspection alone and underscore the need for OCR to recover the actual, user-visible semantics of a page. While OCR is more computationally demanding than HTML parsing, it remains essential for robust detection in adversarial settings.

\subsubsection{Implementation}
\label{sec:ppdet_implementation}



\ppdetabv uses a dual-branch neural network (implemented using PyTorch) that processes visual and textual features in parallel before fusing them for final classification (see Figure~\ref{fig:overview}). Specifically:
\begin{itemize}[leftmargin=*]
  \item The \textbf{visual branch} employs a MobileNetV3-Small~\cite{MobileNets} backbone pre-trained on ImageNet, with the classification head removed and replaced by an identity layer as a visual feature extractor. The output feature vector (576 dimensions) is passed through a dropout layer to reduce overfitting.
  \item The \textbf{text branch} uses a compact transformer encoder based on BERT Mini~\cite{bertmini} (specifically, \texttt{prajjwal1/bert-mini}). The pooled output from the [CLS] token (with dimensionality 312) is processed through a dropout layer followed by a linear projection to a fixed 128-dimensional embedding. 
\end{itemize}

The 576-dimensional visual and 128-dimensional textual embeddings are concatenated to form a unified 704-dimensional feature vector. This joint representation is passed through a feedforward classification head consisting of a 256-unit fully connected layer with ReLU activation, followed by a final linear layer that outputs logits for binary classification. Dropout regularization is applied before the classification head to mitigate overfitting. During inference, logits are passed through a softmax layer to produce class probabilities, with the predicted label selected via \texttt{argmax}. The overall model architecture is illustrated in the \ppdetabv portion of Figure~\ref{fig:overview}. This design allows \ppdetabv to reason jointly over both visual and textual modalities while remaining small ($\sim$30 MB in ONNX format) and efficient enough for real-time deployment on both desktop and mobile platforms.


\begin{algorithm}[b]
  \scriptsize
  \caption{Normalizing Screenshots to Fixed Resolution}
  \label{alg:ss_norm}
  \begin{algorithmic}
  \Require{$I$: Raw screenshot of arbitrary resolution\\
           $T$: Target canvas size (1920$\times$1080)\\
           $S$: Downsampling factor (0.5)}
  \Ensure{$I'$: Normalized image of size 960$\times$540}
  \end{algorithmic}
  \begin{algorithmic}[1]
    \If{$I.\text{width} > 1920$ \textbf{or} $I.\text{height} > 1080$}
      \State scaleFactor = $\min\bigl(1920 / I.\text{width},\,1080 / I.\text{height}\bigr)$
      \State $I_{\text{scaled}}$ = Resize($I$, scaleFactor)
    \Else
      \State $I_{\text{scaled}} = I$
    \EndIf
    \State $C$ = Create zero-valued (RGB [0, 0, 0]) canvas of size 1920$\times$1080
    \State Center $I_{\text{scaled}}$ on $C$
    \State $I'$ = Downsample($C$, factor $S$)
  \end{algorithmic}
\end{algorithm}

To normalize screenshots of arbitrary screen resolutions (to accommodate different device screen formats), we first project each image onto a 1920$\times$1080 frame. For oversized images---those where either dimension exceeds 1920 pixels in width or 1080 pixels in height---we apply isotropic scaling such that the larger dimension fits within the corresponding bound (1920 or 1080), preserving the aspect ratio. For undersized images—those that already fit within 1920$\times$1080—no scaling is applied. In both cases, the scaled image is centered on a 1920$\times$1080 canvas, and any unused margins are padded with zero-valued pixels (RGB [0, 0, 0]). The resulting canvas is then downsampled by a factor of 0.5, empirically chosen to balance classification accuracy, OCR effectiveness, and inference speed, resulting in a final input size of 960$\times$540. A procedural outline is presented in Algorithm~\ref{alg:ss_norm}. This normalization process ensures consistent input dimensions across devices and supports efficient batch inference on the GPU.

To extract text from webpage screenshots, we use the Tesseract OCR engine~\cite{tesseract}. Tesseract was selected for its strong tradeoff between recognition accuracy and performance, particularly in low-resource environments.



\subsection{Pixel Patrol Defense}
\label{sec:ppdef}

To demonstrate the practicality of the \ppabv framework, we develop a proof-of-concept extension for the browser that allows for detecting BMA in real time.

\emph{Initialization.}
When the extension initializes---either at browser launch or when it is re-enabled---it first creates a single off-screen document, which in turn spawns a pool of dedicated Web Workers. One worker hosts the image-text classifier introduced in Section~\ref{sec:ppdet_implementation}, which we implemented by converting the pre-trained model using ONNX~\cite{onnx}. The converted modules are the colored modules shown in the \ppdetabv portion of Figure \ref{fig:overview}. Four additional workers run Tesseract, leveraging its native WebAssembly (WASM) support. This allows all OCR tasks to execute efficiently on the browser, ensuring user privacy (all inference runs locally to the browser with no data leaked to any external API) and delivering consistent performance across both desktop and mobile browsers. All workers are instantiated only once per session, so that model weights and tokenizer tables reside in memory, avoiding repeated loading delays and garbage-collection churn. With the worker pool active, the background extension script starts a non-blocking fixed 5 second page scan timer that drives every subsequent classification cycle.


\emph{Decision Pipeline.}
For each page scan, the extension applies a tiered filtering strategy that minimizes expensive inference. If the current domain is less or equal to rank 100K in the Tranco Top 1M list \cite{tranco}, the page is immediately labelled \emph{benign}, as such high-ranking domains are rarely associated with malicious behavior. This threshold aims to strike a balance between efficiency and coverage, allowing the system to focus resources on less popular and more likely malicious domains. Otherwise, the extension captures a screenshot, computes an 8x8 block mean value-based perceptual hash (BMVBHash)~\cite{phash_algo}, and measures its Hamming distance~\cite{hamming1950error} to the hash from the previous scan.

This yields four mutually exclusive cases:

\begin{description}[leftmargin=1.7em]
  \item[Case~1:] Domain in whitelist $\rightarrow$ verdict of \emph{benign}.
  \item[Case~2:] No previous phash $\rightarrow$ run inference.
  \item[Case~3:] Hamming distance $\,\ge 5$ $\rightarrow$ run inference.
  \item[Case~4:] Hamming distance $\,< 5$ $\rightarrow$ reuse last verdict.
\end{description}

Only Cases~2 and~3 invoke the heavy model, keeping the average per-page overhead low. The hamming distance threshold of 5 was empirically chosen to be conservative, so that an inference result from a previous scan is reused only if the two images are visually very similar, thus justifying the reuse of the previous verdict.


\emph{Inference.}
Whenever a screenshot needs to be analyzed (i.e., in Cases 2 or 3), the scanning loop is temporarily paused. The extension captures a screenshot of the current browser tab and normalizes the image as described in Section~\ref{sec:ppdet_implementation}. This normalized image is then transferred---using zero-copy techniques---to the off-screen document for processing. Within the off-screen document, two types of Web Workers operate sequentially:
\emph{(i)} The OCR workers divide the image into four horizontal slices (1 column $\times$ 4 rows), a layout empirically shown to balance speed and accuracy effectively. Each slice is sent to a separate Tesseract Web Worker, and the resulting plain text outputs are concatenated from top to bottom.
\emph{(ii)} The ONNX worker then receives the resized image along with the extracted OCR text and feeds both into the ONNX-converted model. The model performs inference on the dual inputs and returns a classification label---either \emph{benign} or \emph{malicious}---which is used to update local storage accordingly.

\emph{Handling Dynamic Content.}
To detect dynamically appearing UI components (e.g., delayed popups, scroll-triggered overlays), \ppdefabv employs periodic 5-second scans with perceptual hash comparison. If the Hamming distance between consecutive hashes exceeds 5, indicating significant visual change, the system triggers fresh inference. This captures dynamically appearing BMA elements once they stabilize in the viewport, while the phash-based filtering (Case~4) prevents redundant inference on visually similar states. This design handles common dynamic BMA tactics (delayed scareware popups, notification-stealing prompts) without explicit timeout windows or complex stabilization heuristics.

\emph{User Interaction.}
If the classifier returns \emph{malicious}, the background script opens a notification window presenting a warning that the page is likely malicious, shows the page screenshot, and displays three action choices: \emph{Return to Safety} (redirects the tab to a trusted site), \emph{Ignore Warning}, and \emph{Not Malicious} (records a false-positive override). Scanning pauses while the dialog is active and resumes as soon as the user selects a choice and the window closes. 

\emph{Scope of Interactions.}
\ppdefabv focuses on \emph{click/tap-based} BMAs, which dominate real-world campaigns~\cite{Vadrevu_IMC19,miramirkhani2017,WebPushAds}. Keyboard-based traps, hover-triggered actions, and gesture-based manipulations are currently out of scope. However, if such interactions produce visible page changes, \ppabv can detect the resulting attack state during periodic scans. Future extensions could incorporate additional interaction monitoring, but the current design prioritizes prevalent attack vectors while maintaining a lightweight, privacy-preserving architecture.


\emph{Logging and Configuration.}
The extension also provides a popup UI that allows users to toggle on/off the logging of screenshots and performance metrics, as well as to view the latest classification results.
Execution times, domain names, webpage screenshots, and classification outcomes are buffered in memory and flushed to a timestamped text file on the user's device every two minutes.

\section{Experimental Setup}
We now discuss the setup for evaluating \ppabv framework.

\subsection{Data Collection using Pixel Patrol Discover}

To collect a diverse set of web pages for evaluating \ppdetabv, we use \ppdisabv to crawl (i) seed URLs that use low-tier ad networks that are likely to lead to attack web pages and (ii) an unbiased set of benign URLs, using the setup explained in Appendix~\ref{sec:ppdis_crawler_setup}.

\label{benign_wp_col}

\vspace{3pt}
\noindent
\textbf{Benign Webpage Collection.}~~
To account for variability in the  benign webpage designs and build a representative dataset, we combined seed URLs from two distinct sources: the top 5,000 domains listed by Tranco~\cite{tranco}, representing popular sites, and 10,000 URLs randomly selected from Web Extraction (WET) files spanning one month of the Common Crawl dataset~\cite{common_crawl}, capturing less popular and potentially more diverse sites. Using the crawler to emulate multiple devices and resolutions, we obtained 396,255 distinct screenshots from the popular websites and 386,180 distinct screenshots from the Common Crawl URLs. While some webpages sampled from the Common Crawl dataset may not be explicitly benign, given the random sampling, we consider the likelihood of encountering BMA sites to be minimal, with any resulting noise likely having only minor impacts, such as slightly increased false positives during testing. Further discussion regarding ethical considerations and the negligible impact of our large-scale crawling on benign websites is discussed in Appendix~\ref{sec:ethical_consideration}.




\vspace{3pt}
\noindent
\textbf{BMA Webpage Collection.}~~
To collect seed URLs that may provide a high-yield of BMAs, we follow an approach similar to previous studies~\cite{Vadrevu_IMC19,WebPushAds} and collect URLs of websites that are known to host ads from low-tier ad networks, as they have been observed to often lead to BMA pages. Overall, we collected 24,979 seed URLs associated with 10 low-tier ad networks (listed in Table~\ref{tab:ad_networks}(a), in the Appendix). In addition to this initial collection, we conducted a second BMA gathering effort several months later. This allowed us to evaluate the effectiveness and temporal robustness of our detection system across different collection periods.

\vspace{3pt}
\noindent \textit{Primary BMA Collection: } In this initial collection, we began with {\em 28,923} unfiltered BMA screenshots provided by the authors of~\cite{Vadrevu_IMC19} and expanded the dataset by crawling {\em 24,979} seed URLs. This crawl resulted in {\em 650,255} additional unique screenshots from 55,539 distinct webpages, leading to a total of {\em 679,178} images. We then removed duplicates and applied clustering based on screen resolution and perceptual hashing—following techniques similar to those in~\cite{Vadrevu_IMC19, TRIDENT}—which yielded {\em 8,107} images grouped into {\em 318} clusters.

To identify BMA campaigns, we carried out a systematic labeling process involving three individuals familiar with BMA techniques. This process included multiple rounds of collaborative review followed by independent labeling of the {\em 318} clusters. After independent labeling, we computed the inter-rate reliability score using Krippendorff's alpha score~\cite{krippendorff_computing_2011} among the labelers and obtained $\alpha=0.82$, which indicates a high level of agreement~\cite{de2012calculating}. As a result, we identified {\em 245} clusters containing {\em 7,012} BMA screenshots across 30 screen resolutions.  We then manually grouped these 245 clusters into {\em 74} meta-clusters based on visual and behavioral similarities in attack type, representing distinct BMA campaigns. Table~\ref{Tab:collection1} summarizes the number of campaigns and associated screen resolutions for each attack category in the primary dataset.

\begin{table}[h]
    \captionof{table}{Summary of BMA categories in the initial collection, including the number of campaigns and distinct screen resolutions observed in the collected webpage screenshots.}
    \centering 
    \begin{adjustbox}{width=.75\columnwidth} 
    \small 
\begin{tabular}{@{}lccc@{}}
\toprule
\textbf{BMA Category}          & \textbf{\# Screenshots} & \textbf{\# Resolutions} & \textbf{\# Campaigns} \\ \midrule
\textbf{Fake Software Download}   & 4700                    & 29                      & 29                    \\
\textbf{Notification Stealing}    & 1130                    & 27                      & 7                     \\
\textbf{Service Sign-up Scam}     & 758                     & 24                      & 20                    \\
\textbf{Scareware}                & 213                     & 8                       & 9                     \\
\textbf{Fake Lottery/Sweepstakes} & 194                     & 4                       & 6                     \\
\textbf{Technical Support Scam}   & 17                      & 2                       & 3                     \\ \midrule
\textbf{Total (distinct)}                    & \textbf{7012}           & \textbf{30}             & \textbf{74}           \\ \bottomrule
\end{tabular}
\end{adjustbox} 
\label{Tab:collection1}
\end{table}

\vspace{3pt}
\noindent \textit{Secondary BMA Collection: }
\label{sec:second_bma_collection}
While the primary BMA webpage collection is used for most of our experiments, we collected a secondary set of BMA pages after several months to assess how well \ppdetabv performs against newly observed BMA attack campaigns. By crawling 6,120 seed URLs, we captured 305,411 unique screenshots that led to 137 unique BMA webpage screenshots across 10 new distinct never-before-seen BMA campaigns.

\subsection{Training Pixel Patrol Detect}
\label{sec:ppdet_setup}
To perform all \ppdetabv experiments, we used a dedicated Dell PowerEdge C4140 with 512GB memory and 4x 32GB NVIDIA V100 GPUs, with CUDA 12.2 and PyTorch v2.6.0+cu124.

\label{sec:ppdet_setup_datasets}

\vspace{3pt}
\noindent
\textbf{Dataset Generation.}~~
From the collection runs described above, we construct a master training dataset of {\em 119,536} images comprising both benign and BMA samples.

The benign portion includes {\em 100,000} unaugmented screenshots randomly sampled from our broader benign collection. We ensured proportional representation across all 30 screen resolutions (see Appendix Table~\ref{tab:imres}(b)) and maintained a roughly even split between screenshots obtained from the Tranco Top 5K crawl and those from the Common Crawl.

The BMA portion is based on {\em 6,512} unaugmented screenshot-text pairs from our primary BMA collection (with the remaining labeled BMA samples reserved for testing).
For each screenshot, we used Tesseract OCR~\cite{tesseract} to extract visible on-page text, saving the result as a plain-text file to establish a one-to-one mapping between each image and its OCR output. We then applied both image and text augmentation to this set, resulting in a final BMA training set of {\em 19,536} samples.

We intentionally preserve class imbalance in the training dataset, with a significantly larger number of benign examples. This decision reflects the real-world distribution of web content, where benign pages vastly outnumber malicious ones. Training under a class distribution that reflects the imbalance found in deployment conditions helps the model calibrate its decision boundary more effectively and reduces overfitting to rare malicious patterns. Prior work in anomaly detection and imbalanced learning~\cite{perera2019learning,he2009learning} supports this approach, showing that maintaining class imbalance, when combined with appropriate augmentation or feature learning techniques, can improve generalization and reduce false positives in deployment scenarios.

Test datasets are tailored for each experiment, depending on the evaluated scenario, for example, by excluding specific campaigns or resolutions.


\vspace{3pt}
\noindent
\textbf{Data Augmentation.}~~
To enhance the model's robustness, we applied data augmentation techniques to all training images, targeting both their visual and textual components. For image augmentation, two transformations were randomly selected from a predefined set, which included color inversion, grayscale conversion, random margin cropping, extreme hue shifts, and adjustments to brightness, contrast, saturation, and solarization. Each transformation was applied with randomized parameters to minimize duplication. All selected augmentations were designed to maintain text legibility and the overall visual structure of the webpage.

For text augmentation, we utilized Groq's LLaMA-3.3-70B-Versatile model via their cloud API to perform synonym replacement. This method increases lexical variety while preserving sentence structure and semantic accuracy. Complete prompting details and API settings are provided in Appendix~\ref{sec:text_aug_dets}. As with the image transformations, we ensured that textual modifications did not impair readability or alter the original meaning of the webpage content.





\vspace{3pt}
\noindent
\textbf{Hyperparameter Tuning.}~~
%
Prior to final experimentation with the \ppdetabv model, we performed extensive hyperparameter tuning using 10,000 randomly sampled benign train-test pairs and 2,000 randomly sampled BMA train-test pairs from the master training dataset, supplemented by a validation set comprising 500 benign and 500 BMA test images.


We began the tuning process using the Adam optimizer with an initial learning rate of \(1 \times 10^{-3}\), and explored both higher and lower learning rates. We also evaluated various learning rate schedulers, including cosine decay, one-cycle, and step decay. To assess the impact of regularization, we varied the degree of backbone freezing---evaluating configurations where MobileNetV3 and BERT-Mini were either fully frozen, partially unfrozen, or fully trainable. We additionally tuned dropout rates on the visual branch, textual branch, and the fusion multilayer perceptron.

Other hyperparameters explored included multiple values for weight decay, batch size, and three loss formulations: standard cross-entropy, class-weighted cross-entropy, and focal loss. We also tested different pre-processing configurations, including image downscaling factors and maximum token lengths for the textual input.

The best-performing configuration left all layers fully trainable and used a fixed learning rate of \(2 \times 10^{-6}\) without any scheduler. It applied class-weighted cross-entropy loss, with weights proportional to the observed class imbalance, and a weight decay of \(5 \times 10^{-4}\). Dropout was set to 0.3 for both the visual and textual branches, and 0.6 for the fusion layer. Optimal results were achieved using an image scaling factor of 0.5, a maximum token length of 512, and a batch size of 64.

Although exact composition of the training and test datasets varied across experiments (see Section~\ref{sec:expresults}), all models were trained using this optimized configuration for consistency.

\subsection{Deploying Pixel Patrol Defend}
\label{sec:ppdef_setup}

To quantify both user-visible latency and device-side resource consumption, we evaluated \ppdefabv on representative desktop, tablet, and mobile hardware. Achieving robust results required two complementary steps: (i) choosing test devices that mirror what everyday users own, and (ii) crafting a repeatable measurement protocol.  We outline both below.


\vspace{3pt}
\noindent
\textbf{Device Selection.}~~
To evaluate the performance of our proof-of-concept \ppdefabv browser extension across realistic user environments, we selected representative devices spanning mobile, tablet, and desktop platforms. The mobile and tablet categories are covered by widely available Android devices that balance affordability and performance, reflecting mainstream user hardware. For desktops, we chose a typical enterprise-grade laptop alongside a high-end prosumer device to assess scalability across both resource-constrained and high-performance environments. Detailed device specifications are provided in Table~\ref{Tab:device_specs}.

\begin{table}[h]
    \captionof{table}{Device specifications used in performance evaluation.}
    \centering
    \begin{adjustbox}{width=\columnwidth}
    \small
    \begin{tabular}{@{}lcccc@{}}
    \toprule
    \textbf{Device} & \textbf{Processor} & \textbf{Cores} & \textbf{RAM} & \textbf{OS} \\ \midrule
    Samsung Galaxy A55 & Exynos 1480 & 8 & 8 GB & Android 14 \\
    Samsung Galaxy Tab S9 FE & Exynos 1380 & 8 & 8 GB & Android 14 \\
    Dell Latitude 5420 & Intel Core i7-1165G7 & 4 & 16 GB & Ubuntu 22.04 \\
    MacBook Pro & Apple M4 Max & 16 & 128 GB & macOS Sequoia \\ \midrule
    \end{tabular}
    \end{adjustbox}
    \label{Tab:device_specs}
\end{table}

\label{sec:ppdef_setup_test_strat}

\vspace{3pt}
\noindent
\textbf{Testing Strategy.}~~
As detailed in Subsection~\ref{sec:ppdef}, the \ppdefabv extension supports four execution paths, which give rise to three distinct performance profiles. Our testing strategy is designed to evaluate these three profiles, as well as evaluate the extension initialization step. Additionally, although we developed versions of the extension for both Chrome and Firefox, only the Firefox build is usable across desktop, tablet, and mobile platforms. Accordingly, all procedures described here refer to the Firefox variant.

To systematically evaluate the latency and system resource impact of each profile, we designed the following testing strategy. We selected 15 representative webpages: 5 whitelisted domains (ranked within the top 100,000 of the Tranco Top 1M list) and 10 non-whitelisted domains. 
The larger proportion of non-whitelisted domains reflects our focus on measuring the extension's most resource-intensive behavior---phash comparisons and model inference.

For each device under test, we shuffled the domain list and browsed each domain for two minutes using either Firefox (on desktop) or Firefox Nightly (on tablet and mobile). Throughout the experiment, we relied on our extension's built-in logging functionality to collect timestamps and latency data for each major page processing stage. Note that there is no added latency due to the logging functionality. 

In parallel, we run a custom monitoring script to record system-level CPU and memory usage attributed to the Firefox process. On desktop and tablet environments, we used a Python script that sampled the process list at 500\,ms intervals to extract CPU usage as a percentage of total logical cores (e.g., 800\% = full use of 8 cores) and RAM usage (in MB) using \texttt{ps} and \texttt{/proc/meminfo}. On mobile devices, we used an analogous script that executed \texttt{adb shell top} at regular intervals to retrieve live CPU and memory usage of the Firefox Nightly app (Fenix). In both cases, the monitoring script logged timestamped data and usage as a percentage of total available CPU and RAM on the device.

To establish a performance baseline, we repeated each experiment with the extension uninstalled. This enabled direct comparison of resource utilization with and without \ppdefabv active.

\section{Experimental Results}
\label{sec:expresults}

Our evaluation aims to answer 5 major research questions (RQs) and provide a performance evaluation of our defense. 

\subsection{Pixel Patrol Detect Results}
\label{sec:ppdet_rq1}

\begin{tcolorbox}[left=2pt,right=2pt,top=2pt,bottom=2pt]
  \textbf{RQ1:}  Can \ppdetabv accurately identify {\em new} instances of  BMAs belonging to previously observed campaigns?
\end{tcolorbox}

To answer this question, we evaluate the detection capabilities of our \ppdetabv model. To this end, we randomly select 500 benign examples and 500 BMAs, which are completely separate from the training dataset. Notice that the related web page screenshots were selected across randomly chosen screen sizes and BMA campaigns, and that it is possible that a test BMA sample may be part (as a newly observed attack instance) of a BMA campaign previously seen during training. For the training datasets, we use the master dataset discussed in Section~\ref{sec:ppdet_setup_datasets}. In this simple test setting, \ppdetabv achieves an AUROC score of 1.0 and a detection rate above 99\% at 1\% false positives.

\subsection{Generalization to New Screen Sizes}
\label{sec:ppdet_rq2}

\begin{tcolorbox}[left=2pt,right=2pt,top=2pt,bottom=2pt]
  \textbf{RQ2:}  Can \ppdetabv accurately identify instances of BMAs captured on a {\em new} screen size never seen during training?
\end{tcolorbox}

To answer RQ2, we proceeded as follows. First, we selected 9 different screen resolutions from our master dataset discussed in Section~\ref{sec:ppdet_setup_datasets}. Among these 9 resolutions, 5 are landscape resolutions while the other 4 are portrait resolutions related to mobile devices. Let $r_i$ indicate one of these 9 resolutions. Then, we formed 9 different training datasets, $\{T_1, \dots, T_9\}$, where dataset $T_i$ was formed by selecting images from all resolutions except for $r_i$. The remaining images with resolution $r_i$ were set aside as test dataset $S_i$. We then performed 9 training/test experiment by training on $T_i$ and testing on $S_i$ (i.e., images with the excluded resolution $r_i$).

Notice that, because \ppdisabv visits the same URLs using different browser viewport sizes, the same BMA campaigns can typically be observed under different resolutions. Therefore, $T_i$ and $S_i$ can include BMAs from the same campaigns, but the training and test screenshot images are taken on different device screen sizes. This allows us to focus on the effect of web page screenshots taken on never-before-seen screen sizes, rather than the effects of previously unseen BMA campaigns. Also, to obtain a roughly balanced test dataset, where a particular BMA campaign does not dominate the others, given resolution $r_i$ we selected at most 10 images from each BMA campaign represented under that resolution and 500 random images from benign pages in the same resolution.

The results of this experiment are reported in Table~\ref{Tab:table2}. ``Global'' refers to the overall results computed by first combining the classification scores obtained during the 9 training/test experiments into a single scores set. It is notable that \ppdetabv can generalize very well to previously unseen resolutions, with an AUROC score at or above 0.999 and detection rates at or above 99\% at 1\% false positives.

\begin{table}[h]
  \captionof{table}{Results for generalization to new screen sizes.}
  \centering 
\small 
  \begin{adjustbox}{width=0.65\linewidth}
  \begin{tabular}{@{}ccccc@{}}
  \toprule
  \textbf{Test Res.} & \textbf{\# Benign} & \textbf{\# BMA} & \textbf{AUROC} & \textbf{DR at 1\% FP} \\ \midrule
  \textbf{1366x768}  & 500             & 54       & 1.0        & 1.0               \\
  \textbf{800x1280}  & 500             & 57       & 1.0        & 1.0               \\
  \textbf{1920x998}  & 500             & 41       & 1.0        & 1.0               \\
  \textbf{414x896}   & 500             & 12       & 1.0        & 1.0               \\
  \textbf{1478x837}  & 500             & 34       & 0.999      & 1.0               \\
  \textbf{768x1024}  & 500             & 33       & 1.0        & 1.0               \\
  \textbf{1536x824}  & 500             & 474      & 1.0        & 1.0               \\
  \textbf{360x640}   & 500             & 198      & 0.999      & 0.990             \\
  \textbf{1366x728}  & 500             & 354      & 1.0        & 1.0               \\
  \midrule
  \textbf{Global}    & 4500            & 1257     & 0.999      & 0.998             \\ \bottomrule
  \end{tabular}
\end{adjustbox}
\label{Tab:table2}
\end{table}

\subsection{Generalization to New BMA Campaigns}
\label{sec:ppdet_rq3}

\begin{tcolorbox}[left=2pt,right=2pt,top=2pt,bottom=2pt]
  \textbf{RQ3:}  Can \ppdetabv identify web pages belonging to never-before-seen BMA campaigns?
\end{tcolorbox}

\begin{figure}[h]
  \begin{subfigure}[t]{.15\textwidth}
    \centering
    \frame{\includegraphics[width=\linewidth]{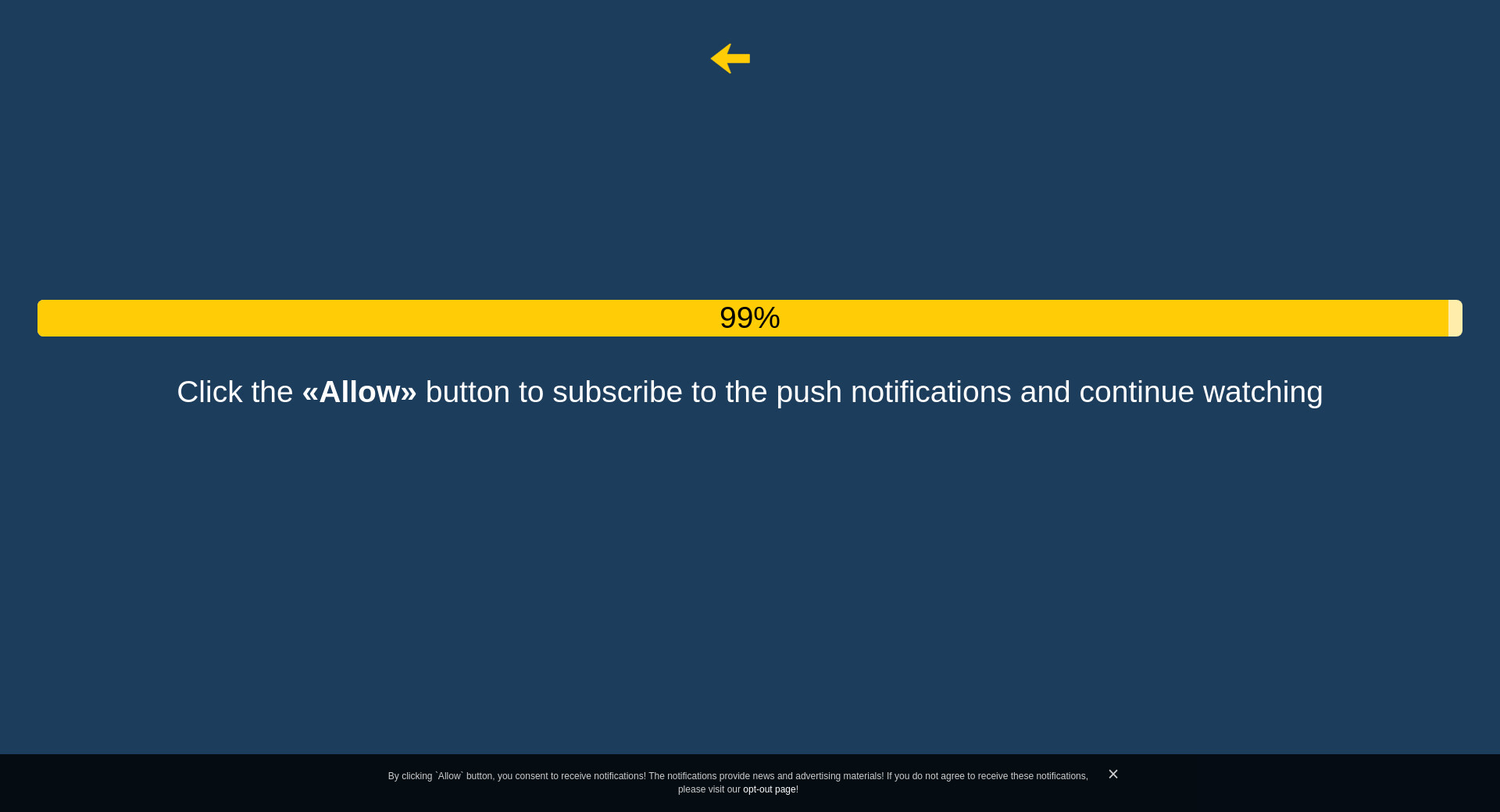}}  
    \caption{\tiny{Notification Stealing}}
    \label{fig:1_SEA}
  \end{subfigure}\hfil 
  \begin{subfigure}[t]{.15\textwidth}
    \centering
    \frame{\includegraphics[width=\linewidth]{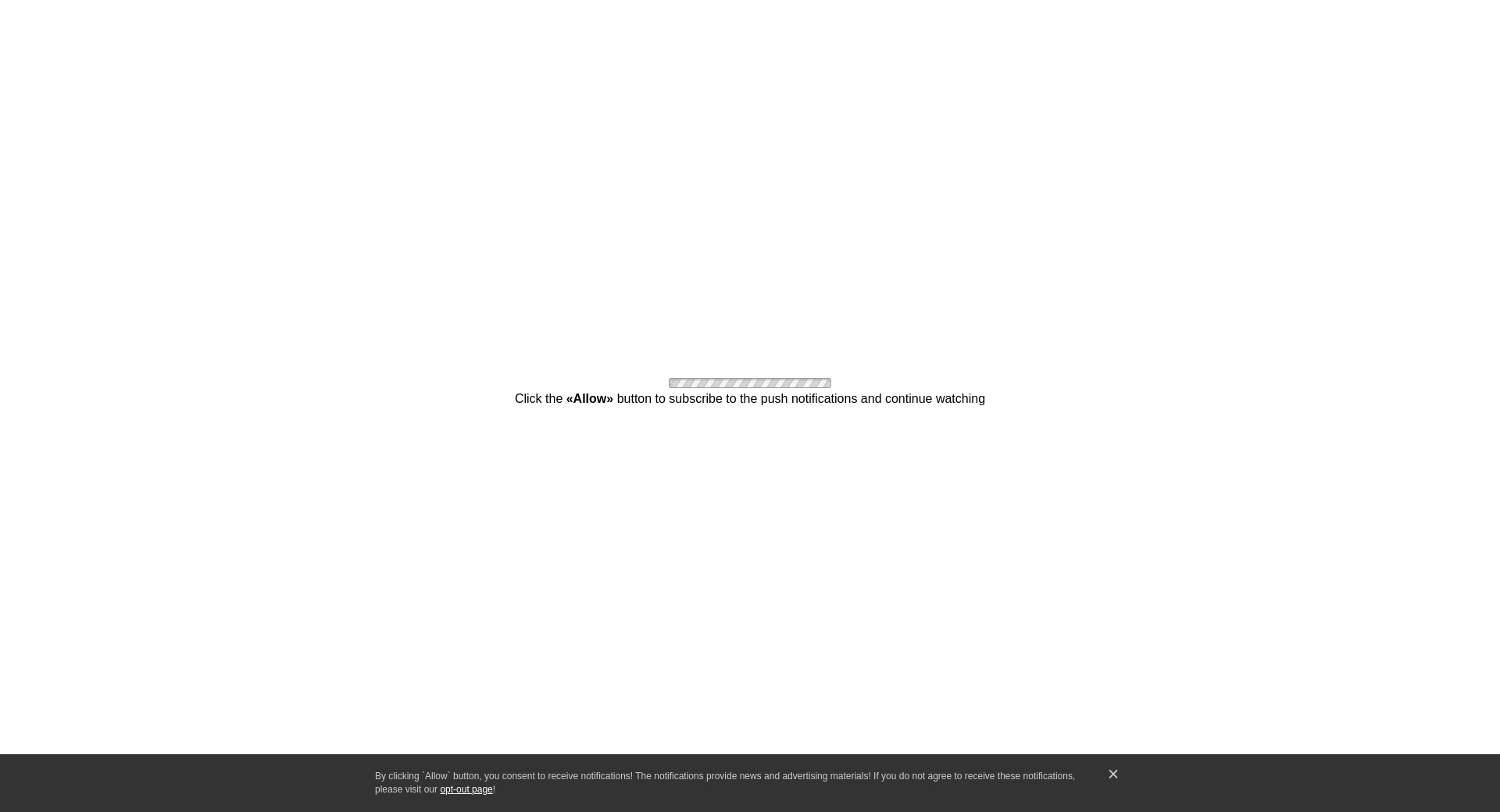}} 
    \caption{\tiny{Notification Stealing}}
    \label{fig:2_SEA}
  \end{subfigure}\hfil 
  \begin{subfigure}[t]{.15\textwidth}
      \centering
      \frame{\includegraphics[width=\linewidth]{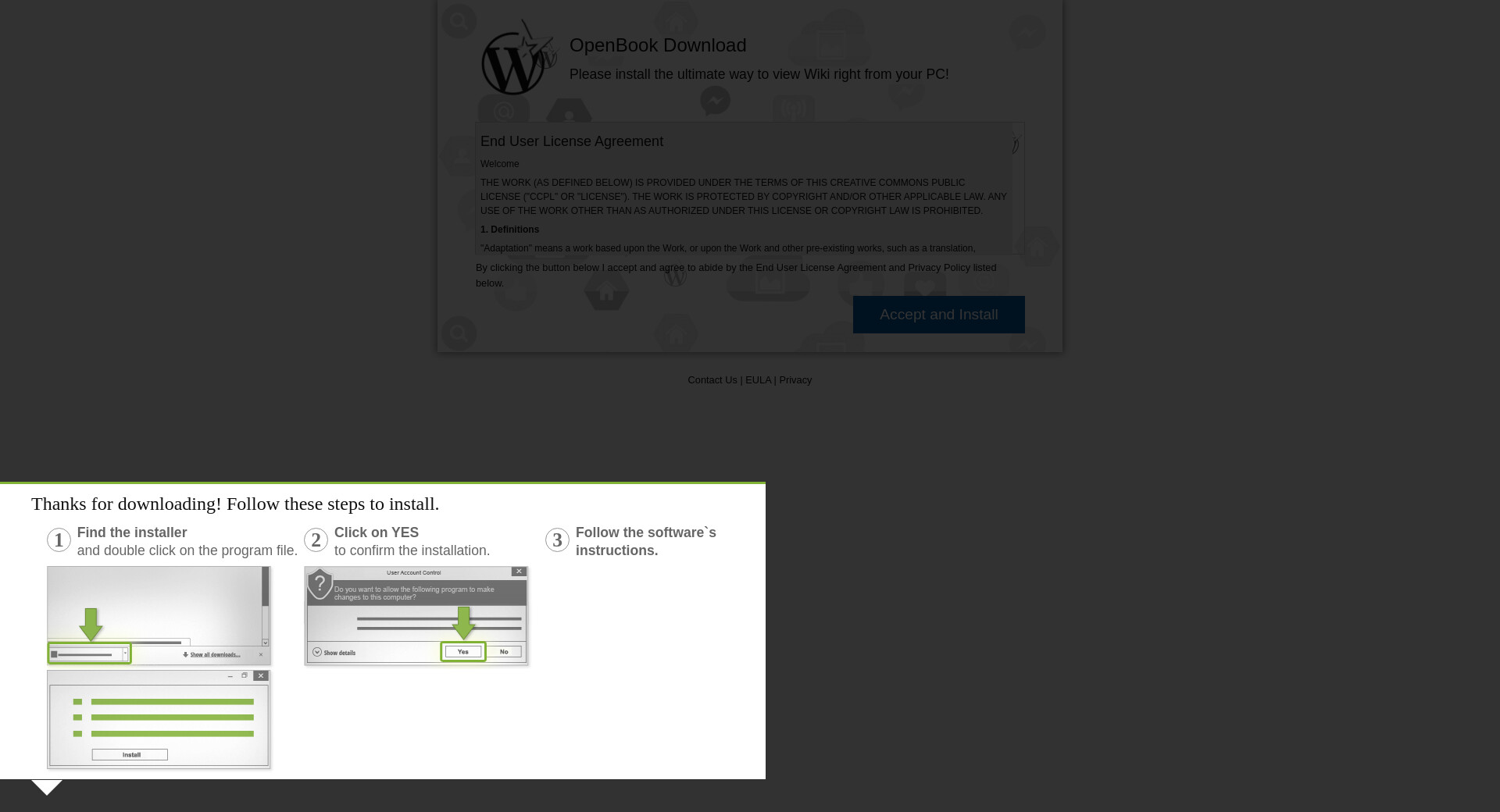}}  
      \caption{\tiny{Fake Software Download}}
      \label{fig:3_SEA}
  \end{subfigure}\hfil
  \vspace{2mm}
  \newline   
  \begin{subfigure}[t]{.15\textwidth}
    \centering
    \frame{\includegraphics[width=\linewidth]{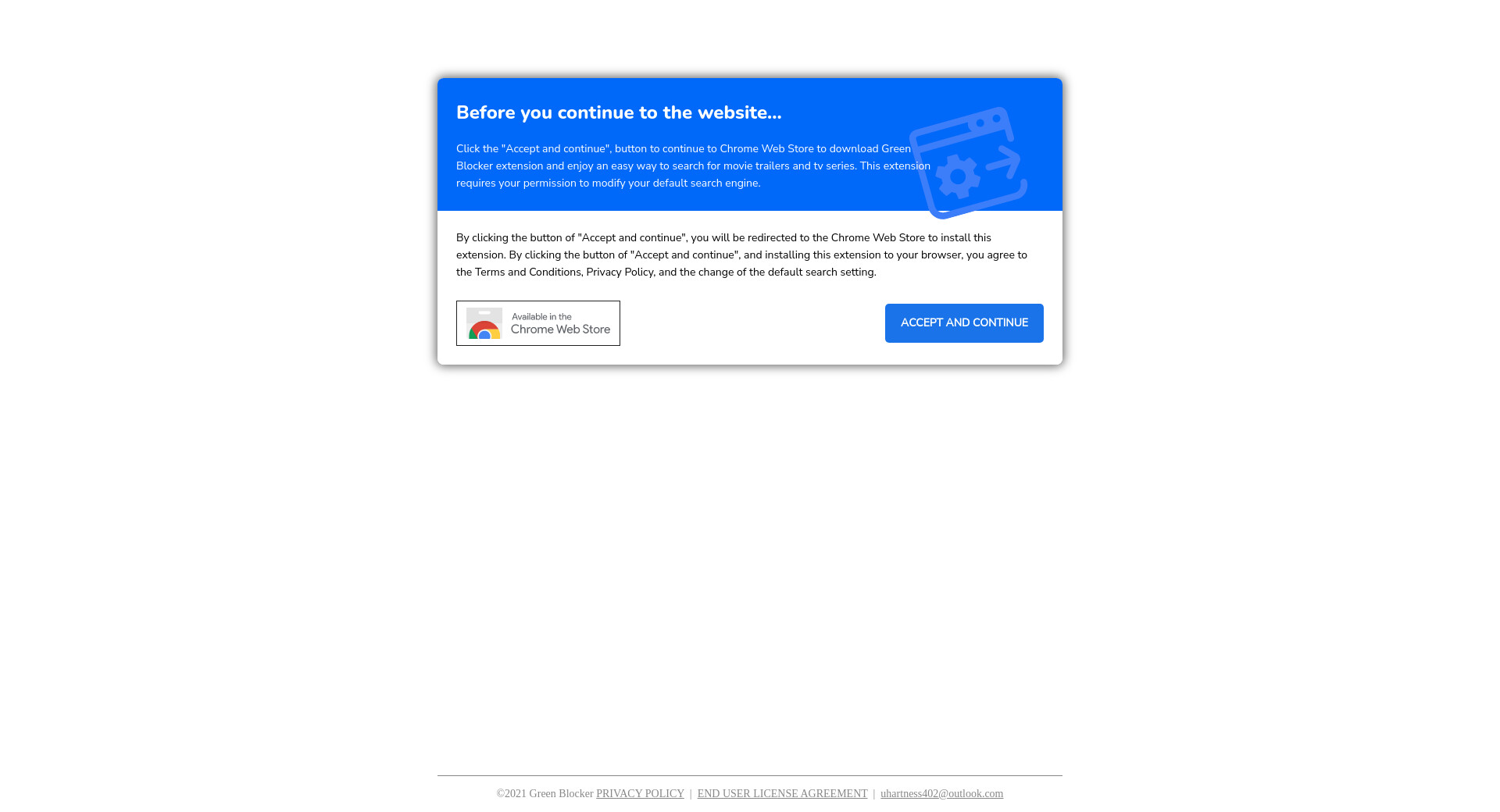}}
    \caption{\tiny{Fake Software Download}}
    \label{fig:4_SEA}
  \end{subfigure}\hfil
  \begin{subfigure}[t]{.15\textwidth}
    \centering
    \frame{\includegraphics[width=\linewidth]{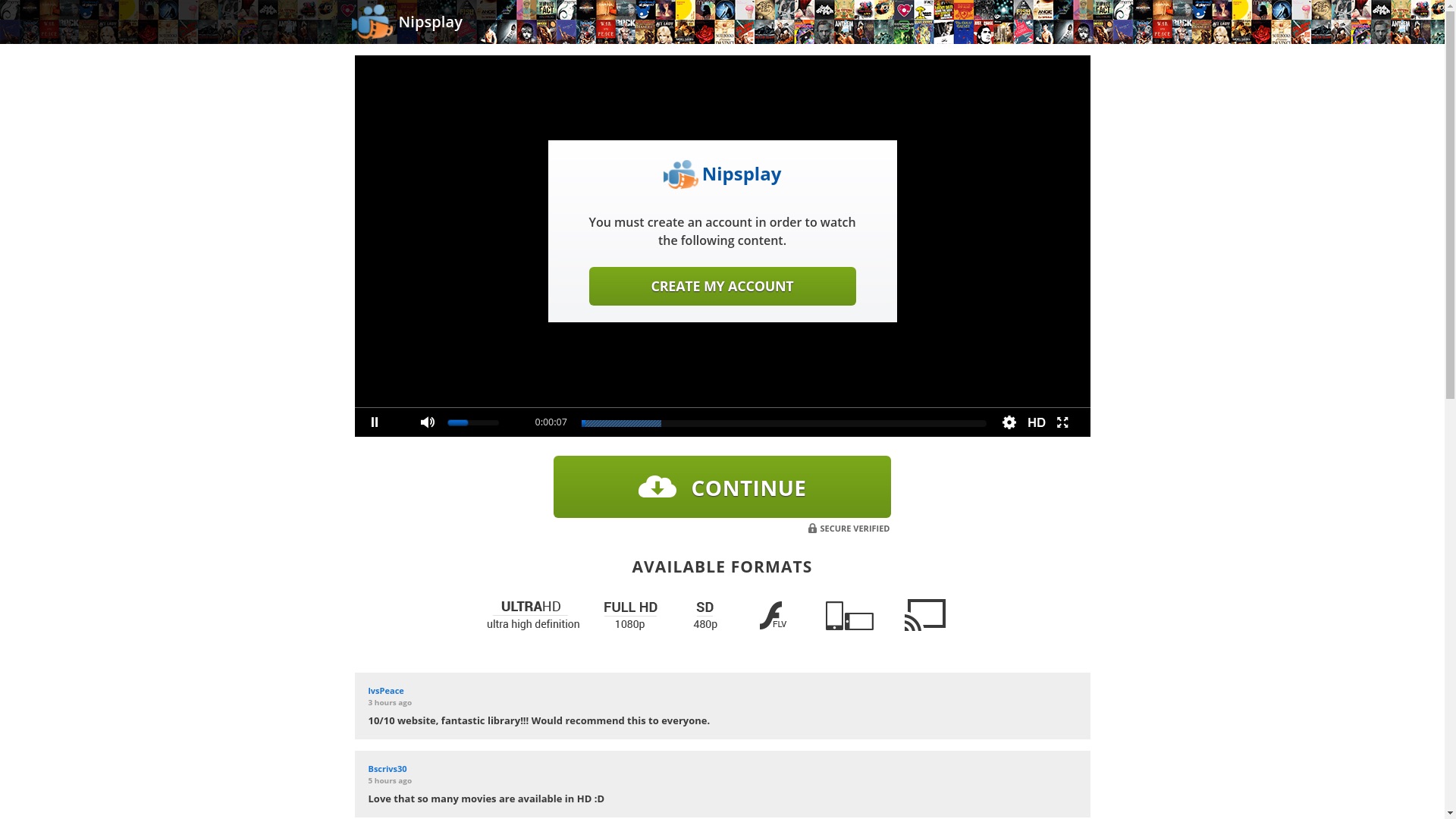}} 
    \caption{\tiny{Service Sign-up Scam}
 }   \label{fig:5_SEA}
  \end{subfigure}\hfil
  \begin{subfigure}[t]{.15\textwidth}
      \centering
      \frame{\includegraphics[width=\linewidth]{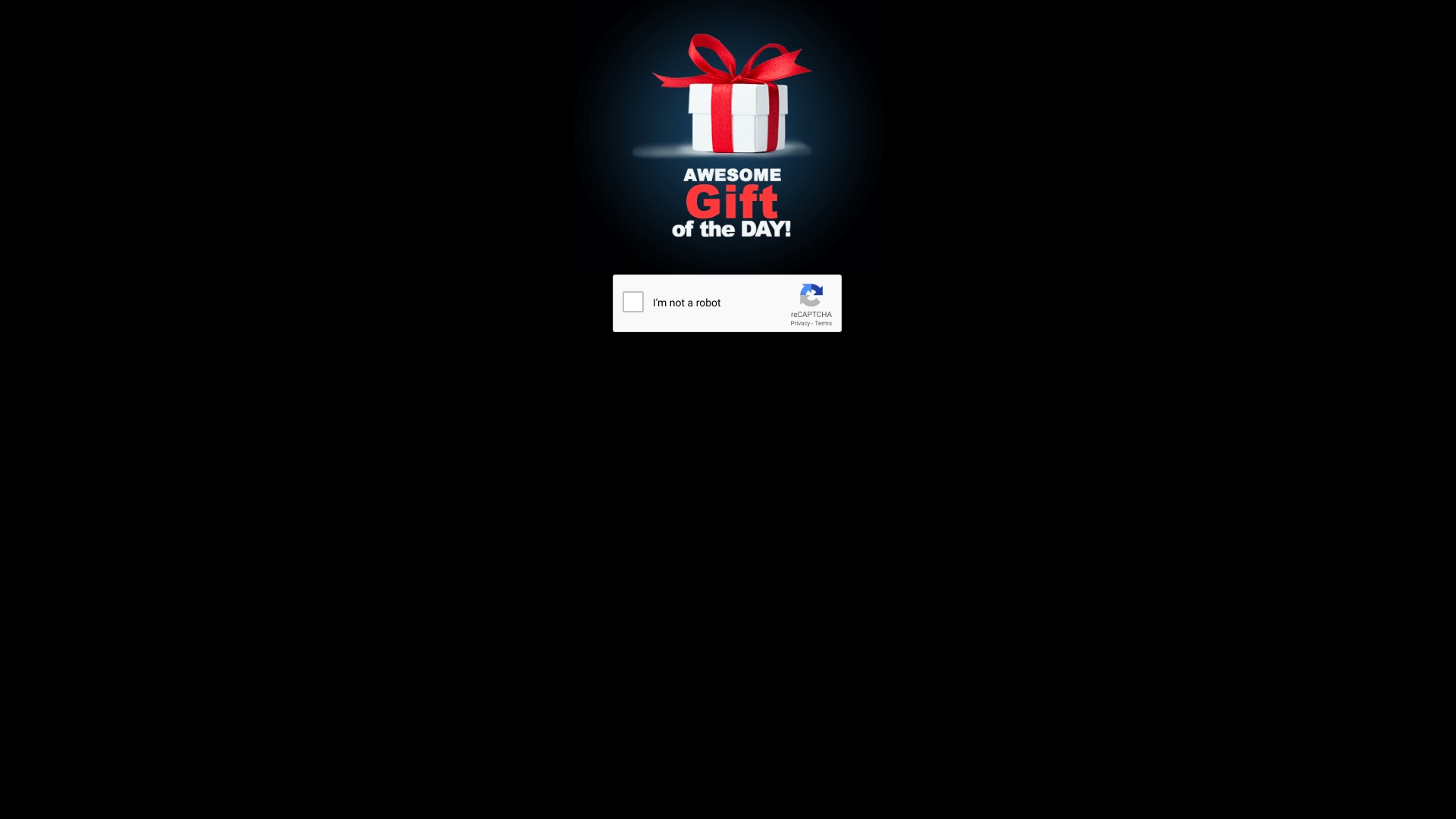}} 
      \caption{\tiny{Fake Lottery / Sweepsta}kes}
      \label{fig:6_SEA}
  \end{subfigure}\hfil
  \vspace{2mm}
  \newline  
  \begin{subfigure}[t]{.11\textwidth}
      \centering
      \frame{\includegraphics[width=\linewidth]{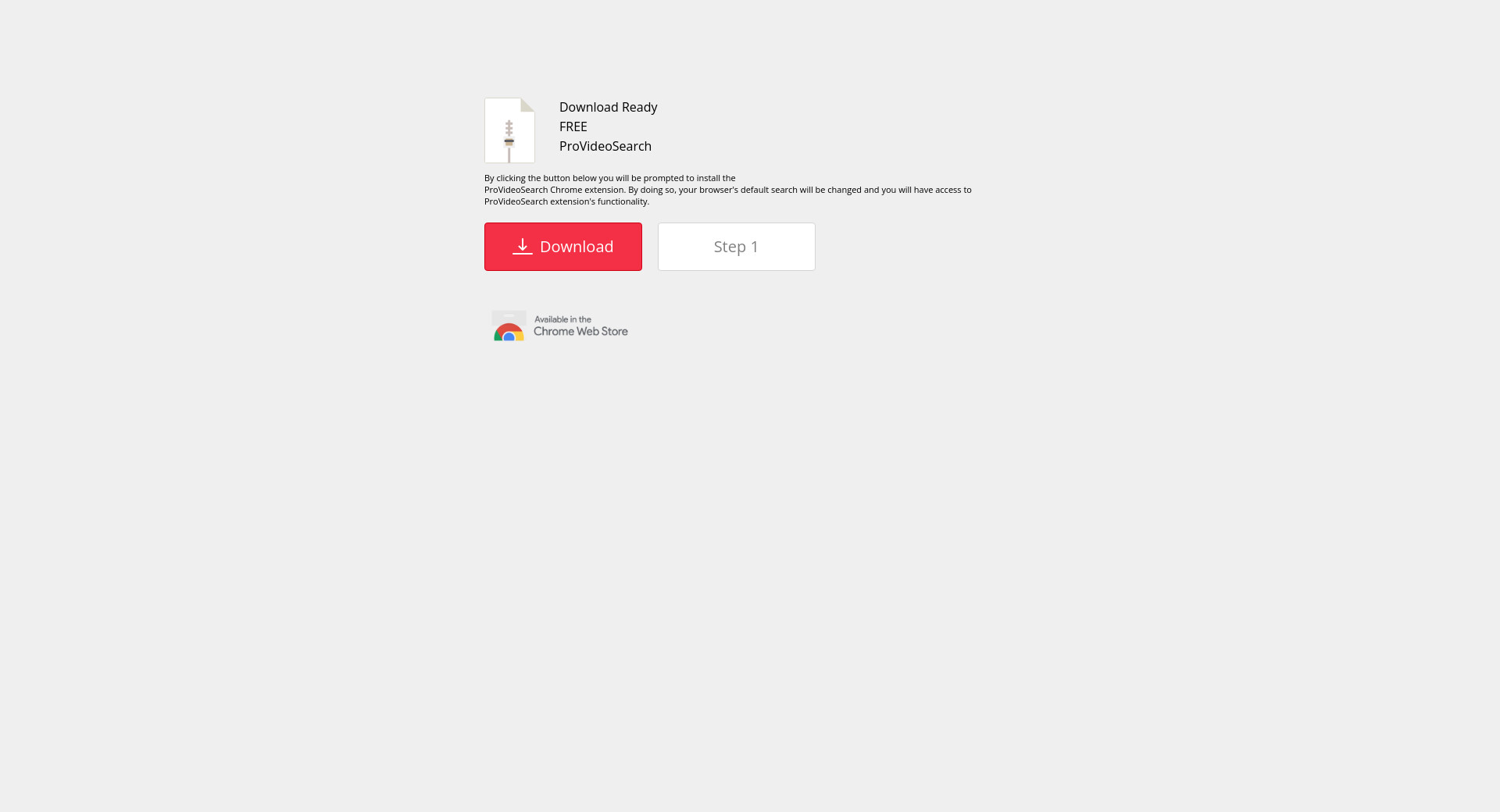}} 
      \caption{\tiny{Software Download}
}      \label{fig:7_SEA}
  \end{subfigure}\hfil
  \begin{subfigure}[t]{.11\textwidth}
      \centering
      \frame{\includegraphics[width=\linewidth]{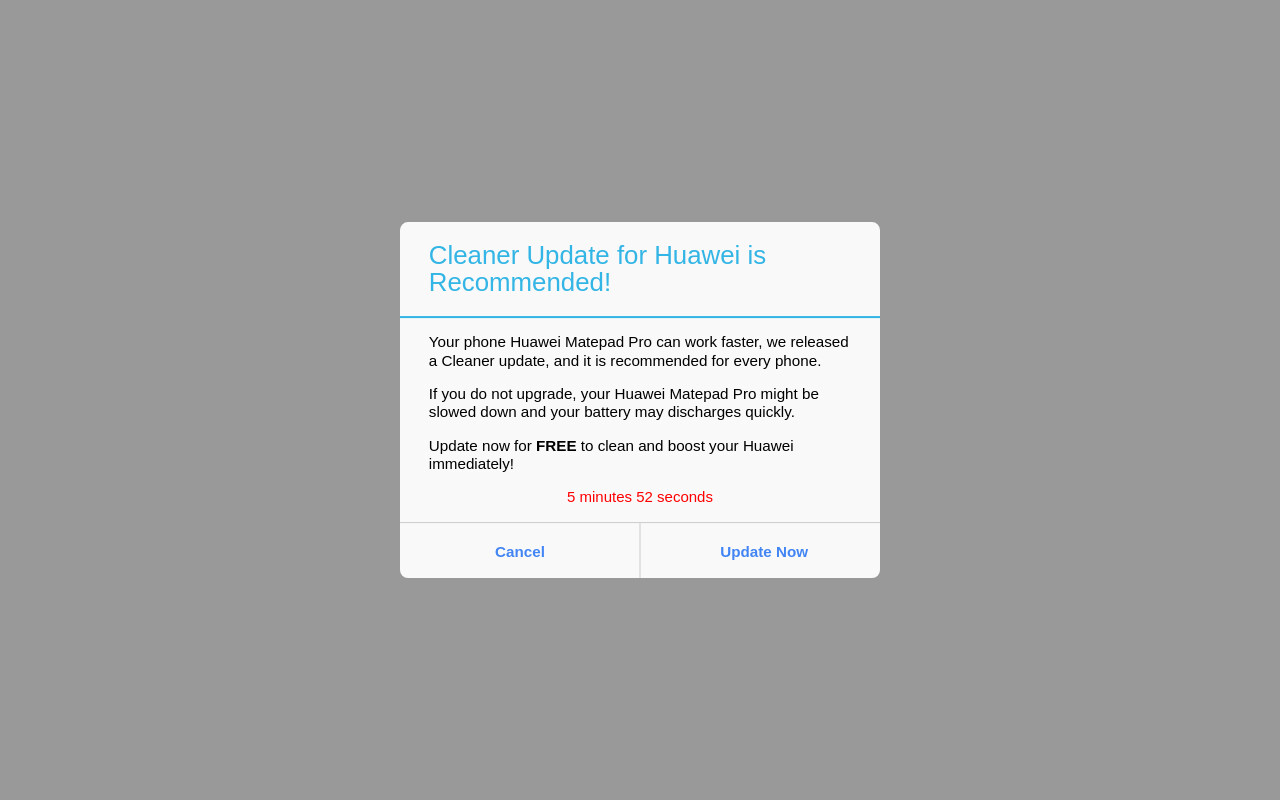}} 
      \caption{\tiny{Software Download}
 }     \label{fig:8_SEA}
  \end{subfigure}\hfil
  \begin{subfigure}[t]{.11\textwidth}
      \centering
      \frame{\includegraphics[width=\linewidth]{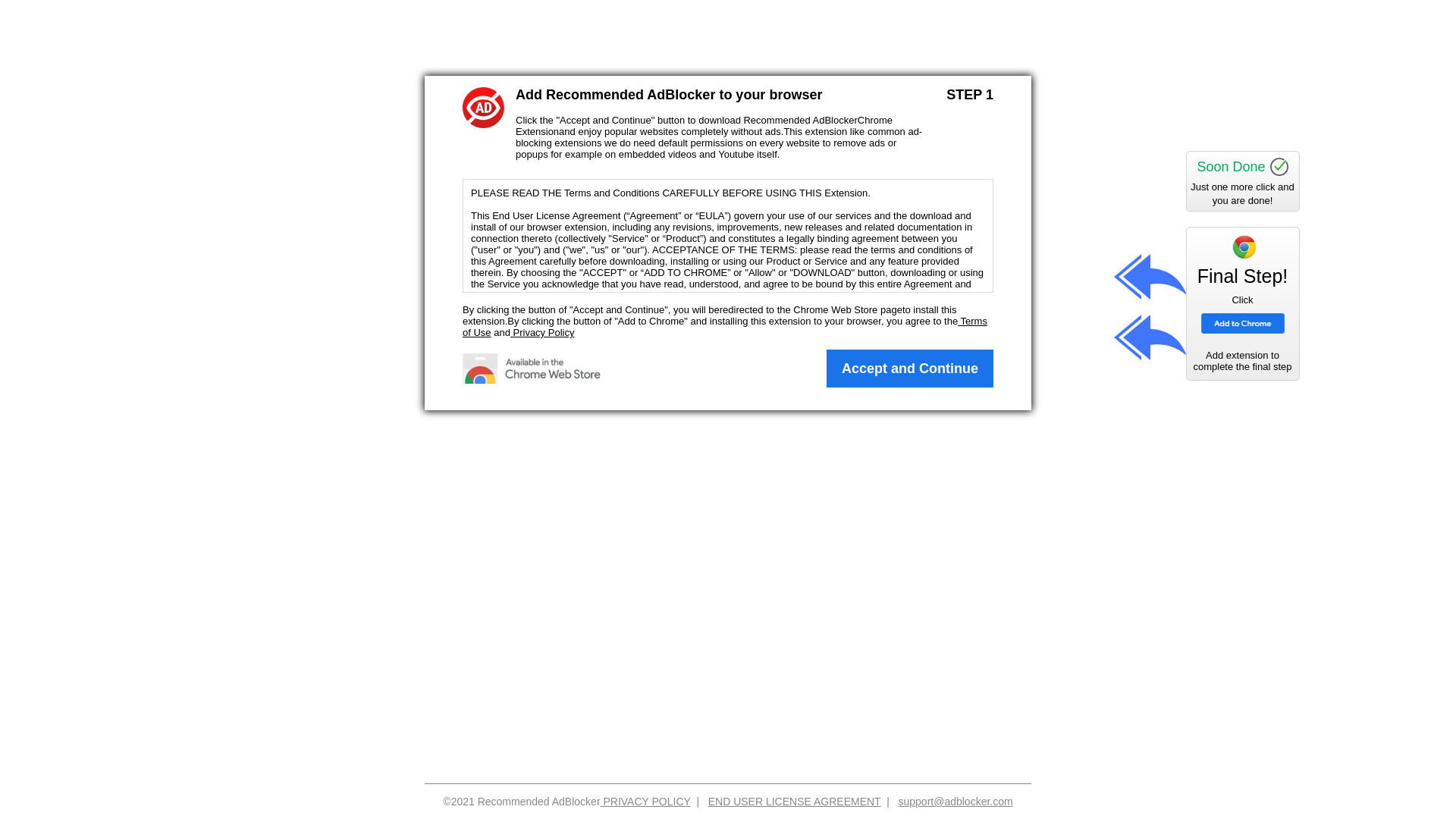}} 
      \caption{\tiny{Software Download}
  }    \label{fig:9_SEA}
  \end{subfigure}\hfil
  \begin{subfigure}[t]{.11\textwidth}
      \centering
      \frame{\includegraphics[width=\linewidth]{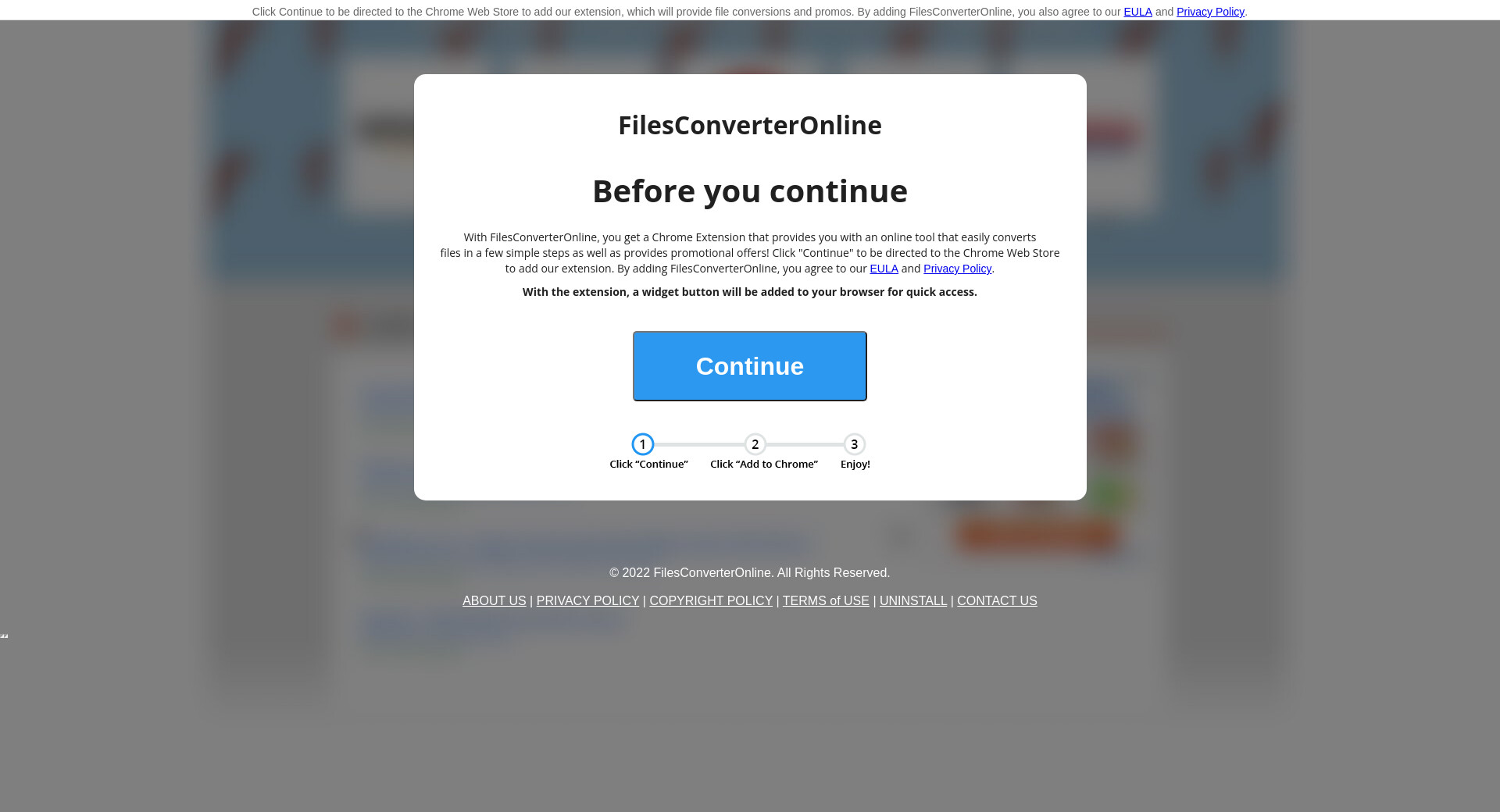}} 
      \caption{\tiny{Software Download}
 }     \label{fig:10_SEA}
  \end{subfigure}\hfil
  \caption{Examples of screenshots from each of the 10 randomly selected test campaigns.}
  \label{fig:MCLUSTERS}
\end{figure}

To setup this experiment, we randomly selected 10 BMA campaigns from our master dataset discussed in Section~\ref{sec:ppdet_setup_datasets}, $\{c_1, \dots, c_{10}\}$, to be excluded (in turn) from training and used for testing, and randomly selected 5,000 benign test pages (not seen during training) from randomly chosen screen resolutions. Example screenshots (one per campaign) for the 10 campaigns $\{c_1, \dots, c_{10}\}$ are shown in Figure~\ref{fig:MCLUSTERS}. Notice that while popular BMA categories, such as Software Download and Notification Stealing are (by chance) represented multiple times, the specific campaigns within those categories are different (e.g., different fake software being distributed and different visual appearance of the attack pages).

We then trained 10 models, $\{T_1, \dots, T_{10}\}$. For each model, $T_i$, we excluded one BMA campaign, $c_i$, and 500 randomly chosen benign images from training. We then tested the model on the excluded BMA campaign and benign images. The results of this experiment are reported in Table ~\ref{Tab:table41}.

We can see that \ppdetabv is able to generalize very well to never-before-seen attack campaigns, with an AUROC score at or above 0.996 and detection rates at or above 99\% at 1\% false positives for all test campaigns.

\begin{table}[h]
  \captionof{table}{Results for the generalization to never-before-seen BMA campaigns.\label{Tab:table41}} 
  \centering 
  \small
  \begin{adjustbox}{width=0.65\linewidth}
  \begin{tabular}{@{}ccccc@{}}
  \toprule
  \textbf{Model name}  & \textbf{\# Benign} & \textbf{\# BMA}            & \textbf{AUROC} & \textbf{DR at 1\% FP}\\ \midrule
  \textbf{Campaign 1}   & 500             & 281    & 1.0          & 0.993            \\
  \textbf{Campaign 2}   & 500             & 229    & 1.0          & 1.0              \\
  \textbf{Campaign 3}   & 500             & 59     & 0.998        & 1.0              \\
  \textbf{Campaign 4}   & 500             & 67     & 1.0          & 1.0              \\
  \textbf{Campaign 5}   & 500             & 910    & 0.999        & 0.998            \\
  \textbf{Campaign 6}   & 500             & 15     & 0.996        & 1.0              \\
  \textbf{Campaign 7}   & 500             & 43     & 1.0          & 1.0              \\
  \textbf{Campaign 8}   & 500             & 83     & 1.0          & 1.0              \\
  \textbf{Campaign 9}   & 500             & 224    & 1.0          & 1.0              \\
  \textbf{Campaign 10}  & 500             & 128    & 1.0          & 1.0              \\
  \midrule
  \textbf{Global}       & 5000            & 2039   & 0.999        & 0.993            \\ \bottomrule
  \end{tabular}
\end{adjustbox}
\end{table}

\subsection{Generalization to Fresh BMA Campaigns}
\label{sec:ppdet_rq4}

\begin{tcolorbox}[left=2pt,right=2pt,top=2pt,bottom=2pt]
  \textbf{RQ4:} Can \ppdetabv accurately identify {\em fresh} BMA attacks that were collected well after the training data?
\end{tcolorbox}

Real-world security systems must contend with the continuous evolution of attack content over time. Adversaries frequently modify or redesign the visual and textual elements of scam pages, rendering previously learned patterns less effective and introducing new features that may lie entirely \emph{out of distribution} with respect to the training data. This phenomenon—known as temporal drift—makes generalization over time a particularly difficult challenge. In such cases, the model must correctly identify BMAs that appear days, weeks, or even months after the training process has concluded.

To evaluate this, we tested the temporal generalization capabilities of the same model used in \textbf{RQ1} (see Subsection~\ref{sec:ppdet_rq1}). Specifically, we assessed its performance on a new BMA test set constructed from all data collected from our second BMA collection (see Section~\ref{sec:second_bma_collection}). To create the benign test set, we take 500 randomly selected images never seen in training from the benign data collection run.

Despite the difficulty of this setting, \ppdetabv achieves an AUROC score above 0.999 and a detection rate of 97.8\% at 1\% false positives. These results suggest that \ppdetabv can effectively generalize to previously unseen and temporally distant attack content.

\subsection{Adversarial Examples Evaluation}

\begin{tcolorbox}[left=2pt,right=2pt,top=2pt,bottom=2pt]
  \textbf{RQ5:}  Can \ppdetabv be strengthened against adversarial examples?
\end{tcolorbox}

Because BMA pages are under the attacker's control, the attacker can attempt to craft attack pages that explicitly attempt to evade \ppdetabv by injecting adversarial noise. At the same time, attackers must operate under two concurrent constraints: (i) attack pages must maintain their visual attack components, and (ii) the level of injected noise should be limited so not to be noticeable to users.

It is well known that adversarial examples forged by adding small perturbations can reduce deep learning models' classification accuracy~\cite{chakraborty2018adversarial} and that adversarial training is a popular approach to make models more robust to them~\cite{ijcai2021bai}, though perfectly defending against adversarial attacks is still a very open research question~\cite{AIGuardian,carlini2023llm}. 

Additionally, our model is unique in that it takes multimodal input in the form of images and their OCR extracted text. This opens another surface to the attacker. They have the ability to perturb the images to not only attempt to throw off the vision-based feature extractor, but also the OCR model. For instance, it has been shown that well-crafted image perturbations can cause OCR systems to misidentify, jumble, or even replace text with its semantic opposite \cite{ocr_post_correction, fooling_ocr_systems}. However, \ppdetabv does not directly train an OCR model and instead uses a pre-trained one (Tesseract~\cite{tesseract}). So, in order to achieve an effect similar to OCR-specific adversarial examples, we directly modify the textual input to our detection model, simulating the kinds of perturbations that may be introduced by adversarial changes to the image itself, according to previous work~\cite{ocr_post_correction, fooling_ocr_systems}.

In this section, we first test \ppdetabv against both white-box adversarial attacks on the image side and black-box adversarial attacks on the text side, and then show that adversarial training can significantly increase its robustness to such attacks.


\vspace{3pt}
\noindent
\textbf{Visual Perturbation.}~~
For visual perturbation, we follow guidance from highly cited prior work~\cite{madry2017towards} and use projected gradient descent (PGD) to construct adversarial images in a white-box attack setting. We implement these attacks using the Foolbox library~\cite{rauber2017foolboxnative,rauber2017foolbox}, targeting the visual feature extractor within our model.

The level of perturbation is controlled by the $\epsilon$ parameter. Larger values of $\epsilon$ produce more visible artifacts to the human eye and increase the likelihood of model evasion. We experiment with five levels of perturbation using $\epsilon \in \{2, 4, 8, 16, 32\}/255$. As shown in Figure~\ref{fig:PGD} in Appendix~\ref{sec:adv_dets}, noise is barely perceptible at $\epsilon = 2/255$, but becomes clearly visible by $\epsilon = 32/255$.

\vspace{3pt}
\noindent
\textbf{Textual Perturbation.}~~
For textual perturbation, we define five levels of increasing perturbation severity, drawing from established techniques in adversarial NLP (e.g., TextAttack~\cite{textattack}) and recent work on adversarial OCR examples~\cite{fooling_ocr_systems,ocr_post_correction}. Each level introduces progressively more aggressive distortions to the text:

\begin{itemize}[leftmargin=*]
  \item Level 1: Minor character-level noise that mimics OCR or typo artifacts.

  \item Level 2: Adds light paraphrasing or synonym substitution while preserving overall meaning.

  \item Level 3: Introduces misleading insertions or reordering that partially alters semantics.

  \item Level 4: Aggressively inverts meaning using antonyms, sentiment flips, or false claims.

  \item Level 5: Heavy semantic and character-level distortion that obscures or breaks meaning.
\end{itemize}

To perform these perturbations at scale, we use OpenAI's \texttt{gpt-4.1-mini} model via the API~\cite{openai_api}. Given the original input text, we prompt the model to generate perturbations corresponding to each of the five defined levels. Additional details on the prompting strategy, system setup, and evaluation of the resulting perturbations are provided in Appendix~\ref{sec:adv_dets}.

\begin{table}[h]
  \renewcommand{\arraystretch}{1.1}
  \captionof{table}{Results of adversarial training experiments\label{Tab:adv_exp}}
  \centering
  \small
  \begin{adjustbox}{width=0.65\linewidth}
    \begin{tabular}{@{}c|c|cc@{}}
      \toprule
      & & \multicolumn{2}{c}{\textbf{DR at 1\% FP}} \\
      \hline
      & & {\em Before adv. training} & {\em After adv. training} \\
      Level & \textbf{$\epsilon$} & ({\em Exp.1}) & ({\em Exp.2}) \\
      \hline
      clean  & clean                & 1.0   & 1.0 \\
      1      & $\nicefrac{2}{255}$  & 0.968 & 0.998 \\
      2      & $\nicefrac{4}{255}$  & 0.816 & 0.998 \\ 
      3      & $\nicefrac{8}{255}$  & 0.840 & 0.998 \\ 
      4      & $\nicefrac{16}{255}$ & 0.564 & 0.982 \\ 
      5      & $\nicefrac{32}{255}$ & 0.040 & 0.994 \\
      \bottomrule
    \end{tabular}
  \end{adjustbox}
\end{table}

The third column of Table~\ref{Tab:adv_exp} ({\em Exp.1}) reports the detection rate of our \ppdetabv model when evaluated on adversarial examples, using the same evaluation setup described in \textbf{RQ1} in Section~\ref{sec:ppdet_rq1}. These adversarial inputs consist of image perturbations generated using PGD with varying $\epsilon$ values, combined with text perturbations applied to the corresponding OCR-extracted content. As the attack strength increases—both in terms of visual distortion ($\epsilon$) and textual manipulation level—the detection rate at a 1\% false positive threshold drops significantly. Notably, at Level 4 and where $\epsilon = 16/255$, the detection rate falls to 0.564, despite the perturbations appearing only slightly noticeable to human users (see Figure~\ref{fig:PGD}, in the Appendix). At the highest attack level (Level 5), performance degrades drastically, with a detection rate of only 0.040, highlighting the model's vulnerability under coordinated multimodal adversarial conditions.

\vspace{3pt}
\noindent
\textbf{Adversarial Training.}~~
To harden \ppdetabv against coordinated multimodal attacks, we retrain the model with an adversarial curriculum. Starting from the clean image--text pairs described in Section~\ref{sec:ppdet_rq1}, we synthesize five tiers of adversarial examples by (i) applying PGD to the screenshots with $\epsilon \in \{2,4,8,16,32\}/255$ and (ii) injecting the corresponding Level 1-5 text perturbations. Each training epoch is built from 10,000 clean pairs augmented with 2,000 adversarial pairs from every tier, yielding a balanced mixture of benign, weak, and strong attacks.  

The adversarially trained model (Table~\ref{Tab:adv_exp}, {\em Exp.2}) maintains a detection rate above 98\% at 1\% FPR even under the most severe Level 5 attack, confirming that the curriculum substantially improves robustness while preserving performance on clean inputs.

\subsection{Pixel Patrol Defend Evaluation}
\label{sec:ppdef_eval}

To evaluate \ppdefabv, we examine its performance along two key dimensions: (1) user-facing latency, with a focus on the time required for inference during page analysis, and (2) device resource utilization, specifically CPU and RAM overhead. These metrics were measured using the setup described in Section~\ref{sec:ppdef_setup} across representative hardware classes, including desktop, tablet, and mobile devices. Our goal is to assess whether \ppdefabv can provide timely and accurate security verdicts without disrupting the user's browsing experience or overloading system resources. Results in this section pertain to the Firefox version of the extension to facilitate comparison across device types as mentioned in Section~\ref{sec:ppdef_setup_test_strat}. 
\footnote{Examples of the extension in action are available at \url{https://pixelpatrol3d.github.io/}.}

Importantly, latency should be interpreted in the context of user reaction time. Recent studies show that users typically take several seconds, often more than eight~\cite{phd_first_nodate}, to cognitively process and make decisions about new or dynamically changing web pages. This provides a practical upper bound for acceptable detection delays and reinforces the feasibility of real-time inference within this window. Moreover, because the extension operates in a non-blocking manner, these background computations do not interfere with or delay the user's browsing experience in any perceptible way. In practice, our objective is to alert the user within a few seconds from the rendering of a BMA page, before they take potentially harmful actions, while remaining unobtrusive to typical browsing behavior.


\label{sec:ppdef_lat_cdf}

\vspace{3pt}
\noindent
\textbf{Latency.}~~
We evaluate the latency experienced by users when interacting with \ppdefabv, focusing on the portion of time attributable to the most computationally intensive component: the inference pipeline. This includes down-sampling the screenshot, executing four parallel OCR workers, running the fused ONNX vision-text model, and returning the final verdict to the background script. 

\begin{figure}[h]
  \centering
  \includegraphics[width=0.8\linewidth]{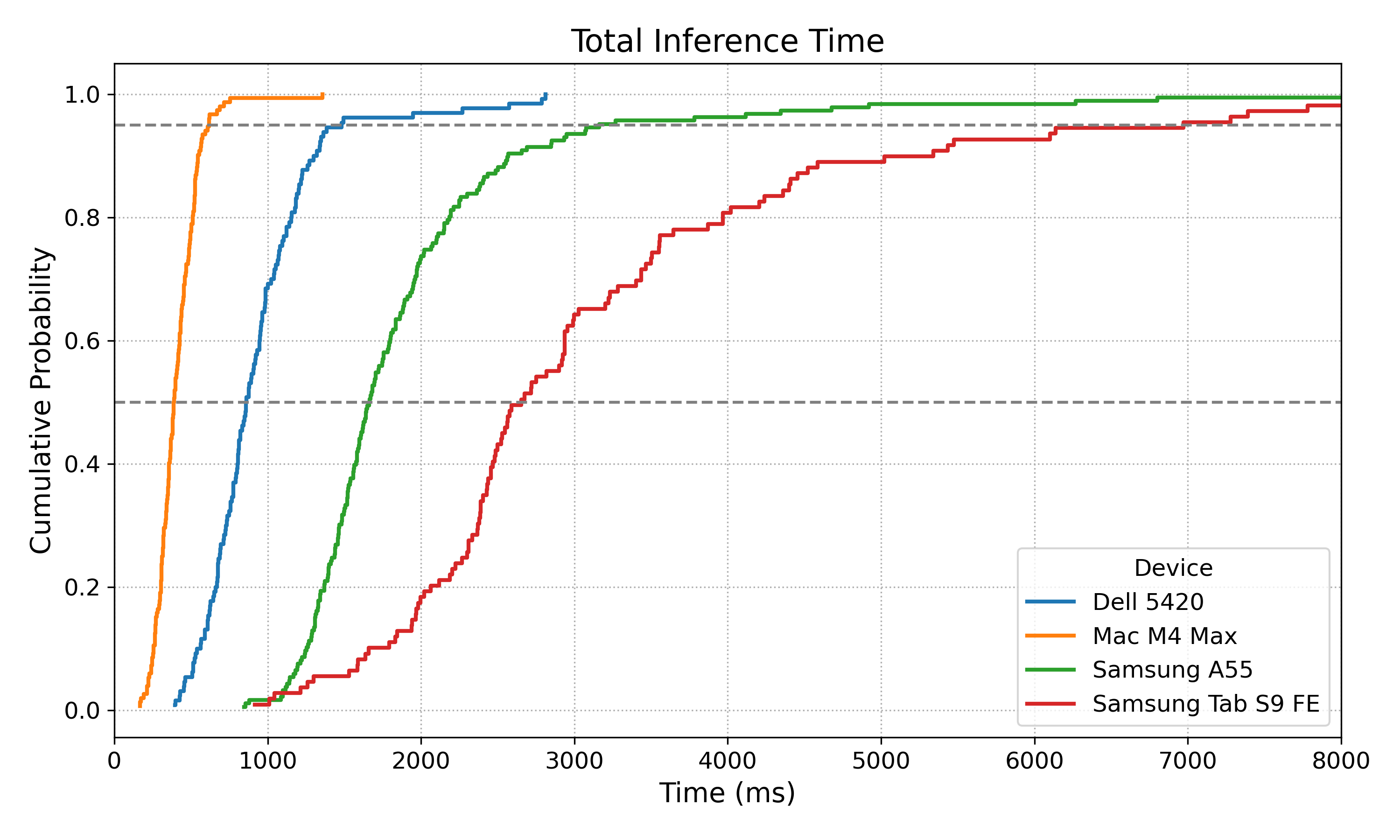}
  \caption{CDF of total inference latency for \ppdefabv. Dashed lines indicate median (P50) and 95\textsuperscript{th}-percentile (P95) thresholds.}
  \label{fig:ppdef_lat_cdf}
\end{figure}

Figure~\ref{fig:ppdef_lat_cdf} presents the CDF of total inference latency across a range of representative devices. As expected, the Apple M4 Max outperforms all others with a median latency of only 388 ms and a 95\textsuperscript{th}-percentile latency of only 618 ms. The Dell 5420 follows closely (median: 859 ms, 95\textsuperscript{th}: 1482 ms), well below the default 5-second scan interval. For mid-range mobile-class devices, the Samsung A55 reaches the P50 and P95 thresholds at 1662 ms and 3162 ms, respectively. The slowest performance was observed on the Samsung Tab S9 FE (median: 2653 ms; 95\textsuperscript{th}: 6972 ms). This is expected, as our extension implementation is a research prototype and the Tab S9 FE is a resource-constrained device with limited CPU and RAM compared to the others. A native implementation of the extension in the browser would likely improve performance on such devices very significantly. We leave this engineering optimization for future work.

It is also worth noting that these latency values are well aligned with human-computer interaction norms. For instance, Nielsen's foundational work~\cite{nielsen_usability_1994} defines 1 second as the upper bound for seamless interaction, and 10 seconds as the limit of user attention. More recent research~\cite{abbas_understanding_2022, arapakis_impact_2021} confirms that users begin to experience noticeable frustration only when delays exceed 7-10 seconds. Our measurements show that even on resource-constrained mobile and tablet devices, inference typically completes well within these bounds. Additionally, user studies suggest that individuals typically do not make immediate navigational decisions following a page load or content change. For example, a recent industry study found that users take an average of 8.5 seconds to make their first click on live webpages~\cite{phd_first_nodate}. Another large-scale analysis of Google user sessions reported an even longer average first click time of 14.6 seconds~\cite{how_people_use_google_search}. These findings indicate that users generally spend several seconds processing a page before interacting with it. Given that \ppdefabv consistently completes inference well within this decision window on most devices, its latency is likely sufficient to warn users before they take potentially harmful actions, while remaining unobtrusive to typical browsing behavior.

Moreover, real-world usage further mitigates user impact. Empirical data shows that a small fraction of domains dominate browsing activity, with over one-third of all web traffic concentrated in just the top 116 sites~\cite{xavier_web_2024}. For such high-confidence domains, our system bypasses full inference entirely, resulting in a worst-case latency of less than 100 ms. Even when encountering non-whitelisted pages, \ppdefabv only triggers inference upon significant page changes. In steady states, it relies on relatively low-cost perceptual hash (phash) checks, which further reduces latency.

Taken together, these results demonstrate that \ppdefabv delivers security-relevant classification with latencies that are both computationally efficient and behaviorally unobtrusive.

\begin{figure}[h]
  \centering

  \includegraphics[width=0.95\linewidth]{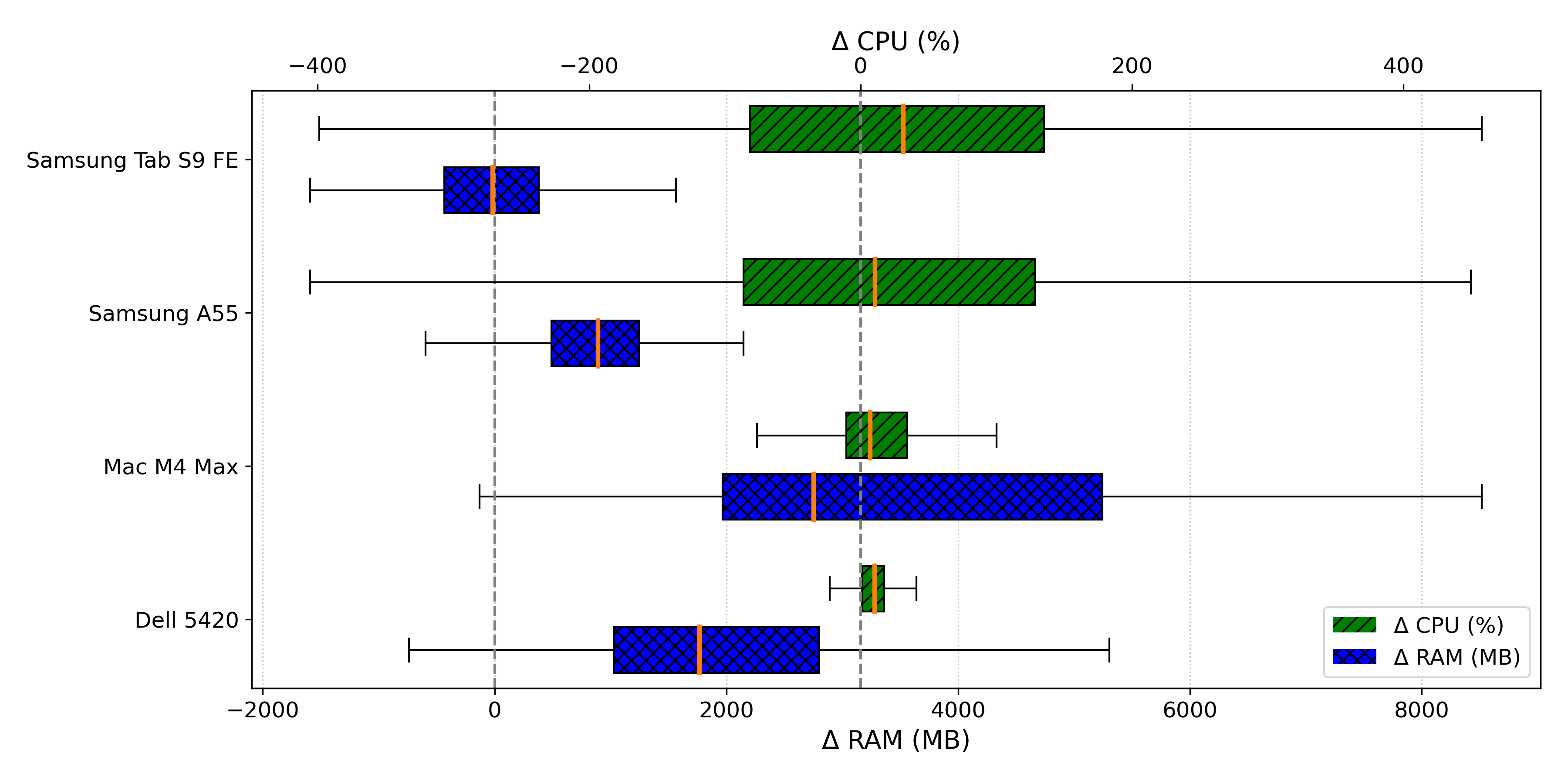}

  \caption{Difference in resource usage compared to the baseline during testing. CPU usage is measured as a percentage of total logical cores (e.g., 800\% = full use of 8 cores).}
  \label{fig:ppdef_ram_cpu_comp}
\end{figure}

\vspace{3pt}
\noindent
\textbf{Device Resource Usage.}~~
To assess the runtime efficiency of \ppdefabv, we also measured changes in device-side CPU and RAM utilization during operation relative to an idle browser baseline. These deltas were computed over the same test sessions described in Section~\ref{sec:ppdef_setup}, allowing us to quantify the additional computational burden incurred by the extension. CPU and RAM utilization deltas were computed by randomly sampling equal-length subsets of CPU and RAM usage values from both the idle (baseline) and extension-enabled (active) logs, and then taking element-wise differences. This ensured balanced comparisons even when log durations varied across sessions. Figure~\ref{fig:ppdef_ram_cpu_comp} summarizes these results. It should be noted again that ours is a research prototype implementation and that a native in-browser implementation of the detection model in the browser would likely improve performance on such devices very significantly. We leave more extensive engineering effort for future releases.




Overall, the increase in CPU usage across devices remained moderate and consistent with expectations for lightweight background inference. On the Dell 5420 laptop, CPU deltas were tightly bounded, with a median increase of 10.4\% and an interquartile range (IQR) of 1.1\% to 17.4\%, indicating stable, moderate overhead. The Mac M4 Max exhibited comparable performance, with a slightly lower median of 7.0\% but a wider IQR from --10.6\% to 34.0\%, reflecting greater variability likely due to background processes or dynamic power scaling.

Mobile-class devices showed greater variance, as expected given their constrained thermal envelopes and aggressive power management. The Samsung A55 had a median CPU delta of 10.75\%, similar to desktop-class systems, but a much broader IQR of --86.1\% to 128.6\%, reflecting larger spikes from OCR or inference activity. The Tab S9 FE showed the highest central CPU usage, with a median increase of 31.5\% and an IQR from --81.6\% to 135.1\%, again highlighting the bursty nature of on-device workloads. These spikes are short-lived, with inference completing quickly (see Section~\ref{sec:ppdef_lat_cdf}), and do not imply sustained load.

Memory usage followed a similar trend. On desktop-class devices, the extension incurred a moderate footprint, with median RAM increases of 1.77~GB (Dell 5420) and 2.75~GB (Mac M4 Max). The IQRs---1.03~GB to 2.80~GB for the Dell and 1.97~GB to 5.24~GB for the Mac---reflect differing levels of memory pressure across devices. Peaks of several gigabytes during inference arise from transient WebAssembly (WASM) heap allocations, intermediate tensors, screenshot/OCR buffers, and extension overhead, with Firefox's WASM memory management further contributing by reserving large regions that are not immediately reclaimed. A notable limitation is that, after extensive testing, we believe this behavior stems from the browser's memory management rather than the extension itself, and current extension APIs do not provide direct control over memory allocation or reclamation. Although these allocations are released after inference, the inability to manage them at the extension level further motivates a native browser integration, which could substantially reduce the footprint through more efficient memory handling and GPU acceleration (e.g., WebGPU).

On mobile devices, memory usage was more constrained. The Samsung A55 showed a median RAM increase of 893~MB, with an IQR from 490~MB to 1.24~GB. In contrast, the Tab S9 FE had a median RAM delta of --17~MB and an IQR from --434~MB to 380~MB, suggesting frequent memory reclamation or reuse by the Android OS under pressure. This behavior is common on mobile systems where memory optimization is tightly integrated into runtime scheduling.

Taken together, these results show that while \ppdefabv introduces measurable CPU and RAM overhead, the impact is transient, device-dependent, and consistent with what is expected from lightweight, on-device inference systems. Crucially, these resource demands do not persist across sessions and are tightly scoped to the moment of inference, ensuring minimal disruption to the user's broader browsing experience. Moreover, with additional engineering effort to integrate the detection system directly into the browser's codebase, its overhead could be reduced significantly.

\section{Discussion and Limitations}
\label{sec:discussion}
In this section, we outline additional considerations and limitations of our framework.

\emph{Generalization Limits.} While \ppdetabv achieves high accuracy on new screen sizes, attack instances, and full campaigns (Sections~\ref{sec:ppdet_rq2} and~\ref{sec:ppdet_rq3}), it may fail when presented with pages whose visual or textual features differ sharply from any seen during training. In practice, most BMAs rely on obvious visual cues or attack-related text designed to entice users~\cite{Vadrevu_IMC19}, so completely novel campaigns are rare. Still, continuous crawling and periodic retraining are necessary to incorporate emerging styles into the training set and gradually close these blind spots.

\emph{Analysis of False Positives.} Although uncommon, some benign pages can be flagged as malicious. For example, the screenshot in Table~\ref{Tab:diff_ex}(1) (Appendix~\ref{sec:bwa_fn_fp_ex}) comes from a benign site (\url{sariascosmetics.com}) but was misclassified because its ``Spin to Win" popup closely resembles the fake-lottery design used in many BMAs. Table~\ref{Tab:diff_ex}(2) shows a legitimate lottery-style popup; the model has learned to associate this pattern with malicious content. Another case, Table~\ref{Tab:diff_ex}(3), depicts a benign captcha page (\url{ozbargain.com.au}) whose layout is similar to ``Notification Stealing" BMAs that use fake ``I'm not a robot" challenges to trick users into clicking ``Allow" on a notifications permission popup (the popup itself is rendered outside the viewport). In both examples, the model mistakenly treated harmless UI elements as BMA indicators.

\emph{Analysis of False Negatives.} Conversely, some malicious pages may slip through as benign if their layout or text diverges significantly from the training examples. Tables~\ref{Tab:diff_ex}(5) and~\ref{Tab:diff_ex}(6) (Appendix) illustrate two such cases from the \emph{Secondary BMA Collection} (Section~\ref{sec:second_bma_collection}): both use unconventional designs that the model had not encountered during training. Despite this, \ppdetabv still detected 97.8\% of BMAs at 1\% false positive rate in the never-before-seen secondary campaigns collected months after training. Regular crawling and retraining remain the best strategy for reducing these blind spots over time.

\section{Related Works}

Most previous research on BMA type attacks focus on specific categories. For instance, Miramirkhani et al.~\cite{miramirkhani2017} performed an analysis of TSS scams and reported websites and phone numbers used by scammers. Stone-Gross et al. ~\cite{FakeAVEconomy} studied how users are persuaded to install fake antivirus software (AVs) by employing scareware attacks. Kharraz et al. ~\cite{KharrazRK18} employed an ML model to detect online survey scams that led victims to reveal personal information for fake prizes or content. Our work is different because we focus on creating an effective and practical in-browser system for detecting \emph{BMAs} more broadly without limitation to a particular category.


Other studies have focused primarily on identifying and measuring the occurrence of BMA campaigns, without offering a detection solution. For instance, Subramani et al. ~\cite{WebPushAds} proposed a system called PushAdMiner to collect and discover web push notification messages that can deliver BMA type campaigns, whereas Vadrevu et al.~\cite{Vadrevu_IMC19} introduced a measurement system that automatically collects examples of BMAs and identifies previously unknown ad networks that promote BMA campaigns.

Recently, Yang et al.~\cite{TRIDENT} proposed a first approach towards detecting and blocking BMA attacks in the browser. The proposed system, named TRIDENT, primarily targets malicious ads injected into publishers' webpages by low-reputation ad networks. Malicious ads are non-traditional ads that themselves utilize SE techniques, such as transparent overlays that perform clickjacking. TRIDENT is able to {\em indirectly} detect these attacks by identifying malicious ads. However, not all BMAs are distributed via such ads. Unlike~\cite{TRIDENT}, our framework directly aims at detecting BMAs by recognizing their visual and semantic traits.

TRIDENT and \ppabv operate at different attack chain levels. TRIDENT blocks JavaScript from low-reputation ad networks (early intervention), potentially preventing redirects to BMAs but risking false positives on legitimate pages reached via low-tier ads and false negatives on BMAs not distributed through monitored networks. \ppabv directly analyzes rendered page content (later intervention), detecting BMAs regardless of distribution mechanism but after navigation occurs.



Direct comparison is very difficult for several reasons: (i) For a side-by-side comparison, we would need to run both systems simultaneously in the same browser. However, at the time of writing, TRIDENT is more than two years old and was released as a Chromium patch, making it incompatible with modern extensions (including ours) and requiring extensive engineering effort to redesign and port to recent browser versions; (ii) TRIDENT's detections depend on the presence of ephemeral ad network code with frequently changing JavaScript-driven redirection chains, which are difficult to reproduce; (iii) The two systems intervene at different stages of the attack chain, which makes direct comparison inherently difficult due to their different detection targets. Taken together, these factors make direct comparison impractical. Therefore, we view the two systems as complementary for layered defense.

\ppabv operates at the \emph{visual level}, analyzing rendered screenshots and OCR text rather than semantic intent or DOM structure. This design exploits BMAs' reliance on overt visual cues (fake dialogs, urgent warnings) but has limitations: \ppabv cannot detect purely semantic deception without visual indicators (e.g., deceptive language without glaring visual manipulation).

\ppabv complements semantic/DOM-based defenses. Systems analyzing HTML forms~\cite{liu2022inferring, aljofey2023webphish} or JavaScript behavior~\cite{cova2010driveby} detect threats \ppabv might miss, while \ppabv identifies visually deceptive BMAs that evade DOM-based detectors via obfuscation (text-to-image, canvas rendering, font remapping). A layered defense combining visual and semantic analysis provides comprehensive coverage.

There is also a large body of research that focuses on detecting IHAs. While these attacks can be considered a subclass of Social Engineering~\cite{syafitri2022social}, they are characterized by different visual traits and attack tactics, as we discussed in Section~\ref{sec:intro}. Recent research has focused on detecting IHA websites using visual cues. For instance, Abdelnabi et al.~\cite{VisualPhishNet} used convolutional networks to detect IHA pages by visual similarity. Lin et al. ~\cite{lin2021phishpedia} detect IHA pages by visually detecting abused company logos. Ji et al. ~\cite{ji2025evaluating} performed a systematic evaluation of such visual similarity-based phishing detection models and showed that they often fail to generalize across large-scale, real-world phishing datasets, revealing key robustness limitations. Liu et al~\cite{liu2022inferring} presented a technique that is a combination of machine learning and browser instrumentation to detect an IHA page by not only using visual cues but also discovering the intention of an IHA webpage that perform credential stealing via web forms. Unlike the above solutions, our system is able to detect BMAs that differ from IHAs, even if no specific benign website/logo is abused or in absence of credential stealing attempts.

Hao et al.'s recent work~\cite{LogoMorph} uses an adversarial diffusion model to morph logos in IHA pages, subtly altering them to successfully evade detection while maintaining their recognizability to an end user. However, applying such techniques to BMA pages that our work focuses on would require altering multiple elements, potentially raising user suspicion due to the extensive changes needed.

\section{Conclusion}


We presented \ppddd, a browser-based framework that detects and defends against behavior-manipulation attacks (BMAs) using large-scale discovery, a resolution-agnostic multimodal model, and a lightweight, privacy-preserving extension. \ppabv achieves high accuracy on unseen screen sizes, new campaigns, and temporally distant data, with low latency and modest resource use. These results highlight its practicality and effectiveness against a largely overlooked class of social engineering threats.

\bibliographystyle{IEEEtran}
\bibliography{references_bib.bib}

\begin{thebibliography}{10}
\providecommand{\url}[1]{#1}
\csname url@samestyle\endcsname
\providecommand{\newblock}{\relax}
\providecommand{\bibinfo}[2]{#2}
\providecommand{\BIBentrySTDinterwordspacing}{\spaceskip=0pt\relax}
\providecommand{\BIBentryALTinterwordstretchfactor}{4}
\providecommand{\BIBentryALTinterwordspacing}{\spaceskip=\fontdimen2\font plus
\BIBentryALTinterwordstretchfactor\fontdimen3\font minus
  \fontdimen4\font\relax}
\providecommand{\BIBforeignlanguage}[2]{{%
\expandafter\ifx\csname l@#1\endcsname\relax
\typeout{** WARNING: IEEEtran.bst: No hyphenation pattern has been}%
\typeout{** loaded for the language `#1'. Using the pattern for}%
\typeout{** the default language instead.}%
\else
\language=\csname l@#1\endcsname
\fi
#2}}
\providecommand{\BIBdecl}{\relax}
\BIBdecl

\bibitem{mann2008hacking}
\BIBentryALTinterwordspacing
I.~Mann, \emph{Hacking the Human: Social Engineering Techniques and Security
  Countermeasures}.\hskip 1em plus 0.5em minus 0.4em\relax Ashgate Publishing,
  Limited, 2008. [Online]. Available:
  \url{https://books.google.com/books?id=1veE2f9rPOgC}
\BIBentrySTDinterwordspacing

\bibitem{KROMBHOLZ2015}
\BIBentryALTinterwordspacing
K.~Krombholz, H.~Hobel, M.~Huber, and E.~Weippl, ``Advanced social engineering
  attacks,'' \emph{Journal of Information Security and Applications}, vol.~22,
  pp. 113--122, 2015, special Issue on Security of Information and Networks.
  [Online]. Available:
  \url{http://www.sciencedirect.com/science/article/pii/S2214212614001343}
\BIBentrySTDinterwordspacing

\bibitem{zhang2021crawlphish}
P.~Zhang, A.~Oest, H.~Cho, Z.~Sun, R.~Johnson, B.~Wardman, S.~Sarker,
  A.~Kapravelos, T.~Bao, R.~Wang \emph{et~al.}, ``Crawlphish: Large-scale
  analysis of client-side cloaking techniques in phishing,'' in \emph{2021 IEEE
  Symposium on Security and Privacy (SP)}.\hskip 1em plus 0.5em minus
  0.4em\relax IEEE, 2021, pp. 1109--1124.

\bibitem{subramani2022phishinpatterns}
K.~Subramani, W.~Melicher, O.~Starov, P.~Vadrevu, and R.~Perdisci,
  ``Phishinpatterns: measuring elicited user interactions at scale on phishing
  websites,'' in \emph{Proceedings of the 22nd ACM Internet Measurement
  Conference}, 2022, pp. 589--604.

\bibitem{FakeAVEconomy}
B.~Stone-Gross, R.~Abman, R.~A. Kemmerer, C.~Kruegel, D.~G. Steigerwald, and
  G.~Vigna, ``The underground economy of fake antivirus software,'' in
  \emph{Economics of Information Security and Privacy III}, B.~Schneier,
  Ed.\hskip 1em plus 0.5em minus 0.4em\relax New York, NY: Springer New York,
  2013, pp. 55--78.

\bibitem{miramirkhani2017}
N.~Miramirkhani, O.~Starov, and N.~Nikiforakis, ``{Dial One for Scam: A
  Large-Scale Analysis of Technical Support Scams},'' in \emph{Proceedings of
  the 24th Network and Distributed System Security Symposium (NDSS)}, 2017.

\bibitem{KharrazRK18}
\BIBentryALTinterwordspacing
A.~Kharraz, W.~K. Robertson, and E.~Kirda, ``Surveylance: Automatically
  detecting online survey scams,'' in \emph{2018 {IEEE} Symposium on Security
  and Privacy, {SP} 2018, Proceedings, 21-23 May 2018, San Francisco,
  California, {USA}}, 2018, pp. 70--86. [Online]. Available:
  \url{https://doi.org/10.1109/SP.2018.00044}
\BIBentrySTDinterwordspacing

\bibitem{liu2023understanding}
J.~Liu, P.~Pun, P.~Vadrevu, and R.~Perdisci, ``Understanding, measuring, and
  detecting modern technical support scams,'' in \emph{2023 IEEE 8th European
  Symposium on Security and Privacy (EuroS\&P)}.\hskip 1em plus 0.5em minus
  0.4em\relax IEEE, 2023, pp. 18--38.

\bibitem{PhishingDetectionSurvey}
M.~{Khonji}, Y.~{Iraqi}, and A.~{Jones}, ``Phishing detection: A literature
  survey,'' \emph{IEEE Communications Surveys Tutorials}, vol.~15, no.~4, pp.
  2091--2121, Fourth 2013.

\bibitem{AfrozG11}
\BIBentryALTinterwordspacing
S.~Afroz and R.~Greenstadt, ``Phishzoo: Detecting phishing websites by looking
  at them,'' in \emph{Proceedings of the 5th {IEEE} International Conference on
  Semantic Computing {(ICSC} 2011), Palo Alto, CA, USA, September 18-21,
  2011}.\hskip 1em plus 0.5em minus 0.4em\relax {IEEE} Computer Society, 2011,
  pp. 368--375. [Online]. Available: \url{https://doi.org/10.1109/ICSC.2011.52}
\BIBentrySTDinterwordspacing

\bibitem{VisualPhishNet}
S.~Abdelnabi, K.~Krombholz, and M.~Fritz, ``Visualphishnet: Zero-day phishing
  website detection by visual similarity,'' in \emph{Proceedings of the 2020
  ACM SIGSAC Conference on Computer and Communications Security}, ser. CCS '20,
  2020, p. 1681–1698.

\bibitem{lin2021phishpedia}
Y.~Lin, R.~Liu, D.~M. Divakaran, J.~Y. Ng, Q.~Z. Chan, Y.~Lu, Y.~Si, F.~Zhang,
  and J.~S. Dong, ``Phishpedia: A hybrid deep learning based approach to
  visually identify phishing webpages.'' in \emph{USENIX Security Symposium},
  2021, pp. 3793--3810.

\bibitem{liu2022inferring}
R.~Liu, Y.~Lin, X.~Yang, S.~H. Ng, D.~M. Divakaran, and J.~S. Dong, ``Inferring
  phishing intention via webpage appearance and dynamics: A deep vision based
  approach,'' in \emph{31st USENIX Security Symposium (USENIX Security 22)},
  2022, pp. 1633--1650.

\bibitem{liu2023knowledge}
R.~Liu, Y.~Lin, Y.~Zhang, P.~H. Lee, and J.~S. Dong, ``Knowledge expansion and
  counterfactual interaction for $\{$Reference-Based$\}$ phishing detection,''
  in \emph{32nd USENIX Security Symposium (USENIX Security 23)}, 2023, pp.
  4139--4156.

\bibitem{liu2024less}
R.~Liu, Y.~Lin, X.~Teoh, G.~Liu, Z.~Huang, and J.~S. Dong, ``Less defined
  knowledge and more true alarms: Reference-based phishing detection without a
  pre-defined reference list,'' in \emph{33rd USENIX Security Symposium (USENIX
  Security 24)}, 2024, pp. 523--540.

\bibitem{li2024knowphish}
Y.~Li, C.~Huang, S.~Deng, M.~L. Lock, T.~Cao, N.~Oo, B.~Hooi, and H.~W. Lim,
  ``Knowphish: Large language models meet multimodal knowledge graphs for
  enhancing reference-based phishing detection,'' in \emph{33rd USENIX Security
  Symposium (USENIX Security 24)}, 2024.

\bibitem{fbi_int_crime_ann_report}
\BIBentryALTinterwordspacing
``\BIBforeignlanguage{en-us}{{FBI} {Releases} {Annual} {Internet} {Crime}
  {Report}}.'' [Online]. Available:
  \url{https://www.fbi.gov/news/press-releases/fbi-releases-annual-internet-crime-report}
\BIBentrySTDinterwordspacing

\bibitem{Vadrevu_IMC19}
P.~Vadrevu and { Roberto Perdisci}, ``What you see is {NOT} what you get:
  Discovering and tracking social engineering attack campaigns,'' in
  \emph{Proceedings of the ACM Internet Measurement Conference}, ser. { IMC},
  2019.

\bibitem{WebPushAds}
K.~Subramani, X.~Yuan, O.~Setayeshfar, P.~Vadrevu, K.~H. Lee, and R.~Perdisci,
  ``When push comes to ads: Measuring the rise of (malicious) push
  advertising,'' ser. IMC '20, 2020, p. 724–737.

\bibitem{ms_edge_scareware_blocker}
\BIBentryALTinterwordspacing
M.~E. Blog and M.~E. Team, ``\BIBforeignlanguage{en-US}{Stand up to scareware
  with scareware blocker, now available in preview in {Microsoft} {Edge}},''
  Jan. 2025. [Online]. Available:
  \url{https://blogs.windows.com/msedgedev/2025/01/27/stand-up-to-scareware-with-scareware-blocker/}
\BIBentrySTDinterwordspacing

\bibitem{GSB}
``Google safe browsing,'' \url{https://safebrowsing.google.com/}.

\bibitem{TRIDENT}
Z.~Yang, J.~Allen, M.~Landen, R.~Perdisci, and W.~Lee, ``{TRIDENT}: Towards
  detecting and mitigating web-based social engineering attacks,'' in
  \emph{{USENIX} Security Symposium}.\hskip 1em plus 0.5em minus 0.4em\relax
  {USENIX}, 2023.

\bibitem{zaoui2024taxonomy}
M.~Zaoui, B.~Yousra, S.~Yassine, M.~Yassine, and O.~Karim, ``A comprehensive
  taxonomy of social engineering attacks and defense mechanisms: Toward
  effective mitigation strategies,'' \emph{IEEE Access}, vol.~12, pp.
  72\,224--72\,241, 2024.

\bibitem{onnx}
\BIBentryALTinterwordspacing
``\BIBforeignlanguage{en}{{ONNX} {Runtime} {\textbar} {Home}}.'' [Online].
  Available: \url{https://onnxruntime.ai/}
\BIBentrySTDinterwordspacing

\bibitem{tesseract}
\BIBentryALTinterwordspacing
``Tesseract.js {\textbar} {Pure} {Javascript} {OCR} for 100 {Languages}!''
  [Online]. Available: \url{https://tesseract.projectnaptha.com/}
\BIBentrySTDinterwordspacing

\bibitem{MobileNets}
\BIBentryALTinterwordspacing
A.~G. Howard, M.~Zhu, B.~Chen, D.~Kalenichenko, W.~Wang, T.~Weyand,
  M.~Andreetto, and H.~Adam, ``Mobilenets: Efficient convolutional neural
  networks for mobile vision applications,'' \emph{CoRR}, vol. abs/1704.04861,
  2017. [Online]. Available: \url{http://arxiv.org/abs/1704.04861}
\BIBentrySTDinterwordspacing

\bibitem{bertmini}
\BIBentryALTinterwordspacing
I.~Turc, M.-W. Chang, K.~Lee, and K.~Toutanova, ``Well-{Read} {Students}
  {Learn} {Better}: {On} the {Importance} of {Pre}-training {Compact}
  {Models},'' Sep. 2019, arXiv:1908.08962 [cs]. [Online]. Available:
  \url{http://arxiv.org/abs/1908.08962}
\BIBentrySTDinterwordspacing

\bibitem{tranco}
``{A} research-oriented top sites ranking hardened against manipulation -
  {T}ranco --- tranco-list.eu,'' \url{https://tranco-list.eu}, [Accessed
  13-Jun-2023].

\bibitem{phash_algo}
\BIBentryALTinterwordspacing
``commonsmachinery/blockhash,'' Mar. 2025, original-date: 2014-09-02T17:46:34Z.
  [Online]. Available: \url{https://github.com/commonsmachinery/blockhash}
\BIBentrySTDinterwordspacing

\bibitem{hamming1950error}
R.~W. Hamming, ``Error detecting and error correcting codes,'' \emph{Bell
  System Technical Journal}, vol.~29, no.~2, pp. 147--160, 1950.

\bibitem{common_crawl}
\BIBentryALTinterwordspacing
``\BIBforeignlanguage{en}{Common {Crawl} - {Open} {Repository} of {Web} {Crawl}
  {Data}}.'' [Online]. Available: \url{https://commoncrawl.org/}
\BIBentrySTDinterwordspacing

\bibitem{krippendorff_computing_2011}
\BIBentryALTinterwordspacing
K.~Krippendorff, ``\BIBforeignlanguage{en}{Computing {Krippendorff}'s
  {Alpha}-{Reliability}},'' no.~43, Jan. 2011. [Online]. Available:
  \url{https://repository.upenn.edu/handle/20.500.14332/2089}
\BIBentrySTDinterwordspacing

\bibitem{de2012calculating}
K.~De~Swert, ``Calculating inter-coder reliability in media content analysis
  using krippendorff's alpha,'' \emph{Center for Politics and Communication},
  vol.~15, pp. 1--15, 2012.

\bibitem{perera2019learning}
P.~Perera and V.~M. Patel, ``Learning deep features for one-class
  classification,'' in \emph{Proceedings of the IEEE/CVF Conference on Computer
  Vision and Pattern Recognition (CVPR)}, 2019, pp. 802--811.

\bibitem{he2009learning}
H.~He and E.~A. Garcia, ``Learning from imbalanced data,'' \emph{IEEE
  Transactions on Knowledge and Data Engineering}, vol.~21, no.~9, pp.
  1263--1284, 2009.

\bibitem{chakraborty2018adversarial}
A.~Chakraborty, M.~Alam, V.~Dey, A.~Chattopadhyay, and D.~Mukhopadhyay,
  ``Adversarial attacks and defences: A survey,'' \emph{arXiv preprint
  arXiv:1810.00069}, 2018.

\bibitem{ijcai2021bai}
\BIBentryALTinterwordspacing
T.~Bai, J.~Luo, J.~Zhao, B.~Wen, and Q.~Wang, ``Recent advances in adversarial
  training for adversarial robustness,'' in \emph{Proceedings of the Thirtieth
  International Joint Conference on Artificial Intelligence, {IJCAI-21}}, Z.-H.
  Zhou, Ed.\hskip 1em plus 0.5em minus 0.4em\relax International Joint
  Conferences on Artificial Intelligence Organization, 8 2021, pp. 4312--4321,
  survey Track. [Online]. Available:
  \url{https://doi.org/10.24963/ijcai.2021/591}
\BIBentrySTDinterwordspacing

\bibitem{AIGuardian}
H.~Zhu, S.~Zhang, and K.~Chen, ``Ai-guardian: Defeating adversarial attacks
  using backdoors,'' in \emph{2023 IEEE Symposium on Security and Privacy
  (SP)}, 2023, pp. 701--718.

\bibitem{carlini2023llm}
N.~Carlini, ``A llm assisted exploitation of ai-guardian,'' 2023.

\bibitem{ocr_post_correction}
N.~Imam, V.~Vasilakis, and D.~Kolovos, ``\BIBforeignlanguage{en}{{OCR}
  post-correction for detecting adversarial text images}.''

\bibitem{fooling_ocr_systems}
\BIBentryALTinterwordspacing
C.~Song and V.~Shmatikov, ``Fooling {OCR} {Systems} with {Adversarial} {Text}
  {Images},'' Feb. 2018, arXiv:1802.05385 [cs]. [Online]. Available:
  \url{http://arxiv.org/abs/1802.05385}
\BIBentrySTDinterwordspacing

\bibitem{madry2017towards}
A.~Madry, A.~Makelov, L.~Schmidt, D.~Tsipras, and A.~Vladu, ``Towards deep
  learning models resistant to adversarial attacks,'' \emph{arXiv preprint
  arXiv:1706.06083}, 2017.

\bibitem{rauber2017foolboxnative}
\BIBentryALTinterwordspacing
J.~Rauber, R.~Zimmermann, M.~Bethge, and W.~Brendel, ``Foolbox native: Fast
  adversarial attacks to benchmark the robustness of machine learning models in
  pytorch, tensorflow, and jax,'' \emph{Journal of Open Source Software},
  vol.~5, no.~53, p. 2607, 2020. [Online]. Available:
  \url{https://doi.org/10.21105/joss.02607}
\BIBentrySTDinterwordspacing

\bibitem{rauber2017foolbox}
\BIBentryALTinterwordspacing
J.~Rauber, W.~Brendel, and M.~Bethge, ``Foolbox: A python toolbox to benchmark
  the robustness of machine learning models,'' in \emph{Reliable Machine
  Learning in the Wild Workshop, 34th International Conference on Machine
  Learning}, 2017. [Online]. Available: \url{http://arxiv.org/abs/1707.04131}
\BIBentrySTDinterwordspacing

\bibitem{textattack}
\BIBentryALTinterwordspacing
``{QData}/{TextAttack},'' May 2025, original-date: 2019-10-15T00:51:44Z.
  [Online]. Available: \url{https://github.com/QData/TextAttack}
\BIBentrySTDinterwordspacing

\bibitem{openai_api}
\BIBentryALTinterwordspacing
``\BIBforeignlanguage{en-US}{Overview - {OpenAI} {API}}.'' [Online]. Available:
  \url{https://platform.openai.com}
\BIBentrySTDinterwordspacing

\bibitem{phd_first_nodate}
\BIBentryALTinterwordspacing
J.~S. PhD, W.~S. PhD, E.~Short, and J.~L. PhD,
  ``\BIBforeignlanguage{en-US}{First {Click} {Times} on {Websites} {Versus}
  {Images} – {MeasuringU}}.'' [Online]. Available:
  \url{https://measuringu.com/first-click-times-on-websites-versus-images/}
\BIBentrySTDinterwordspacing

\bibitem{nielsen_usability_1994}
J.~Nielsen, \emph{Usability {Engineering}}.\hskip 1em plus 0.5em minus
  0.4em\relax San Francisco, CA, USA: Morgan Kaufmann Publishers Inc., Oct.
  1994.

\bibitem{abbas_understanding_2022}
\BIBentryALTinterwordspacing
T.~Abbas, U.~Gadiraju, V.-J. Khan, and P.~Markopoulos, ``Understanding {User}
  {Perceptions} of {Response} {Delays} in {Crowd}-{Powered} {Conversational}
  {Systems},'' \emph{Proc. ACM Hum.-Comput. Interact.}, vol.~6, no. CSCW2, pp.
  345:1--345:42, Nov. 2022. [Online]. Available:
  \url{https://dl.acm.org/doi/10.1145/3555765}
\BIBentrySTDinterwordspacing

\bibitem{arapakis_impact_2021}
\BIBentryALTinterwordspacing
I.~Arapakis, S.~Park, and M.~Pielot, ``\BIBforeignlanguage{en}{Impact of
  {Response} {Latency} on {User} {Behaviour} in {Mobile} {Web} {Search}},''
  Jan. 2021. [Online]. Available: \url{https://arxiv.org/abs/2101.09086v1}
\BIBentrySTDinterwordspacing

\bibitem{how_people_use_google_search}
\BIBentryALTinterwordspacing
``\BIBforeignlanguage{en-US}{How {People} {Use} {Google} {Search} ({New} {User}
  {Behavior} {Study})},'' Aug. 2020. [Online]. Available:
  \url{https://backlinko.com/google-user-behavior}
\BIBentrySTDinterwordspacing

\bibitem{xavier_web_2024}
\BIBentryALTinterwordspacing
H.~S. Xavier, ``\BIBforeignlanguage{en}{The {Web} unpacked: a quantitative
  analysis of global {Web} usage},'' Apr. 2024. [Online]. Available:
  \url{https://arxiv.org/abs/2404.17095v2}
\BIBentrySTDinterwordspacing

\bibitem{aljofey2023webphish}
A.~Aljofey, X.~Li, Z.~Cheng, X.~Fu, and X.~Xu, ``Webphish: Detecting phishing
  web pages by exploiting raw url and html content with deep learning,''
  \emph{Expert Systems with Applications}, vol. 225, p. 120230, 2023.

\bibitem{cova2010driveby}
M.~Cova, C.~Kruegel, and G.~Vigna, ``Detection and analysis of
  drive-by-download attacks and malicious {JavaScript} code,'' in
  \emph{Proceedings of the 19th International Conference on World Wide Web
  (WWW)}.\hskip 1em plus 0.5em minus 0.4em\relax ACM, 2010, pp. 281--290.

\bibitem{syafitri2022social}
W.~Syafitri, Z.~Shukur, U.~Asma’Mokhtar, R.~Sulaiman, and M.~A. Ibrahim,
  ``Social engineering attacks prevention: A systematic literature review,''
  \emph{IEEE Access}, vol.~10, pp. 39\,325--39\,343, 2022.

\bibitem{ji2025evaluating}
\BIBentryALTinterwordspacing
F.~Ji, K.~Lee, H.~Koo, W.~You, E.~Choo, H.~Kim, and D.~Kim, ``Evaluating the
  effectiveness and robustness of visual similarity-based phishing detection
  models,'' in \emph{Proceedings of the 34th USENIX Security Symposium (USENIX
  Security 2025)}, 2025, short Presentation. [Online]. Available:
  \url{https://www.usenix.org/conference/usenixsecurity25/presentation/ji}
\BIBentrySTDinterwordspacing

\bibitem{LogoMorph}
``It doesn’t look like anything to me: Using diffusion model to subvert
  visual phishing detectors,''
  \url{https://qingyinghao.web.illinois.edu/files/USENIX24-visual-phish.pdf}.

\bibitem{phishtank}
``Phishtank: Phishing intelligence,'' \url{https://phishtank.com/}.

\bibitem{openphish}
``Openphish: Phishing intelligence,'' \url{https://openphish.com/}.

\bibitem{puppeteer}
``{P}uppeteer | {P}uppeteer --- pptr.dev,'' \url{https://pptr.dev}, [Accessed
  13-Jun-2023].

\bibitem{stealth}
``puppeteer-extra-plugin-stealth --- npmjs.com,''
  \url{https://www.npmjs.com/package/puppeteer-extra-plugin-stealth}, [Accessed
  13-Jun-2023].

\bibitem{statcounter-resolutions}
{StatCounter}, ``{Desktop Screen Resolution Stats Worldwide, December 2020},''
  \url{https://gs.statcounter.com/screen-resolution-stats/desktop/worldwide},
  [Accessed 21-Jun-2023].

\bibitem{rafique2016s}
M.~Z. Rafique, T.~Van~Goethem, W.~Joosen, C.~Huygens, and N.~Nikiforakis,
  ``It's free for a reason: Exploring the ecosystem of free live streaming
  services,'' in \emph{Proceedings of the 23rd Network and Distributed System
  Security Symposium (NDSS 2016)}.\hskip 1em plus 0.5em minus 0.4em\relax
  Internet Society, 2016, pp. 1--15.

\bibitem{cpm_rate}
``The best cpm ad networks for publishers,''
  \url{https://publishergrowth.com/category/ad-networks/cpm-ad-networks}.

\bibitem{levenshtein1966binary}
V.~I. Levenshtein, ``Binary codes capable of correcting deletions, insertions,
  and reversals,'' \emph{Soviet physics doklady}, vol.~10, no.~8, pp. 707--710,
  1966.

\bibitem{reimers2019sentence}
N.~Reimers and I.~Gurevych, ``Sentence-bert: Sentence embeddings using siamese
  bert-networks,'' in \emph{EMNLP}, 2019.

\bibitem{wang2020minilm}
W.~Wang, F.~Wei, L.~Dong, H.~Bao, N.~Yang, and M.~Zhou, ``Minilm: Deep
  self-attention distillation for task-agnostic compression of pre-trained
  transformers,'' \emph{NeurIPS}, 2020.

\bibitem{lin2004rouge}
C.-Y. Lin, ``Rouge: A package for automatic evaluation of summaries,'' in
  \emph{Text summarization branches out}, 2004, pp. 74--81.

\end{thebibliography}

\appendices
\label{sec:appendix}

\section*{Appendix}
\setcounter{section}{0}
\renewcommand\thesection{\Alph{section}}

\makeatletter
\newcommand{\appsection}[1]{%
  \refstepcounter{section}
  \par\addvspace{1ex}
  {\noindent\normalfont\bfseries\thesection.\ #1\par}\nobreak\medskip
}
\makeatother

\appsection{Challenging BMA Examples}
\label{sec:bwa_fn_fp_ex} 


Table~\ref{Tab:diff_ex} presents examples of misclassified pages that illustrate the challenges of generalizing across evolving BMA attack strategies (discussed in Section~\ref{sec:discussion}). 


\begin{table}[ht]
  \caption{Examples of misclassified BMA pages.}
  \label{Tab:diff_ex}
  \centering
  \small
  \begin{tabular}{@{}cc@{}}
    \begin{tabular}{c}
      \frame{\includegraphics[width=0.22\textwidth, height=2.75cm, keepaspectratio]{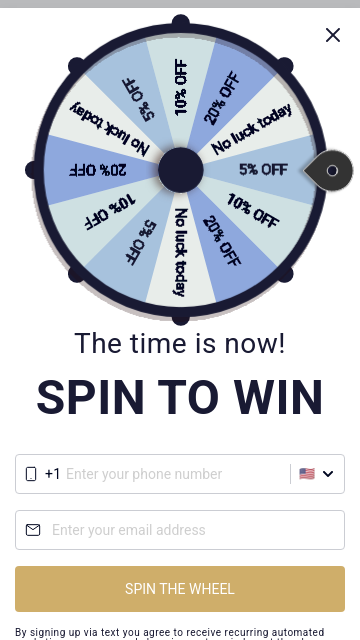}} \\
      (1) - False Positive \\
    \end{tabular}
    &
    \begin{tabular}{c}
      \frame{\includegraphics[width=0.22\textwidth, height=2.75cm, keepaspectratio]{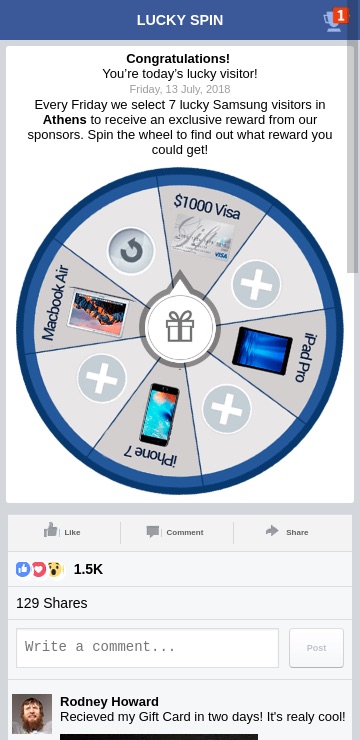}} \\
      (2) -  Malicious Example
    \end{tabular}
    \\
    \begin{tabular}{c}
      \frame{\includegraphics[width=0.22\textwidth, height=2.75cm, keepaspectratio]{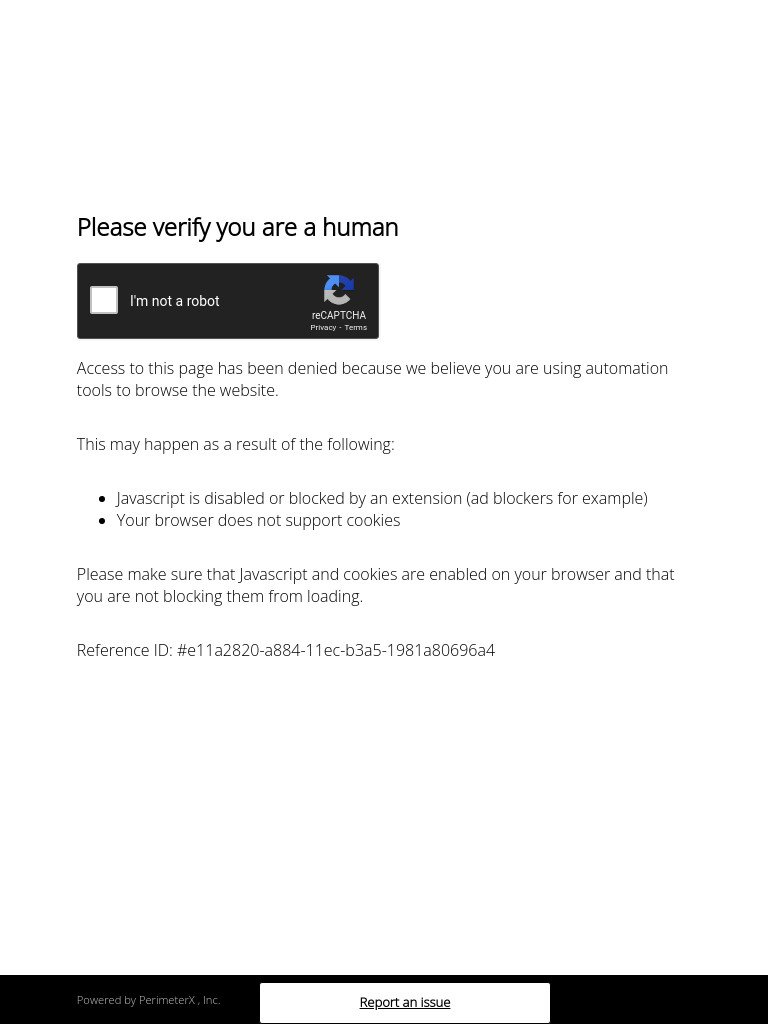}} \\
      (3) - False Positive \\
    \end{tabular}
    &
    \begin{tabular}{c}
      \frame{\includegraphics[width=0.22\textwidth, height=2.75cm, keepaspectratio]{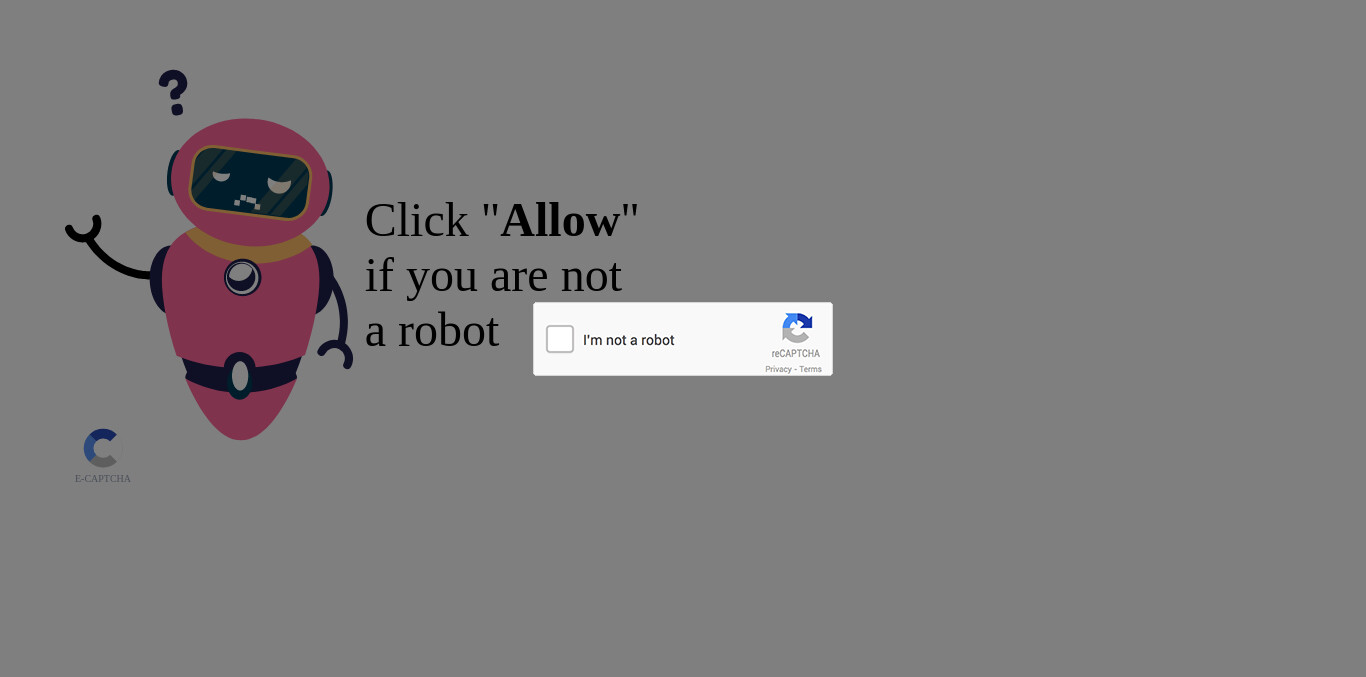}} \\
      (4) - Malicious Example
    \end{tabular}
    \\
    \begin{tabular}{c}
      \frame{\includegraphics[width=0.22\textwidth, height=2.75cm, keepaspectratio]{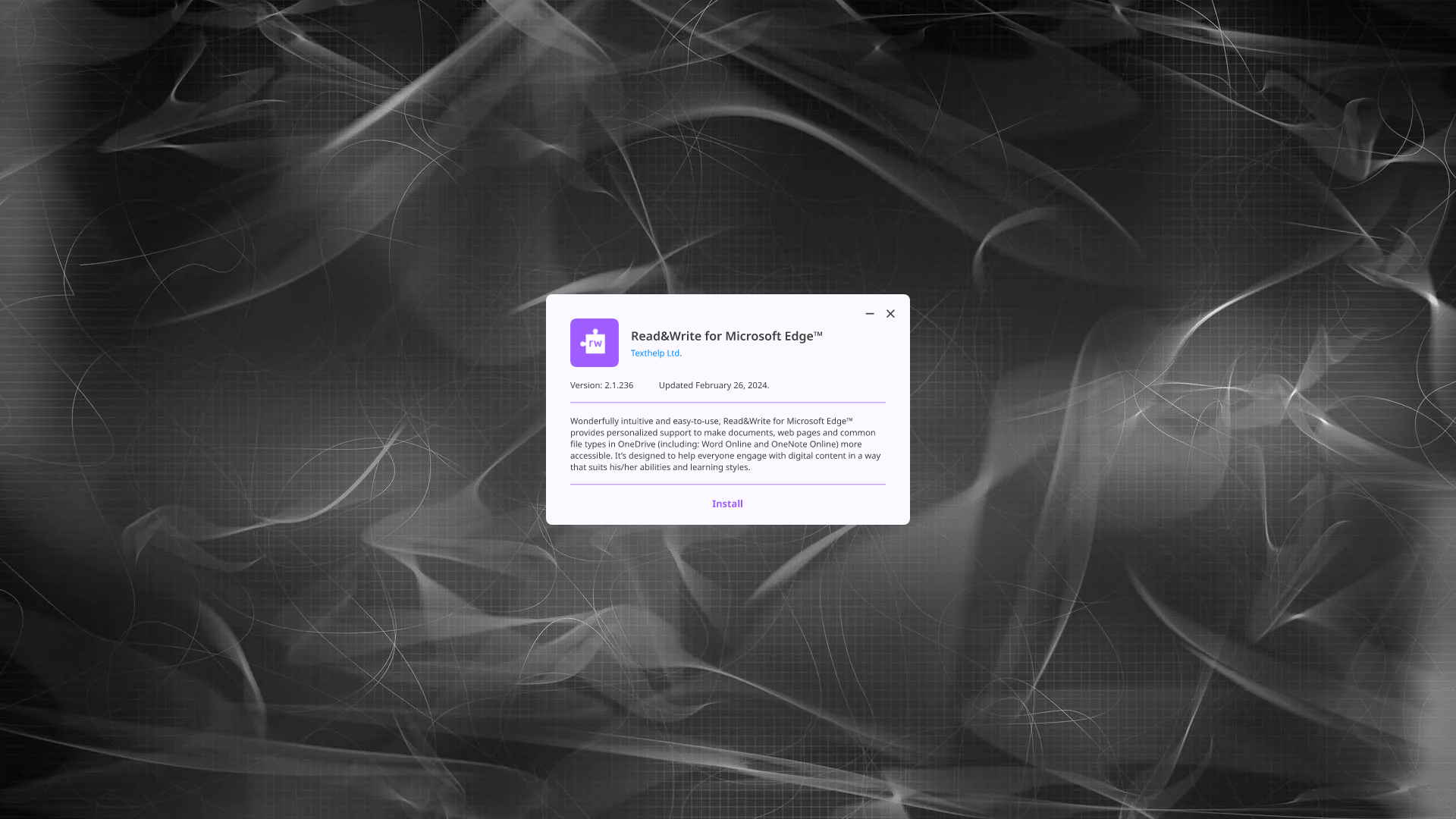}} \\
      (5) - False Negative \\
    \end{tabular}
    &
    \begin{tabular}{c}
      \frame{\includegraphics[width=0.22\textwidth, height=2.75cm, keepaspectratio]{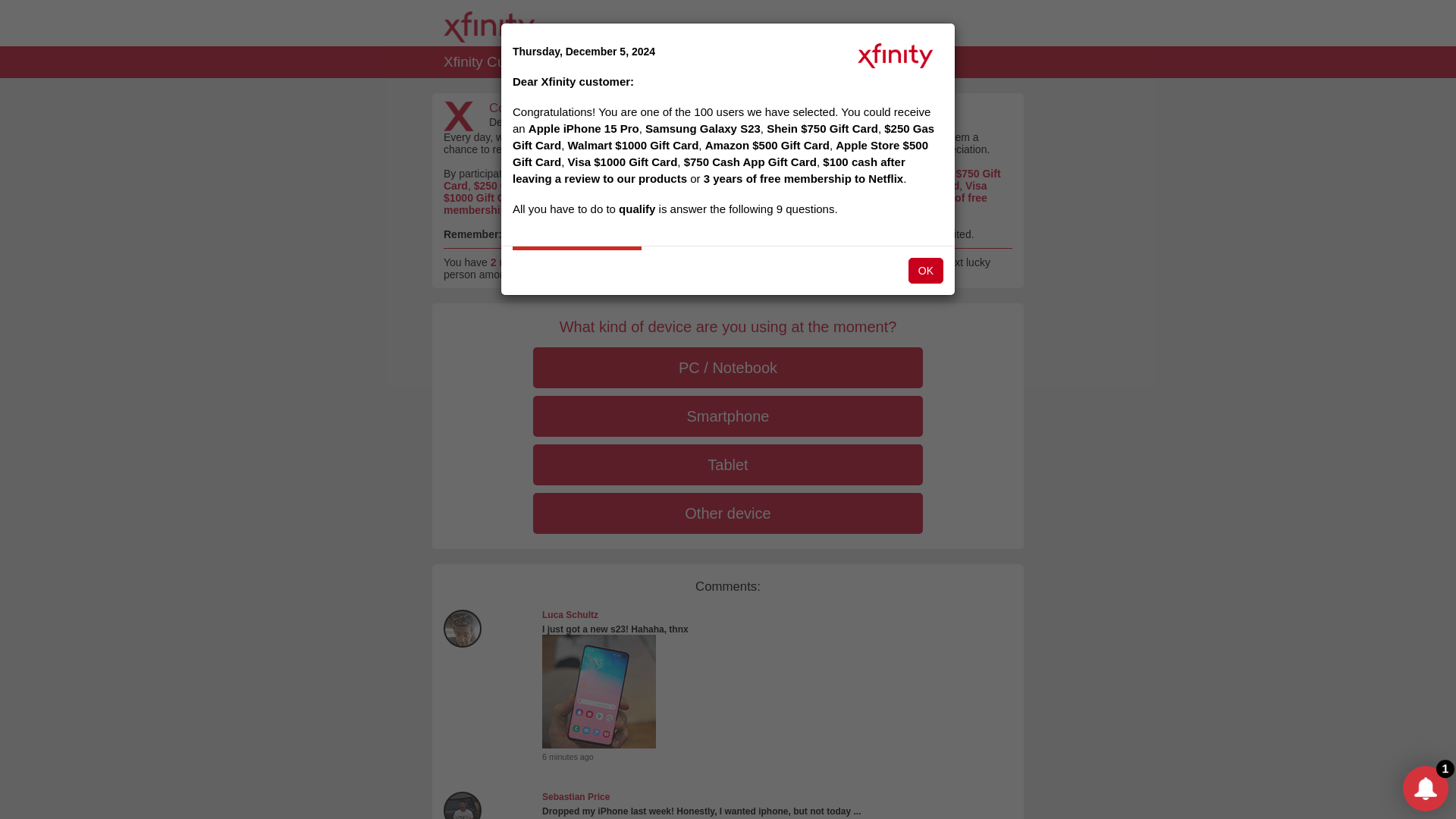}} \\
      (6) - False Negative \\

    \end{tabular}
  \end{tabular}
\end{table}


\appsection{Pixel Patrol Discover Details}
\label{sec:ppdis}

While research on phishing defenses can leverage URL feeds such as PhishTank~\cite{phishtank} and OpenPhish~\cite{openphish} to collect ground truth data, to our best knowledge, there exist no analogous feeds or repositories that would allow us to readily collect fresh samples of BMA. Therefore, we had to build our own system, called \ppdis (\ppdisabv), for discovering and collecting recent examples of such attacks in the wild, by extending the BMA campaigns discovery approach proposed in~\cite{Vadrevu_IMC19}. Although \ppdisabv is not the main contribution of our work, we introduce important improvements that enable better data collection and more systematic labeling, which we describe below.

Briefly, the approach we use is based on a {\em crawler farm} that is seeded with URLs that are likely to lead to BMAs. A crawler consists of an instrumented browser that loads a seed URL and automatically interacts with web pages in an attempt to trigger a redirection to a BMA. A number of heuristics are used to pilot the browser to collect {\em landing pages} that have a high likelihood of being related to BMAs. As these landing pages are visited, the crawler records the related URL and a screenshot. Then, to discover BMA campaigns and filter out noise (i.e., non-BMA pages), a webpage screenshot clustering algorithm is used, which is based on the intuition that BMAs are typically orchestrated into BMA campaigns that distribute the same (or very similar) attacks across multiple malicious domain names.

We re-implement~\cite{Vadrevu_IMC19} using Puppeteer~\cite{puppeteer} and extended it to add the following improvements: 

\noindent -- \textit{Improved crawler heuristics:} In~\cite{Vadrevu_IMC19}, ``click-worthy'' page elements were prioritized by sorting them simply in decreasing element size order (i.e., based solely on the size of an element on the rendered webpage). To allow \ppdisabv to more quickly reach BMA content, we empirically derived additional heuristics. For instance, we prioritize interacting with page elements (e.g., buttons, images, links, etc.) that include keywords such as {\em update}, {\em download}, {\em play}, etc. (we used a total of 38 keywords often found on pages that lead to BMAs). 
Additionally, we added the use of~\cite{stealth} to improve the ``stealthiness'' of our crawlers (i.e., make it more difficult for pages to detect browser automation) and yield a larger variety of BMAs.

\noindent -- \textit{Screen size diversity:}
Because one of our main goals is to develop a practical defense that can be deployed on devices with widely different screen resolutions, we significantly extended the number of devices and screen sizes emulated by our crawlers, from 5 to 30; To select the screen sizes and devices to be emulated, we relied on external statistics~\cite{statcounter-resolutions} as well as results from a prior IRB-approved user study (from an unrelated research project) involving 400 MTurk users that allowed us to derive the most popular viewport sizes and screen aspect ratios to be embedded in our \ppdisabv system.

\noindent -- \textit{\label{labeling1}Systematic labeling process:}
Unlike in~\cite{Vadrevu_IMC19}, we use a systematic labeling approach. Namely, we label potential BMA campaigns and remove non-BMA pages by using three different human labelers with expertise on web security and social engineering attacks who were tasked with labeling each cluster of webpage screenshots as either {\em BMA Campaign} or {\em benign} (i.e., not BMA).  

After the labelers independently completed their tasks, they met to discuss and
attempt to resolve possible label disagreements. In the rare cases in which
consensus on the label to be assigned could not be found, they marked the
related cluster as {\em unknown} and excluded it from further consideration.
This systematic data labeling process allowed us to collect higher quality
BMA examples. We computed the inter-rate reliability score using Krippendorff's alpha score~\cite{krippendorff_computing_2011} among the labelers and obtained $\alpha=0.82$, which indicates a high level of agreement~\cite{de2012calculating}.
Besides collecting and labeling examples of BMAs, we also used our \ppdisabv to collect benign webpages under 30 different screen resolutions by seeding it with urls as described in Section~\ref{benign_wp_col}.

\appsection{Crawler Setup}
\label{sec:ppdis_crawler_setup}
We use Docker to run many parallel instances of
\ppdisabv's instrumented browser. Each instance represents an isolated and
clean environment that does not retain any session information related
to crawling previous seed URLs. We deploy our crawlers on a server with 24 CPU
cores running Ubuntu Linux, where we run 15 crawler instances at a time. For any
given seed URL, we visit it using multiple instances of our instrumented
browser, each configured to emulate one of 10 different browser/device
combinations, including different smartphones, tablets, laptops, etc., and
browsers that render pages with different viewport sizes (the browser/OS
combinations we use include Firefox on Windows, Safari on Mac, Edge on Windows,
Chrome on Windows, Chrome on Linux, Chrome on Mac, Chrome on an Android Phone,
Safari on iPhone, Safari on iPad, and Chrome on different Android tablets).
Additionally, to make our crawling as stealthy as possible and circumvent the
anti-crawler mechanisms implemented by some ad networks, we used
Puppeteer's ``stealth'' libraries~\cite{stealth}.


\appsection{Ethical considerations}
\label{sec:ethical_consideration}
\ppdisabv uses an approach similar to recent research~\cite{Vadrevu_IMC19,WebPushAds,TRIDENT,rafique2016s}. Like previous works, it is possible that our crawler inadvertently clicked on ads belonging to legitimate websites. As a result, genuine advertisers might face a small expense due to our system's data collection activity, since they are likely to compensate a third-party publisher and an ad network for their services (it is important to note that we do not profit financially from these activities). This scenario is not unique to our study but is also observed in other similar works. Consequently, we have explored the ethical implications of our research by drawing on established precedents~\cite{Vadrevu_IMC19,WebPushAds,TRIDENT}. 

To ensure that our actions do not adversely affect legitimate third parties, we assessed the financial impact our crawler might have on advertisers and determined that the effect is minimal. Specifically, we calculated the number of clicks that led our crawler to navigating to new domain names that were different from the visited domain where the click occurred. As it is difficult to determine if the navigation was related to clicking on an ad, we {\em overestimate} the number of ad-related navigations by considering every navigation to a third-party domain as an ad click. We then calculated the cost associated with visiting each landing domain by taking into account the number of times our system navigated to that domain, by applying the maximum Cost Per Mille (CPM) rate of \$5~\cite{cpm_rate}. On average, each domain was visited 56 times by our system, resulting in an (overestimated) average expense of \$0.28 per domain (or per advertiser). Given these minimal costs, we consider the impact on advertisers to be acceptable, especially in light of the research benefits: \ppdisabv enabled the collection of a large, in-the-wild BMA dataset and facilitated the development of a novel and effective in-browser defense. We will also share the collected dataset with the research community to foster further research.

\appsection{Additional data}

Table~\ref{tab:ad_networks} presents additional details
about our data collection. Table~\ref{tab:ad_networks}(a) lists the ad networks we
used to find seed URLs, whereas Table~\ref{tab:imres}(b) shows a detailed breakdown
of our datasets in terms of number of images per screen resolution.

\begin{table}[H]
  \caption{Additional dataset details. Left: Seed URLs by ad network. Right: Samples per screen resolution for the master dataset.}
  \begin{center}
  \begin{minipage}[t]{0.45\linewidth}
      \centering
      \small
      \caption*{(a)}
      \begin{adjustbox}{width=\textwidth}
      \begin{tabular}{c|c}
          \toprule
          Ad Network & Seed URLs \\ \midrule
          Adroll     & 21716 \\
          Adblade    & 2161 \\
          Adpushup   & 373 \\
          Adsupply   & 205 \\
          Pop Ads    & 178 \\
          Adcash     & 147 \\
          AdMaven    & 118 \\
          Ad4Game    & 42 \\
          AdReactor  & 24 \\
          Adsense    & 15 \\
          \bottomrule
      \end{tabular}
      \end{adjustbox}
      \label{tab:ad_networks}
  \end{minipage}
  \hspace{0.10\linewidth}
  \begin{minipage}[t]{0.25\linewidth}
      \centering
      \small
      \caption*{(b)}
      \begin{adjustbox}{width=\textwidth}
      \begin{tabular}{@{}ccc@{}}
          \toprule
          Resolution & Benign & BMA \\ \midrule
          360x640    & 8730  & 642   \\
          360x740    & -     & 4062  \\
          414x896    & 8660  & 54    \\
          750x1334   & 1476  & -     \\
          768x1024   & 8885  & 123   \\
          800x1280   & 8653  & 135   \\
          1024x768   & 840   & 54    \\
          1280x800   & 740   & 87    \\
          1200x803   & 1481  & -     \\
          1366x677   & 52    & 21    \\
          1366x720   & 213   & 42    \\
          1366x724   & 60    & 12    \\
          1366x728   & 2800  & 891   \\
          1366x738   & 214   & 45    \\
          1366x741   & 96    & 21    \\
          1366x768   & 351   & 111   \\
          1478x837   & 167   & 75    \\
          1536x816   & 253   & 147   \\
          1536x824   & 2757  & 1200  \\
          1536x826   & 138   & 54    \\
          1536x834   & 394   & 198   \\
          1536x864   & 149   & 51    \\
          1785x993   & -     & 8901  \\
          1858x1053  & 164   & 60    \\
          1858x1080  & 144   & 45    \\
          1920x998   & 384   & 90    \\
          1920x1032  & 292   & 111   \\
          1920x1040  & 3162  & 1251  \\
          1920x1050  & 473   & 156   \\
          1920x1052  & 145   & 36    \\
          1920x1080  & 47918 & 774   \\
          1920x1097  & 209   & 87    \\ \midrule
          \textbf{Total} & \textbf{100000} & \textbf{19536} \\
          \bottomrule
      \end{tabular}
      \end{adjustbox}
      \label{tab:imres}
  \end{minipage}
  \end{center}
\end{table}

\appsection{Text Augmentation Details}
\label{sec:text_aug_dets}

To augment the OCR-extracted text data used in training, we employed Groq's cloud API with their LLaMA-3.3-70B-Versatile model to perform large-scale, batched synonym-based data augmentation. This approach aimed to increase the diversity of the training text while preserving the original intent and structure of each sample.

\emph{Batch Augmentation Pipeline.}
We implemented a fully asynchronous Python pipeline using the \texttt{asyncio} and \texttt{aiofiles} libraries to maximize throughput and minimize I/O bottlenecks. Text files were grouped into batches of ten and sent in parallel to Groq's API while respecting a rate limit of 100 requests per minute. A queue-based locking mechanism ensured compliance with Groq's rate-limiting policies. Each batch request was constructed by inserting up to ten input texts into a single structured prompt. The prompt instructed the model to perform keyword-based synonym replacement (focused on verbs, nouns, and adjectives) while explicitly avoiding cleanup of potentially nonsensical or adversarial content. The prompt also requested that outputs be labeled clearly as ``Text 1:", ``Text 2:", etc., to facilitate automated parsing of the response.

\emph{Prompt Format.}
The core prompt used in each batch was structured as follows:

\begin{tcolorbox}[colback=gray!5!white, colframe=black!75!white,
  boxrule=0.5pt, arc=2pt, fontupper=\ttfamily\mytiny, width=\linewidth]
\raggedright
Text 1: <original text 1>\\
Text 2: <original text 2>\\
...\\
Text 10: <original text 10>\\

\vspace{3mm}

Please perform data augmentation on the texts above. Modify each text while preserving its overall meaning and structure. Use keyword-based synonym replacement, focusing on semantically important words (e.g., verbs, nouns, adjectives). Do not clean up or correct any nonsensical phrases-leave them unchanged. Each augmented response should be unique and written in plain text. Please do not add any other information, notes, or text other than what is described below. Clearly label each output as 'Text \#:' matching the input numbering.\\

\vspace{3mm}

Example: Text 1: This is an example response.
\end{tcolorbox}

The model's response was parsed using regular expressions to extract the modified text for each numbered input. This method enabled scalable, semantic-preserving augmentation of BMA texts, significantly increasing lexical variability.

\appsection{Adversarial Setup and Evaluation Details}
\label{sec:adv_dets}

\textbf{Adversarial Text Generation.} To generate adversarial textual examples, we use OpenAI's \texttt{gpt-4.1-mini} model via the official API. Our generation script reads raw OCR-extracted text files and produces five adversarial versions per input, each targeting a specific perturbation level. The five-tier perturbation scheme is designed to progressively degrade model performance while remaining plausible to a human reader.

We format each API prompt to include a detailed description of the five perturbation levels. Each batch includes one text sample (batch size = 1) to remain within the model's token limits and ensure precise control over the output. Responses are parsed and written to disk by level.

\textbf{Prompt Format.} The core prompt (split across 4 prompt boxes to conserve space) used in each batch was structured as follows:

\vspace{5pt}
\begin{tcolorbox}[colback=gray!5!white, colframe=black!75!white,
  boxrule=0.5pt, arc=2pt, fontupper=\ttfamily\mytiny, width=\linewidth]
\raggedright
You~are~an~adversarial~text~generator. Your task is to perturb the input text to fool downstream models such as classifiers or OCR engines. Generate five versions of the input text, each with a different level of perturbation as defined below.
\end{tcolorbox}

\begin{tcolorbox}[colback=gray!5!white, colframe=black!75!white,
  boxrule=0.5pt, arc=2pt, fontupper=\ttfamily\mytiny, width=\linewidth]
\raggedright
Perturbation Levels:
Level 1 - Minimal (light noise): Minor character-level changes. Mostly readable and retains original meaning. Mimics accidental typos or OCR noise. (e.g., swap adjacent letters, lowercase/uppercase inconsistency, "O" to "0", "l" to "1") \\
Level 2 - Mild (readable but noisier): Combine character-level noise with small paraphrases or word substitutions. Still semantically faithful but noticeably altered. \\
\end{tcolorbox}

\begin{tcolorbox}[colback=gray!5!white, colframe=black!75!white,
  boxrule=0.5pt, arc=2pt, fontupper=\ttfamily\mytiny, width=\linewidth]
\raggedright
Level 3 - Moderate (semantics shift): Introduce paraphrasing, antonyms, misleading insertions, or reordered phrases. Meaning is partially changed; models may misclassify it.\\
Level 4 - Strong (semantic inversion): Aggressively alter meaning through word replacements, phrase reversals, and misleading information. Still somewhat readable. \\
Level 5 - Extreme (highly adversarial): Severe character and semantic distortion. Text is hard to read or nonsensical, and original meaning is obscured or inverted completely. \\
\end{tcolorbox}

\begin{tcolorbox}[colback=gray!5!white, colframe=black!75!white,
  boxrule=0.5pt, arc=2pt, fontupper=\ttfamily\mytiny, width=\linewidth]
\raggedright

Instructions:

\vspace{3mm}

1. For each level, label the output clearly as Level 1:, Level 2:, etc. \\
2. Each version must be unique. \\
3. Do not explain your changes. \\
4. Return the output in plain text only. \\
\end{tcolorbox}

This structure ensures that the model returns five labeled outputs corresponding to Levels 1 through 5, with each level representing an increasing degree of adversarial noise.

\begin{table}[b]
  \caption{
    Average perturbation metrics across different perturbation levels. Higher Levenshtein Distance indicates greater character level changes. In contrast, lower scores for Semantic Similarity and ROUGE-L F1 indicate more significant semantic and structural deviations from the original input text.
  }  
  \small
  \centering
  \begin{adjustbox}{width=0.75\linewidth}
  \begin{tabular}{@{}cccc@{}}
    \toprule
    \textbf{Level} & \textbf{Levenshtein Distance} & \textbf{Semantic Similarity} & \textbf{ROUGE-L F1} \\
    \midrule
    1 & 33.96  & 0.931  & 0.940 \\
    2 & 150.31 & 0.934  & 0.753 \\
    3 & 194.96 & 0.884  & 0.684 \\
    4 & 211.80 & 0.815  & 0.600 \\
    5 & 273.61 & 0.264  & 0.124 \\
    \bottomrule
  \end{tabular}
  \end{adjustbox}
  \label{tab:adv_text_metrics}
\end{table}

\begin{figure}[b]
  \captionsetup[subfigure]{labelformat=empty}
  \begin{subfigure}[t]{.15\textwidth}
    \centering
     \frame{\includegraphics[width=\linewidth]{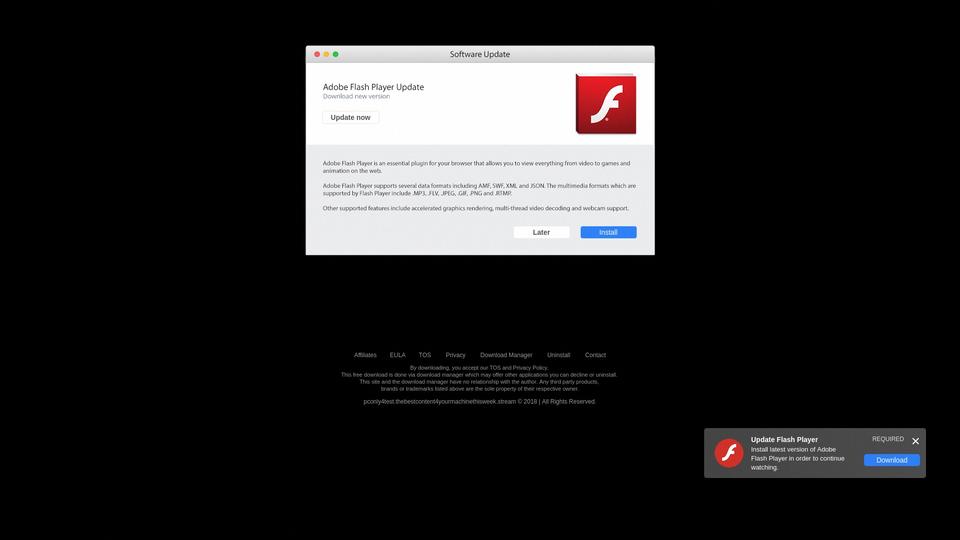}}  
    \caption{$\epsilon = \nicefrac{2}{255}$}
    \label{fig:1_PGD}
  \end{subfigure}\hfil 
  \begin{subfigure}[t]{.15\textwidth}
    \centering%
    \frame{\includegraphics[width=\linewidth]{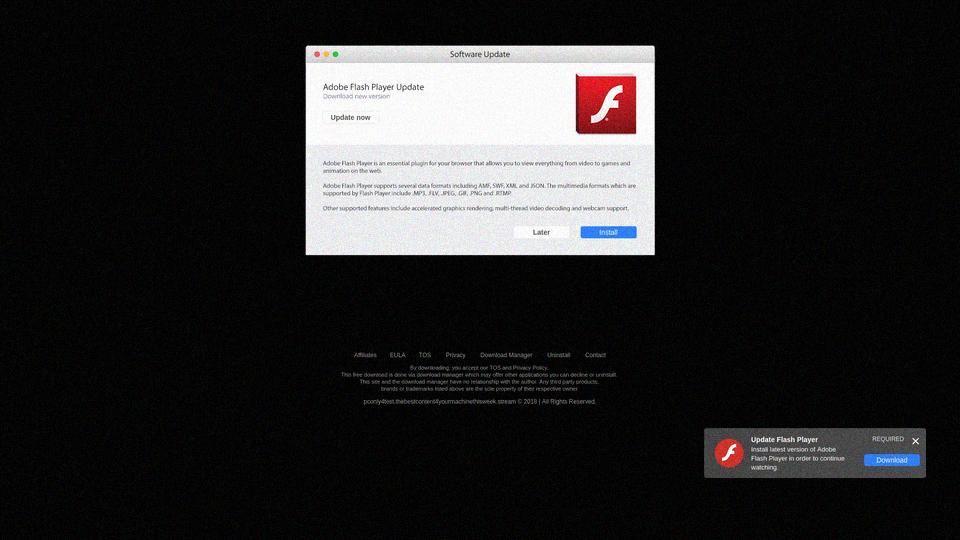}} 
    \caption{$\epsilon = \nicefrac{16}{255}$}
    \label{fig:2_PGD}
  \end{subfigure}\hfil 
  \begin{subfigure}[t]{.15\textwidth}
      \centering
      \frame{\includegraphics[width=\linewidth]{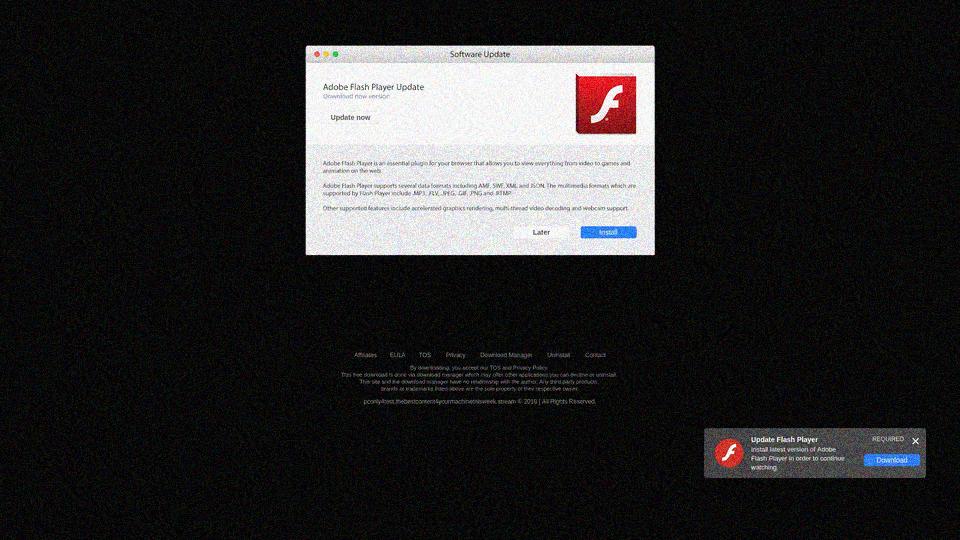}}  
      \caption{$\epsilon = \nicefrac{32}{255}$}
      \label{fig:3_PGD}
  \end{subfigure}\hfil 
  \newline   
  \begin{subfigure}[t]{.15\textwidth}
    \centering
    \frame{\includegraphics[width=\linewidth]{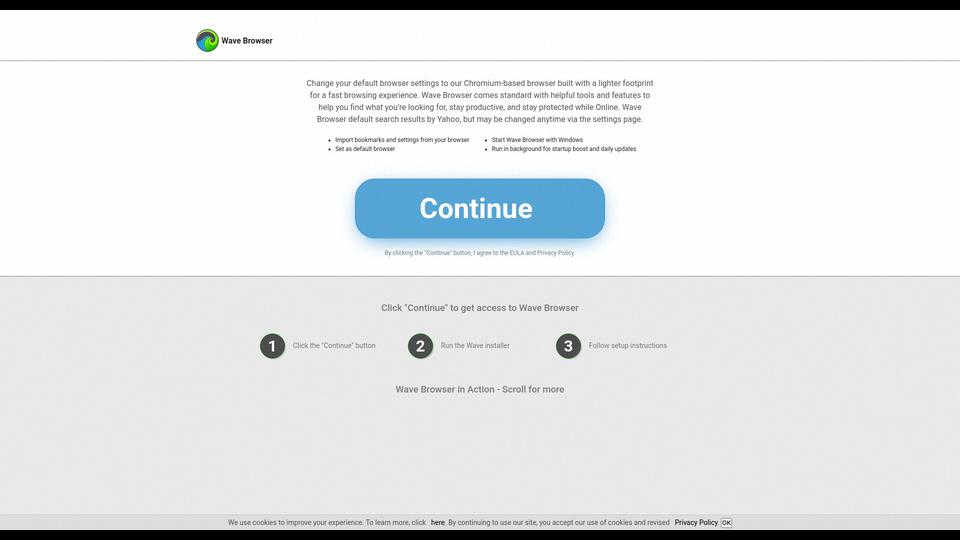}}
    \caption{$\epsilon = \nicefrac{2}{255}$}
    \label{fig:4_PGD}
  \end{subfigure}\hfil 
  \begin{subfigure}[t]{.15\textwidth}
    \centering
    \frame{\includegraphics[width=\linewidth]{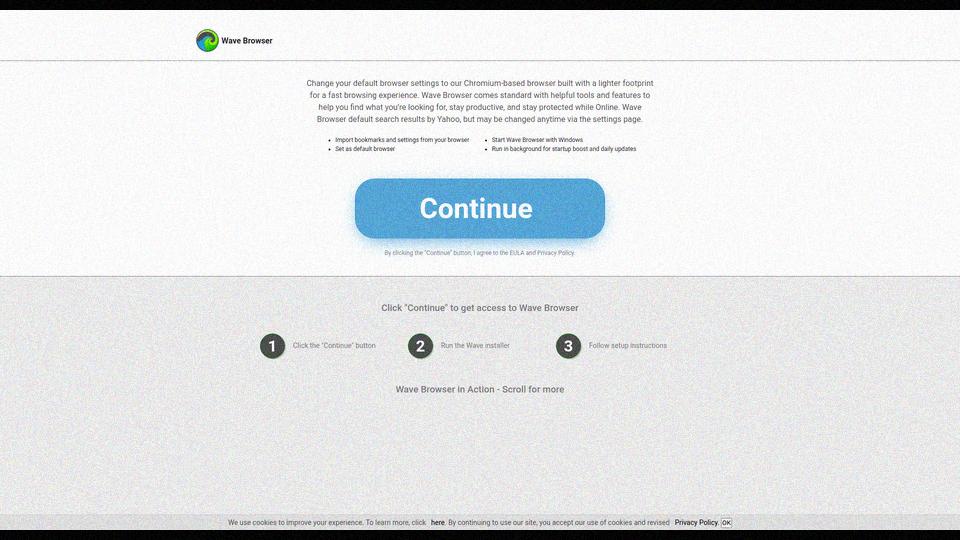}} 
    \caption{$\epsilon = \nicefrac{16}{255}$}
    \label{fig:5_PGD}
  \end{subfigure}\hfil 
  \begin{subfigure}[t]{.15\textwidth}
      \centering
      \frame{\includegraphics[width=\linewidth]{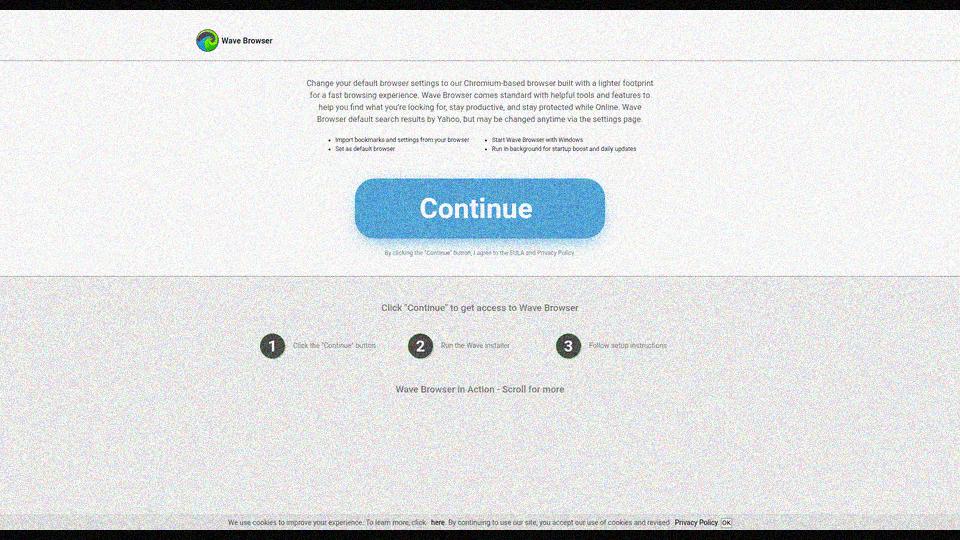}} 
      \caption{$\epsilon = \nicefrac{32}{255}$}
      \label{fig:6_PGD}
  \end{subfigure}\hfil 

  \caption{Adversarial examples generated using PGD at increasing perturbation strengths ($\epsilon$).}
  \label{fig:PGD}
\end{figure}

\textbf{Evaluation Metrics.} To quantify the perturbations, we compute three metrics between each adversarial version and its original text:

\begin{itemize}
  \item Levenshtein Distance~\cite{levenshtein1966binary} — A character-level edit distance that measures how many insertions, deletions, or substitutions are required to transform the original text into the perturbed version.
  \item Semantic Similarity~\cite{reimers2019sentence, wang2020minilm} — Cosine similarity between sentence embeddings generated using the \texttt{all-MiniLM-L6-v2} model from \texttt{SentenceTransformers}, measuring semantic distortions between text variants.
  \item ROUGE-L F1~\cite{lin2004rouge} — A token-level metric commonly used in summarization, capturing overlap between longest common subsequences.
\end{itemize}

\textbf{Evaluation Pipeline.} We sample 1,000 shared text examples across all perturbation levels. For each example, we compute the Levenshtein distance, semantic similarity, and ROUGE-L score between the original and each adversarial variant.

\textbf{Quantitative Results.} Table~\ref{tab:adv_text_metrics} summarizes the average perturbation metrics for each level. These results confirm that perturbation strength increases consistently with each level. Level 1 introduces minor noise with minimal semantic degradation, while Level 5 causes substantial semantic drift and textual distortion. This quantitative validation supports the use of these adversarial levels for training and evaluation of robustness in \ppdetabv.

\textbf{Adversarial Image Generation Examples.} Figure~\ref{fig:PGD} shows examples of images perturbed using PGD at various levels of~$\epsilon$. At $\epsilon = 16/255$, the introduced noise becomes clearly perceptible to the human eye.

\end{document}